\documentclass[11pt,a4paper]{JHEP3}

\usepackage{amsmath,epsfig}
\usepackage{amssymb,amsfonts}
\usepackage{latexsym}
\usepackage[latin1]{inputenc}
\usepackage{slashed}
\usepackage{empheq}
\numberwithin{equation}{section}
\usepackage{dsfont}
\usepackage{cancel}
\usepackage{overpic}
\usepackage{subcaption}
\usepackage{subeqnarray}
\usepackage{xcolor}
\let\normalcolor\relax
\usepackage{graphicx}
\usepackage{longtable}
\usepackage{color}
\usepackage{multirow}
\usepackage{marvosym}
\usepackage{bbm}
\usepackage[normalem]{ulem}

\newcommand{\exclude}[1]{}
\def\a#1{\alpha_{#1}}

\def\beq{\begin{equation}}
\def\eeq{\end{equation}}
\def\be{\begin{equation}}
\def\ee{\end{equation}}
\def\bea{\begin{eqnarray}}
\def\eea{\end{eqnarray}}
\def\bal{\begin{align}}
\def\eal{\end{align}}

\def\2b2[#1,#2][#3,#4]{\left( \begin{array}{cc} #1 & #2 \\ #3 & #4 \end{array}
\right)}
\def\3b3[#1,#2,#3][#4,#5,#6][#7,#8,#9]{\left( \begin{array}{ccc} #1 & #2 #3 \\
#4 & #5 & #6\\#7&#8&#9\end{array} \right)}

\newcommand\fverb{\setbox\pippobox=\hbox\bgroup\verb}
\newcommand\fverbdo{\egroup\medskip\noindent%
                        \fbox{\unhbox\pippobox}\ }
\newcommand\fverbit{\egroup\item[\fbox{\unhbox\pippobox}]}

\newcommand{\bear}{\begin{eqnarray}}

\newcommand{\eear}{\end{eqnarray}}

\newcommand{\bsea}{\begin{subeqnarray}}
\newcommand{\esea}{\end{subeqnarray}}
\newbox\pippobox

\def\f{\varphi}

\def\6{\partial}
\def\a{\alpha}

\def\ff{\phi}

\def\m{\mu}
\def\n{\nu}
\def\r{\rho}
\def\s{\sigma}

\def\sp{\;\;\;,\;\;\;}

\def\z{\zeta}
\def\sq
\def\a{\alpha}

\def\ff{\varphi_f}
\def\ft{\varphi_t}
\def\Vz{V_0}
\def\>{\rangle}
\def\<{\langle}

\title{Revisiting Coleman-de Luccia transitions in the AdS regime using holography}

\author{Jewel K.~Ghosh$^{a,b}$, Elias Kiritsis$^{c,d}$, Francesco Nitti$^c$, Lukas T.~Witkowski$^{e}$
~\\
$^a$ \href{http://www.iub.edu.bd/}{Independent University Bangladesh (IUB)}, \\
Plot 16, Block B, Aftabuddin Ahmed Road, Bashundhara R/A, Dhaka, Bangladesh\\
~\\
$^b$ \href{https://www.icts.res.in/}{International Centre for Theoretical Sciences}, Tata Institute of Fundamental Research, \\
Shivakote, Bengaluru 560089, India\\
~\\
$^c$ \href{http://www.apc.univ-paris7.fr}{APC, AstroParticule et Cosmologie}, Universit\'e de Paris, CNRS/IN2P3, CEA/IRFU,
Observatoire de Paris,\\
10, rue Alice Domon et L\'eonie Duquet, 75205 Paris Cedex 13, France\\
~\\
$^d$ \href{http://hep.physics.uoc.gr}{Crete Center for Theoretical Physics}, Institute for Theoretical and Computational Physics,
Department of Physics,  Voutes University Campus, \\
University of Crete, 70013, Heraklion, Greece\\
~\\
$^e$ \href{http://www.iap.fr/}{Institut d'Astrophysique de Paris}, GReCO, UMR 7095 du CNRS et de Sorbonne Universit\'e, 98bis boulevard Arago, 75014 Paris, France
}

\preprint{CCTP-2021-01
\\ ITCP-IPP-2021/1
}

\abstract{Coleman-de Luccia processes for AdS to AdS decays in Einstein-scalar theories are studied. Such tunnelling processes are interpreted as vev-driven holographic RG flows of a quantum field theory on de Sitter space-time. These flows do not exist for generic scalar potentials, which is the holographic formulation of the fact that gravity can act to stabilise false AdS vacua. The existence of Coleman-de Luccia tunnelling solutions in a potential with a false AdS vacuum is found to be tied to the existence of  exotic RG flows in the same potential. Such flows are solutions where the flow skips possible fixed points or reverses direction in the coupling. This connection is employed to construct explicit potentials that admit Coleman-de Luccia instantons in AdS and to study the associated tunnelling solutions. Thin-walled instantons are observed to correspond to dual field theories with a parametrically large value of the dimension $\Delta$ for the operator dual to the scalar field, casting doubt on the attainability of this regime in holography. From the boundary perspective, maximally symmetric instantons describe the probability of symmetry breaking of the dual QFT in de Sitter. It is  argued that, even when such instantons exist, they do not imply an instability of the same theory on flat space or on $R\times S^3$.}

\begin{document}

\maketitle 

\section{Introduction and summary of results} \label{sec:intro}
Systems with multiple ground states are common in many areas of
physics. The  transitions  from a higher energy ground state
(false vacuum) to a lower energy one (true vacuum)   can
proceed via tunnelling, with the
formation of a bubble which subsequently expands. As was understood by
Coleman, this process is described by  instanton solutions in the
Euclidean theory \cite{Coleman:1977py}.  The Euclidean instanton provides
initial conditions for the subsequent Lorentzian  evolution. In the absence of gravity, such
solutions always start in the false vacuum in the far past  and end in
the true vacuum in the far future.

The story becomes richer and more complex in the presence of
gravity. In
\cite{CdL}, the geometry mediating vacuum decay
in the context of General Relativity was first studied. The corresponding
solution is generally referred to as the Coleman-de Luccia (CdL)
instanton.
When coupling a field theory to general relativity, the
values of the ground state energies (and not only the energy
difference) become important.  In particular, it makes a big
difference if the end point of the vacuum decay process has a negative
vacuum energy, compared to the case of a positive or zero vacuum
energy. In the latter case, the true vacuum space-time has the geometry
of de Sitter or Minkowski, whereas in the former case it  has anti-de
Sitter (AdS) geometry.

As already noticed in \cite{CdL}, decays to de Sitter
or Minkowski space are not qualitatively very different from the
corresponding processes in the absence of gravity. This is not the case
for decays to AdS for at least two reasons: 1) Tunnelling is not always
allowed, i.e.~gravity can stabilize the false vacuum, and, 2) when tunnelling
is allowed, the generic endpoint of the process (when continued to Lorentzian
signature) is not the true vacuum AdS space-time, but it is rather
an open Friedmann-Robertson-Walker (FRW)  universe which
 initially expands but is expected to
undergo a big crunch singularity.

The problem of vacuum decay from/to AdS space-time has received a
renewed attention in recent years, for two main reasons. On the one
hand, the string theory landscape of vacua displays a (very) large
number of AdS solutions with different values of the cosmological
constant. If one thinks of each of these vacua as a local extremum of a
(loosely defined) potential energy landscape, the question whether these vacua can decay to lower and lower ones by
bubble nucleation, and the corresponding decay rates, is very
important for phenomenology.

On the other hand, the AdS/CFT correspondence relates space-times with
asymptotically AdS geometry to lower-dimensional asymptotically
conformal quantum field theories (QFTs). One can then try
to use the duality to understand vacuum decay from the perspective of
the dual QFT. Furthermore, attempts were made to give a QFT
description of the singularities in the Lorentzian ``crunching AdS''
space-time, with the hope of giving a non-perturbative resolution to
cosmological singularities (see e.g.~\cite{Hertog:2004rz,Hertog:2005hu,Barbon:2011ta,Marolf:2010tg}).

At the same time, the AdS/CFT interpretation poses a
few puzzles to the problem of vacuum decay. As emphasized in \cite{Banks:2002nm,Banks:2012hx}, the gravitational path integral must be performed with
fixed boundary conditions at the AdS boundary. If these correspond to
the false vacuum AdS, then a complete vacuum decay can never occur,
since it would have to be mediated by a solution  which in the future
approaches asymptotically to the true vacuum AdS. This indicates that
CdL-like solutions in AdS cannot always be given an interpretation in
terms of vacuum decay, but rather have to be interpreted as
holographic RG flows defined on curved space-times, a point of view
which is also put forward in \cite{Maldacena:2010un}.

From the technical standpoint, computing the decay rate of the false
vacuum is often a difficult task. Given the Euclidean field equations  for the metric $g_{ab}$ and the other
dynamical  fields $\f_i$,  one first has to find a solution
$\{\bar{g}_{ab}, \bar{\f}_i\}$ which interpolates between the false vacuum
and the true vacuum.\footnote{Or rather, a field value close to the
  true vacuum.}
The tunnelling rate is then given semiclassically\footnote{This semiclassical expression is valid approximately when $S_E[\{\bar{g}_{ab}, \bar{\f}_i\}]\gg 1$.} by
\be \label{intro1}
\Gamma = A e^{-S_E[\{\bar{g}_{ab}, \bar{\f}_i\}]}
\ee
where the exponent is given by the Euclidean action evaluated on the
solution, and $A$ is a pre-factor (which is a  one-loop determinant).

In order to perform this task, two hypotheses were made in
\cite{CdL}, and are often used in the literature since then.
The first one is that the tunnelling is described by a
solution with maximal symmetry. In the Euclidean case, for a
$D$-dimensional space-time, this is an $O(D)$-invariant instanton of the form
\be \label{intro2}
ds^2  = d\xi^2 + a^2(\xi) d\Omega_{d}^2 \, , \qquad \f_i = \f_i(\xi) \, ,
\qquad \xi\in [0, +\infty] \, ,
\ee
where $d= D-1$ and $d\Omega_d^2$ is the metric on the $d$-dimensional
unit sphere. The solution must be regular at the center $\xi=0$, where
$a(\xi \rightarrow 0) \rightarrow  \xi$, and has to approach the false vacuum metric as
$\xi \to +\infty$. This solution describes a bubble of the (near-) true vacuum
 (interior region) inside a space-time which asymptotically approaches the false vacuum.

The second assumption  which is widely used is the so-called {\em
  thin-wall} approximation: this means that the region  where the
solution departs significantly from either vacuum is
a thin shell around some value $\xi_0$, whose thickness is small
compared to its radial size, $a(\xi_0)$. Under this assumption, it is
often possible to obtain an analytic expression  for the decay
rate, \cite{CdL}. Moreover, in this approximation, it is common practice to replace
the original problem (with dynamical variables including various fields
beyond gravity) by a much simpler problem in which one freezes the
fields to their vacuum values on either side of a codimension-one
hypersurface (the bubble wall). The problem is now that of finding a
pure gravity solution on each side of the wall, gluing them using
Israel's junction condition, then extremizing the action with respect
to the remaining variable, i.e.~the position of the wall. In certain
examples, in top-down AdS/CFT models related to $\mathcal{N}=4$ super Yang-Mills
theory, the thin wall was identified with a D$_3$-brane separating AdS
solutions with different units of RR flux \cite{Horowitz:2007pr,Barbon:2010gn}.

While the thin-wall approximation is very useful in practice, it is
not always evident under which conditions it can be applied. In the
absence of gravity, the thin-wall approximation holds when the
potential energy difference between the true and false vacua is small
compared to the height of the barrier separating them. However in the
presence of gravity, and particularly when tunnelling to AdS, things
are not so clear, and this is one of the issues that we shall analyze in this work.

In this paper, we define the gravitational theory using the holographic correspondence. We shall therefore consider transitions in the AdS regime.
We perform a detailed study of CdL  solutions without
relying on  the thin-wall limit, in an
AdS/CFT setup in $D=d+1$ dimensions, which consists of Einstein
gravity minimally coupled to  a
single\footnote{This is for simplicity. Although the multiscalar case can be involved, its structure of solutions has been studied in detail, \cite{multi}, and our qualitative conclusions are expected to hold more generally.}  scalar field $\f$.
The  scalar  field  potential $V(\f)$  has several
extrema where the potential itself takes negative values, each
corresponding to  an AdS solution. This gravitational theory,  we assume to be holographically dual to a QFT. In this theory  we focus on Euclidean $O(D)$-invariant solutions of the form (\ref{intro2}), which describe tunnelling
towards the lowest energy AdS minimum, starting either from another
(higher) AdS minimum or from an AdS maximum.

Starting from an AdS false vacuum has the advantage that one can in
principle give a holographic interpretation to both the starting point
and the end point of the tunnelling process.  In particular, this makes
the boundary problem at $\xi =\infty$ in the metric (\ref{intro2})
well defined.

In the dual, $d$-dimensional field theory, each
extremum corresponds to a different conformal field theory.  The usual
holographic interpretation of a gravity solution interpolating between
two AdS extrema along the radial direction is that of a field theory
RG flow between conformal fixed points. Near each fixed point, the
geometry is well approximated by the AdS boundary region (UV fixed
point) or the AdS interior geometry (IR fixed point). The scalar field
corresponds to an operator responsible for breaking conformal
invariance, either because its (relevant) coupling is non-zero, or
because it acquires a vacuum expectation value.

The above interpretation  usually refers to RG flows of
field theories defined in flat (Euclidean or Lorentzian)
space-time. Here however the situation is different.  Since the
constant-$\xi$ slices are spheres all the way to the boundary of
AdS space-time at $\xi=+\infty$, according to the holographic dictionary the
dual UV field theory is defined on  a $d$-dimensional sphere ($S^d$) in the Euclidean case, and on
$d$-dimensional de Sitter space (dS$_d$) in the Lorentzian case. The solutions
(\ref{intro2}) can therefore be interpreted as holographic RG flows
of QFTs defined on a $d$-dimensional space-time of constant positive curvature, with
the UV fixed point corresponding to the false vacuum theory. In this
case, as we shall review below, the flow cannot reach the true vacuum
(i.e.~the lower AdS minimum) but at the endpoint  $\xi \to 0$, it
stops at a field value $\f_0$ where $V'(\f_0) \neq 0$, \cite{Ghosh:2017big}. In the
subsequent evolution in real time, the scalar field continues to roll
towards the true minimum but it is not guaranteed (and in fact, as we
shall see, generically it does not occur) that it settles in the true
minimum.

In order to describe {\em vacuum} tunnelling, the false vacuum theory
must correspond to a conformal fixed point, without any relevant
couplings turned on. This is automatically the case when we consider
tunnelling from an AdS minimum: in such cases the scalar corresponds to
an irrelevant operator in the dual field theory, whose QFT coupling  must
necessarily be set to zero in the UV. If one starts from an
AdS maximum instead, the scalar field is dual
 to a relevant operator and in general one has the choice
of turning on the corresponding  coupling. If the coupling is
non-zero, we are in the presence
of a source-driven RG flow, in which conformal invariance is
explicitly broken in the UV. If instead  the coupling is set to zero,
one can still have a non-trivial flow (which we call a vev-driven
flow), where the breaking of conformal invariance is spontaneous,
and the boundary conditions are the same as in the fixed-point
theory. It is only the latter case which corresponds to vacuum tunnelling, since
these solutions are in the same class (satisfy the same boundary
conditions\footnote{It should be clear that when we talk about AdS boundary conditions,
 we mean the value of the source term in the asymptotic expansion near the AdS boundary.}) as the pure AdS solution which sits at the fixed point.

Based on the discussion above, studying AdS to AdS vacuum decay via $O(D)$
CdL instantons is equivalent to constructing vev-driven holographic
RG flows on constant positive curvature space-times. Such solutions
were recently  studied in detail in
\cite{Ghosh:2017big}.

Vacuum decay of/to AdS, in connection to the AdS/CFT correspondence, has been widely considered in the literature \cite{CdL,Hertog:2003xg,Hertog:2004rz,Hertog:2005hu,Horowitz:2007pr,Barbon:2010gn,Harlow:2010az,Maldacena:2010un,Barbon:2011ta}.
In  this paper,  we  make  use of the results  of
\cite{Ghosh:2017big},  to draw general  results about AdS to
AdS vacuum decay in Einstein-scalar theories, without relying on the thin-wall
approximation.  In this context, we shall address the question of the
existence (or not) of CdL bubbles in theories with generic scalar
potentials, the existence of  a thin-wall limit, and the fate of the
real-time solution inside the bubble. We shall analyse these questions
in general, and with the help of numerical analyses in a few explicit
(but generic) examples.

In  the rest of this section, we briefly summarize our results.\footnote{All the results listed below apply to the case we analyse here, i.e.~tunnelling via maximally symmetric instantons. It is possible that considering instantons with a lower degree of symmetry will lead to different conclusions. We leave this as an open question.}

\paragraph*{When is AdS vacuum decay possible?}
A question we address is under what circumstances $O(D)$-instanton solutions mediating AdS decay exist in a potential.

\begin{figure}[t]
\centering
\begin{subfigure}{.5\textwidth}
 \centering
   \begin{overpic}
[width=1.0\textwidth]{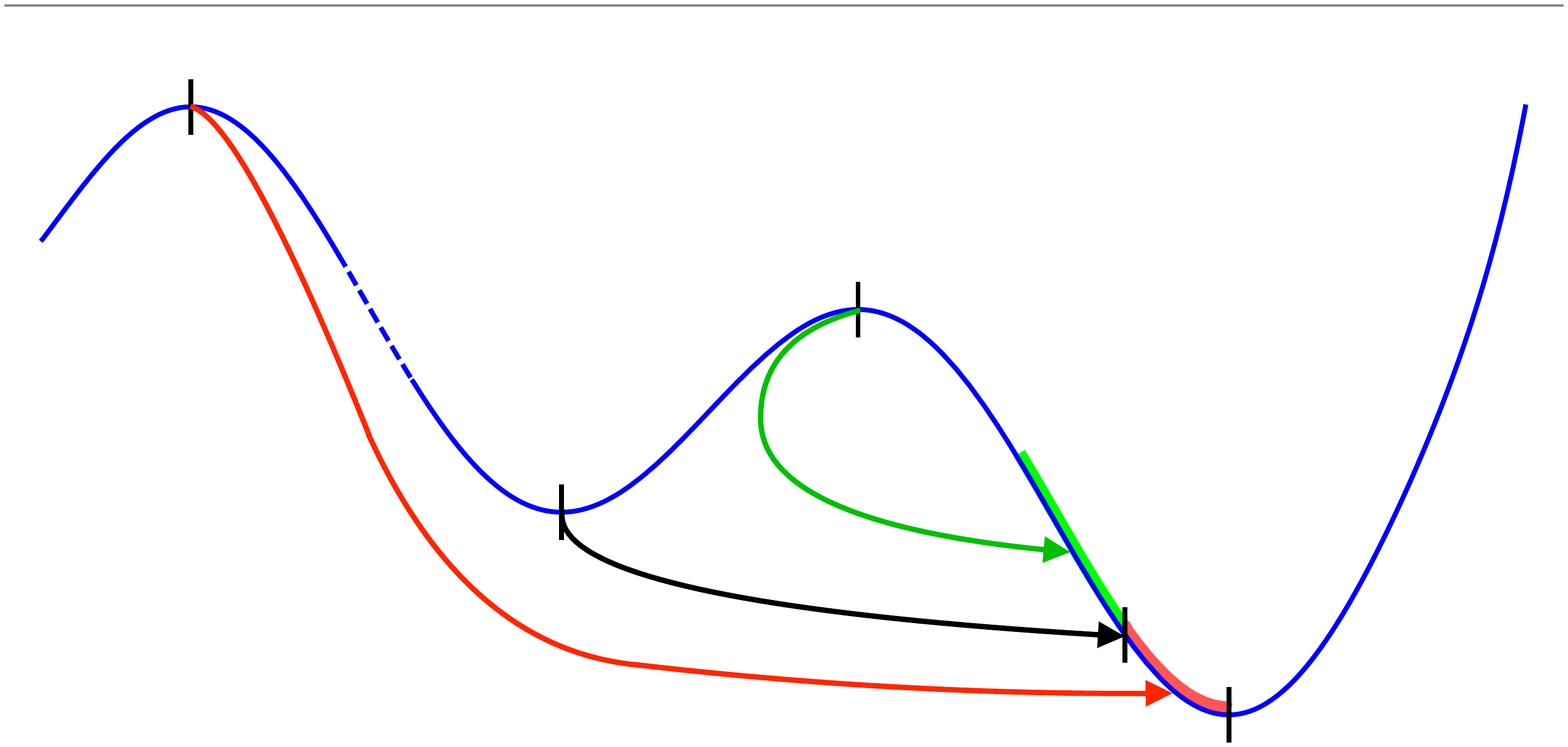}
\put(-2,56){$0$}
\put(7.5,44){$\f_{{\textsc{uv}_1}}$}
\put(34,30.5){$\f_f$}
\put(52,42){$\f_{{\textsc{uv}_2}}$}
\put(70,23){$\f_0$}
\put(75.25,17.5){$\f_t$}
\put(82,48){$V(\f)$}
\end{overpic}
\caption{\hphantom{A}}
\label{fig:skipintro}
\end{subfigure}%
\begin{subfigure}{.5\textwidth}
 \centering
   \begin{overpic}
[width=1.0\textwidth]{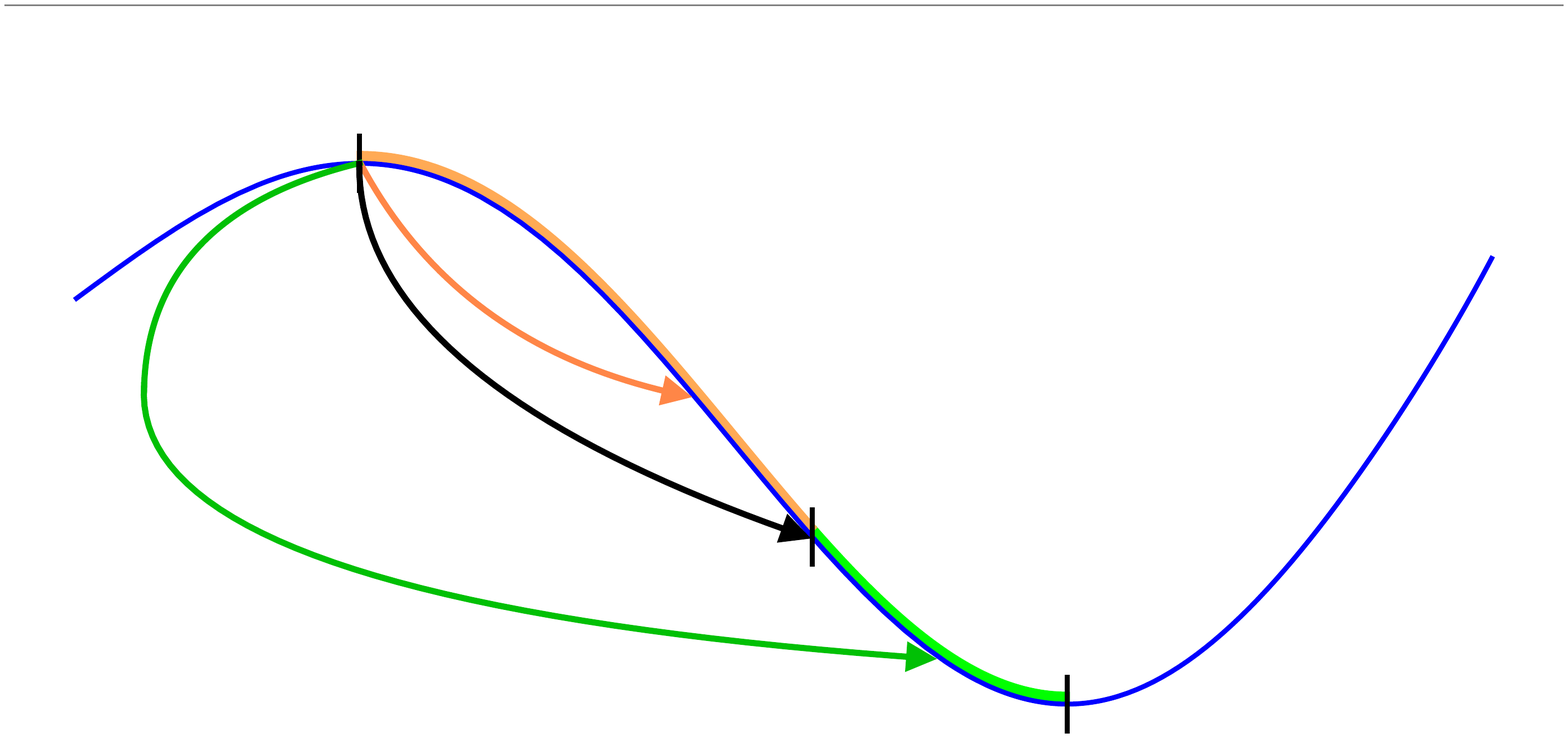}
\put(100,56){$0$}
\put(21.3,52){$\f_f$}
\put(49.5,30){$\f_0$}
\put(65,19){$\f_t$}
\put(80,44){$V(\f)$}
\end{overpic}
\caption{\hphantom{A}}
\label{fig:bounceintro}
\end{subfigure}%
\caption{\textbf{(a)} Potential $V(\f)$ with two minima at $\f_f$ (false vacuum) and $\f_t$ (true vacuum) and two maxima at $\f_{\textsc{uv}_1}$ and $\f_{\textsc{uv}_2}$. The potential admits an $O(D)$-instanton describing tunnelling from $\f_f$ to $\f_0$ (black arrow). This potential will then also admit holographic RG flow solutions from the UV fixed point at $\f_{\textsc{uv}_1}$ to an end point in the red region that skips past the other maximum at $\f_{\textsc{uv}_2}$. This is a so-called skipping flow of \cite{Kiritsis:2016kog,Ghosh:2017big} (red arrow). In addition, the potential will allow for flows leaving $\f_{\textsc{uv}_2}$ to the left before changing direction and flowing to an end point in the green region. This is a so-called bouncing flow of \cite{Kiritsis:2016kog,Ghosh:2017big} (green arrow). The holographic RG flows are for a QFT defined on $S^d$. \textbf{(b)} Potential $V(\f)$ with a maximum at $\f_f$ and a minimum at $\f_t$ permitting an $O(D)$-instanton describing tunnelling from $\f_f$ to $\f_t$ (black arrow). This potential will also exhibit bouncing holographic RG flows (green arrow) from $\f_f$ to the green region, in addition to standard source-driven holographic RG flows (orange arrow). Again, the RG flows are for theories defined on $S^d$.}
\label{fig:skipbounceintro}
\end{figure}

Although it is hard to single out exactly which  {\em local} features of a
potential allow for instanton solutions (i.e.~height and curvature of
the extrema), we are able to connect the existence of CdL instantons to
another feature of the scalar potential:
the existence of {\em exotic} holographic RG flows,
\cite{Kiritsis:2016kog}. These are RG flows which are characterized by a
non-monotonic behavior of the coupling, also referred to as a `bounce'\footnote{This is not related to Coleman's ``bounce".}, and/or by `skipping'
of intermediate fixed points. Such a behavior is possible in holographic
theories, but it is not allowed in perturbative QFTs. These solutions
where found and analyzed in detailed in  \cite{Kiritsis:2016kog} in the case of
a flat-sliced bulk theory (corresponding to the dual QFT defined in
Minkowski space), and in \cite{Ghosh:2017big} and
\cite{Gursoy:2018umf,Bea:2018whf} in the case of theories on constant curvature spaces and at finite temperature,
respectively. Exotic flows exist for special classes of
potentials. Their existence is a global feature of the
potential and it is likely not possible to relate it to local
features of the extrema.\footnote{Our observation is that potentials with an AdS false vacuum in general do not admit exotic RG flows and hence tunnelling from AdS is non-generic. In contrast, in the limit where gravity decouples from the scalar dynamics (while keeping the scalar potential fixed) $O(D)$-symmetric tunnelling solutions always exist as follows from the overshoot/undershoot argument in \cite{Coleman:1977py}. This implies that as the gravitational coupling constant is `decreased', the space of potentials that admit $O(D)$-instantons is expected to increase. We thank Daniel Harlow for pointing this out to us. However, note that in this limit the AdS radius becomes parametrically large and a holographic interpretation of tunnelling along the lines of AdS/CFT becomes questionable.} 

Here, we show that $O(D)$-instantons exist if and only if
the theory admits such exotic flows. More precisely:
\begin{itemize}

\item When the false vacuum is a minimum of the potential, a CdL
  instanton exists \emph{if and only if} the potential also permits exotic RG flows for theories on $S^d$ of both the skipping and bouncing type, see fig.~\ref{fig:skipintro}. Flat-sliced RG flows also play an important role: If the potential admits a flat-sliced RG flow of skipping type, this is a sufficient condition for the potential to also permit an $O(D)$-instanton.

\item When the false vacuum is a maximum of the potential, a CdL
  instanton exists \emph{if and only if} the theory admits exotic RG flows for the theory on $S^d$ which display a bounce, see fig.~\ref{fig:bounceintro}.\footnote{The potential depicted in fig.~\ref{fig:skipintro} also permits a CdL instanton describing tunnelling from $\f_{\textsc{uv}_2}$ in virtue of the existence of the bouncing flows. We refrained from depicting this, for simplicity.}
\end{itemize}

We also find that the condition for the minima to be nearly degenerate is not sufficient for an instanton solution to exist.\footnote{Here near-degenerate means that the energy separation between the two vacua is small compared to the height of the potential barrier between them.} The requirement of near-degenerate minima was introduced in \cite{CdL} as a condition for the applicability of the thin-wall approximation. However, the near-degeneracy of minima does not guarantee that such a solution exists, which we observe here for AdS decays. This confirms and generalizes previous observations using the thin-wall approximation. Such observations presuppose that minima are nearly
degenerate and then claim that gravity tends to stabilize a metastable vacuum with
negative curvature \cite{CdL,Barbon:2010gn,Harlow:2010az}.

\paragraph*{Tunnelling rate.}When a solution of the form (\ref{intro2}) exists, we compute the
 corresponding semiclassical tunnelling rate by evaluating the bulk
 action on the solution, see (\ref{intro1}). We show on general grounds that, for
 solutions corresponding to  normalizable boundary asymptotics
 (i.e.~zero source in the dual field theory), the tunnelling rate is
 finite. In contrast, the decay rate vanishes for
 non-normalizable scalar field asymptotics due to uncancelled near-boundary
 divergences. This confirms the expectation that source-driven
 solutions do not describe vacuum decay, but only have an
 interpretation in terms of holographic RG flows. This expectation
 is completely natural from the dual field theory point of view:
to consider spontaneous decay of the fixed point CFT, all
 conformal symmetry-breaking couplings  must to be set to zero.

\paragraph*{How to construct bulk potentials which support CdL
  instantons.} Having found that AdS vacuum decay is non-generic, we describe a
method which allows to  design bottom-up bulk potentials which permit instanton
solutions. We start with a fine-tuned potential, which allows for a {\em
  flat
domain wall-solution}  interpolating between two extrema, in which the
metric is sliced by flat radial hypersurfaces,
\be \label{intro3}
ds^2 = d\xi^2 + a^2(\xi) \eta_{\mu\nu} dx^\mu dx^\nu \, , \qquad \xi\in (-\infty,+\infty)\;.
\ee
The bulk scalar obeys special  asymptotics approaching
the false vacuum in the UV:
\be \label{intro4}
\f(\xi) \simeq \f_f + \f_+ e^{\Delta \xi} \, , \qquad \xi \to -\infty \, ,
\ee
where $\f_f$ is the scalar field  value corresponding to the false
vacuum, and $\Delta >0$ is the dimension of the dual operator in the UV
CFT.   In the dual field
theory, these solutions correspond to holographic RG flows with the
source of the operator usually driving the flow set to zero.  Solutions
of this kind, that at the same time are non-singular in the interior, do not exist for generic potentials, since they have less
free parameters than the source-driven flows.\footnote{For source-driven flows the scalar
field approaches the false vacuum as $\exp(d-\Delta)\xi$, with $\Delta
< d$. They  exist  only if the false vacuum is a maximum of the potential.} This makes it generically impossible to impose the regularity conditions where the scale factor vanishes, i.e.~at the locus $a(\xi) \to 0$.  However,
it is possible to fine-tune the bulk potential in such a way that
solutions like (\ref{intro3}-\ref{intro4}), that are also regular in the interior, exist.

Starting from such a tuned potential, we show that,  by deforming it
 in an appropriate way (therefore relaxing the fine-tuning),
the resulting potential admits spherical $O(D)$-instantons. In this way,
one can obtain continuous families of potentials in which one of the AdS
extrema is unstable to $O(D)$-invariant bubble nucleation, provided we have the curved boundary conditions for the asymptotic metric.

\paragraph*{The thin-wall approximation.}
In this work we also assess the validity of the
the thin-wall approximation.

In the limit of an infinitely thin wall with non-zero tension, the approximation
consists of taking  the scalar field to be constant and
equal to the value extremizing the potential on each side of the wall,
and the metric to be exactly AdS,  with the curvature
scale determined by the value of the potential on each side. The two
AdS space-times are then connected using Israel's junction conditions
at the wall location, 
see e.g.~\cite{Barbon:2010gn,Harlow:2010az,Banerjee:2018qey}.

One of the main points of our analysis is that, independently of the
specific form of the bulk potential, there is {\em no limit} of the
Einstein-scalar setup in which
the above thin-wall setup is strictly valid, i.e.~where the CdL bubble reduces to
an infinitely thin-wall with finite tension separating two AdS vacua,
such that 1) the  two vacua have both finite (negative) potential energy
and 2) they have a finite separation in field space. The reason is
that in Einstein's equations, it is impossible to match  the Dirac
distributions that arise in this limit.\footnote{This is unlike
what happens in the absence of gravity, where the thin-wall
approximation can be understood as a consistent limiting procedure. In
this case only the jump equation in the scalar field must be
satisfied.}

As a consequence,  although in specific  examples the wall can be
parametrically thin and the transition between two AdS
space-times may be quite fast,  the scalar field dynamics
may never be completely neglected and can be important for obtaining
correct results. This is a serious caveat that has to be kept in mind
when one considers models which use this simplified picture, which are
very common in the literature in various contexts
\cite{Charmousis:2007ji,Barbon:2010gn,Harlow:2010az,Banerjee:2018qey,Banerjee:2019fzz}.

This analysis does not apply to cases when the bubble has a
different origin than a bulk scalar field domain wall, e.g.~when  the
object mediating vacuum decay is genuinely
lower-dimensional. Examples include nucleation of probe D3-branes in type
IIB string theory \cite{Barbon:2010gn}, which do not have a simple interpretation  in terms
of bulk scalars interpolating between two minima of a potential. More
generally, this applies to  cases where vacuum decay is mediated by
excitations in the open string sector.

It is nevertheless interesting to analyze in
more detail, in which cases the bubble wall is
parametrically thin (although infinite thinness cannot
be achieved in any consistent limit). To this end,  we consider a
class of analytically-engineered potentials constructed as explained
in the previous paragraph, and which depend on a few controllable
parameters:  the values of the potential at the minima and their
curvatures there, which in turn control the dimension $\Delta$ of the
operator dual to the bulk scalar in the two CFTs living at the
extrema. In this parametrization,  we study  examples
based on a bulk potential given by a sextic polynomial.
Numerical analyses of the solutions lead to the following observations:
\begin{itemize}
\item Most importantly, for a solution to describe a thin-walled bubble, the dimension $\Delta$ of the dual CFT operator both in the false and true minimum CFTs must become parametrically large.  This is needed for the interpolation between false and (near-) true minimum to be sufficiently `rapid' as characteristic for a thin wall. Hence we expect this finding to hold beyond the family of potentials studied.

\item If a thin-walled CdL instanton solution exists, we find that the
  corresponding potential has a large barrier between the two minima
  compared to the energy difference between the minima. This is in
  agreement with the heuristic observations of Coleman and de Luccia
  in their original work \cite{CdL}.

\item Computing the solution and the on-shell action numerically, we
  observe that the decay rate is faster
  for thick walls than for thin walls, as suggested in \cite{CdL}.

\item CdL instantons describing tunnelling from an AdS maximum can also be found. However, these are observed to always be thick-walled solutions. This is consistent with our previous finding as the dimension $\Delta$ of the dual CFT operator describing deformations from a maximum is bounded as $\Delta<d$ and cannot become parametrically large.

\end{itemize}
These results show that Coleman's thin-wall approximation (for
which there is always a solution based on Israel's junction condition,
if the domain wall tension is taken to be sufficiently small) may
sometimes underestimate the decay rate. Also, the fact that the
thin-walled solutions require the dimension $\Delta$ of the corresponding
CFT operator in the false and true minimum to be very large is quite
problematic: it is believed that there are upper bounds on the
lowest dimension irrelevant operator in a CFT (as suggested from results from the conformal bootstrap program, see e.g.~\cite{Dymarsky:2017yzx}). If, on the other hand, the vacuum CFT contains an
operator of lower dimension, then this corresponds to an additional
bulk scalar of lower mass, whose dynamics must be included in the
analysis, invalidating the results obtained in the single-scalar
model we are using.
We can, however, give a qualitative argument that extrapolates our results to the multiscalar case.
At a generic saddle point, one has both positive mass scalar directions (dual to irrelevant operators) and negative mass ones (dual to relevant directions).
For an instanton to exist, regular curved vev-flow solutions must exist.
Generically, several instanton solutions may be also possible. The larger the slope in the direction of the associated flow, the thinner the domain wall, and the larger the instanton action will be.
However, we expect that in the presence of several low-dimension operators, the associated vevs and flow will be mostly in their direction and the domain wall will be thick.
A detailed analysis is, however, necessary in order to establish whether this is a generic phenomenon, or if it always happens.

\paragraph*{Bubble interior and crunch singularity.}

We analyze the fate of the solution in real
time, after the bubble has nucleated.
As mentioned earlier, it has
been observed  (starting with \cite{CdL} and later with \cite{Abbott}) that, in the thin-wall
limit, the space-time inside the bubble is a cosmological space-time
which grows at first, but then crunches into a singularity. This is based on the fact that,
in the thin-wall limit, the space-time in the interior is   AdS,
which  has  a  big-crunch coordinate  singularity in the  bubble
interior. This  is expected to become a physical singularity
when deviations from AdS due to the scalar field dynamics are
included, \cite{CdL}.

Here, we show that the big crunch is {\em generically}
unavoidable.\footnote{Although it may be in principle possible
  to construct a fine-tuned potential with regular interior
  geometries. This is left as an open question.} This can be traced
to the fact that, inside the bubble, the scalar field does not reach
the true vacuum, but in real time it starts forced oscillations which
eventually destabilize the system towards a singularity.  The latter
however is hidden behind a horizon (the  cone swept by the real-time evolution  of
the center of the bubble, $\xi=0$)  which cloaks it from
the AdS boundary. Both the horizon and
the singularity reach the boundary in the  infinite future  as
measured by the boundary QFT living on de Sitter space-time. Therefore, from the
boundary QFT point of view, the solution looks like an
RG flow on $d$-dimensional de Sitter space-time at all times, and the interior of
the bubble is akin to the interior of an eternally expanding  black
hole, which however   cannot send signals to the
boundary.

 Although this qualitative picture is well established,\footnote{Based on this picture, it was suggested in
  \cite{Maldacena:2010un} that one consider
  the bubble wall as an IR (broken) CFT which encodes the interior
     degrees of freedom.} here we show
 that this is the case generically and independently of the details of
 the potential and of the thin-wall approximation.

\paragraph*{The interpretation of the solution from the dual CFT point of view.}

In the Lorentzian patch of the geometry, the near-boundary solution is that of the (false vacuum) CFT$_d$ defined on dS$_d$  space.
This is clear from the structure of the sources. The structure of the
Lorentzian solutions outside the bubble does {\em not}  correspond to
the ground state of the CFT$_d$ on dS$_d$, in which the bulk scalar
is constant. When the scalar is constant, the state is dual to the
AdS$_{d+1}$ solution in the bulk, and has the full $O(2,d)$
symmetry\footnote{Despite
the conformal anomaly, a CFT on dS$_d$ has the same amount of symmetry
than a CFT in flat $d$-dimensional Minkowski space, since dS$_d$ has
the same number of conformal vector fields as flat space (see e.g. \cite{Tod}). This is
clear from the bulk perspective, as  the flat and dS slicings of
AdS$_{d+1}$ result by a different choice of coordinates of the same
$d+1$-dimensional manifold in the embedding space $R^{2,d}$, which has
symmetry group $O(2,d)$. } of AdS$_{d+1}$.
The CdL solution corresponds  instead to a state with a scalar vev, dual to a
non-trivial (non-AdS) geometry in the bulk. The  running scalar is sourced by a non-zero vev of an irrelevant
operator and breaks (spontaneously) the full symmetry $O(2,d)$  to
$O(1,d)$, the isometry group of dS$_d$.

The instanton instability in this context is interpreted as follows.
There is a special state in the CFT$_d$ on dS$_d$ that is essentially a Hartle-Hawking state. It is defined by initial conditions on $S^{d-1}$ given by doing the CFT path-integral on a half $S^d$ with $S^{d-1}$ as a boundary.
Such a state, for weakly-coupled theories corresponds to the
Bunch-Davis vacuum.

On the gravity side, this state is dual to the analogous bulk
Hartle-Hawking state, \cite{Hartle:1983ai,HHR},  whose wave-function
is defined by the path-integral over Euclidean geometries with EAdS
boundary conditions \cite{eternal}.

Moreover, there are two semiclassical states for the CFT$_d$ on dS$_d$.
The first is the  maximal symmetry  state in which all scalar vevs  vanish.
The other is the reduced symmetry state in  which the vev of the
operator dual to the bulk scalar is non-trivial  and the symmetry is
broken as $O(2,d)\to O(1,d)$. The latter has a higher Euclidean free energy than the uniform state.

Since the Euclidean path integral is dominated by
 the two semiclassical solutions above, the Hartle-Hawking state has a
 non-zero overlap with both semiclassical states. It is mostly the
 highest symmetry,
zero vev  state, and  it has a small admixture of the non-zero vev, lower
 symmetry state. The amplitude for this admixture is given by the instanton amplitude.
Therefore, for such a ``false vacuum'' CFT$_d$, there is a small
probability for the scalar obtaining a vev and for the breaking of the
$O(2,d)$ symmetry.

\paragraph*{Vacuum decay of different boundary QFT space-time geometries.}
As we have stressed earlier, maximally symmetric instantons of  the
form (\ref{intro2}) can mediate AdS vacuum decay only when the asymptotic
boundary has the geometry of $S^d$. The dual boundary QFT is
therefore defined on a constant positive curvature space-time and from
the boundary perspective, the decay takes an infinite time, as we
discussed in the previous paragraph.

The natural question is whether the existence of the maximally
symmetric instanton allows to draw conclusions about the  same QFT on flat space, or on the cylinder $R_t\times S^{d-1}$, which in the
gravity dual would correspond to an asymptotic boundary of
Poincar\'e-AdS$_{d+1}$ and Global AdS$_{d+1}$, respectively.

 It is important to stress here that, in AdS/CFT (or more generally when dealing with semiclassical gravity in
asymptotically AdS space-times) the boundary conditions on the metric  must be included in the
 definition of  the (gravitational) path integral: only solutions
 satisfying the same set of asymptotic boundary conditions contribute
 to the same quantum path integral. Changing the boundary conditions
 amounts to turning on fluctuations which are non-normalizable near the
 AdS boundary, and correspond to a change in the external sources
 (e.g.~the background metric) of
 the dual field theory.
If we were to investigate  vacuum decay in the flat space
QFT, the $O(D)$ instanton would not be the appropriate solution,
because it does not obey the appropriate boundary
conditions.

One can nevertheless hope to say something about the theory
on  other boundary geometries, because sometimes one can change the
boundary conditions by a (large) diffeomorphism, and this changes the
dual theory by changing the sources, while still having a solution to
the bulk Einstein equation. This procedure therefore provides a semiclassical
saddle point to a {\em different} path integral.

Now, it is possible to find such a large coordinate transformation
that maps the CdL instanton (with near boundary leading behavior of
the metric  corresponding to dS$_d$) to a geometry whose metric has
leading behavior corresponding  to $R_t\times S^{d-1}$. This can be
seen as a map between the QFT on de Sitter and the QFT on the
cylinder.\footnote{We carry out the detailed analysis for the boundary
 geometry $R \times S^{d-1}$ but similar considerations also hold for a flat Minkowski boundary.} Through this map, the infinite de Sitter future is mapped to
finite global-AdS time. In particular, when mapped to the cylinder,  the crunch singularity of the CdL solution  reaches the boundary in finite global
time, at which point it is not possible to extend the solution
further.  One can think of this as being caused by an instantaneous
source  at global boundary time $t = \pi/2$,  which displaces the scalar
field by a finite amount from the false-vacuum value.

It is important to notice that the solution described above, obtained by a
coordinate transformation from the CdL solution, should not be
considered as a state in the same theory as the global AdS solution
residing at the extremum of the bulk potential. States in the same
dual field theory are identified by imposing the same asymptotic
boundary conditions on the bulk metric and scalar field. One
possibility is to choose boundary conditions compatible with the state
described by global AdS with the scalar field fixed at the
false-vacuum value. This theory corresponds to the
false-vacuum CFT on the cylinder, and has as its asymptotic boundary
geometry the whole cylinder $R_t \times S^{d-1}$. Geometries
corresponding to  states in these
theory are such  that the scalar field
 asymptotes to the false-vacuum value (with zero source term) at all times  $-\infty < t < +\infty$.

In contrast, in the solution resulting from mapping the CdL instanton
to cylinder slicing,  the scalar
field asymptotes to the false-vacuum value on the boundary (with zero source term) only for a finite time interval but then its value jumps by a finite amount instantaneously at global time
$t=\pi/2$.  One can imagine modifying the boundary conditions on the gravity
  side  in order to accommodate the cylinder-CdL geometry as an allowed
  solution. Provided this can be done
consistently, this would  define a {\em different} field theory on the
  cylinder, which does not admit the global AdS solution as one its
  states. This theory can only be extended up to a finite global time,
  at which point the time-evolution reaches a singularity. This point
  of view is the one taken in
  \cite{Barbon:2010gn}, where the CdL geometry was  interpreted in
terms of an unstable dual field theory with an unbounded Hamiltonian.

In conclusion,  the existence of the maximally symmetric instanton
solution does not  imply an  instability of the boundary
CFT on  $R_t\times S^{d-1}$. This does not necessarily imply that the theory on the
whole cylinder is absolutely stable. Decay may occur via other 
solutions with less symmetry. 

\subsection{Discussion and open questions} \label{sec:outlook}
Our results show that tunnelling by bubble nucleation between AdS vacua,
mediated by maximally symmetric instantons is non-generic. When it does occur, the
thin-walled bubble limit seems to be very problematic:  it requires a
very large leading irrelevant operator  dimension in the endpoint CFT,
and additionally, a very small
separation in field space in our examples. Therefore, decay by
thin-walled bubbles is the exception rather than the norm, when
dealing with AdS to AdS vacuum tunnelling, which as we just stated, is
rather non-generic in and of itself.

Furthermore,  although one can  define this approximation
parametrically, the thin-wall
approximation is {\em never} a mathematically consistent limiting procedure in a model with
gravity and scalars.  This leads one
to question the very widespread use of the exact thin-wall limit
to treat in a simplified way systems with gravity and scalars \cite{Charmousis:2007ji,Barbon:2010gn,Harlow:2010az,Banerjee:2018qey,Banerjee:2019fzz}. This
leaves the exact thin-wall limit to  the nucleation of lower
codimension objects such as D-branes in string theory.

From the dual QFT perspective, the maximally symmetric instantons we have
considered here necessarily require the QFT to be defined on de Sitter
space (or on a Euclidean sphere). Therefore, vacuum decay in other
geometries cannot be mediated by these processes. This is the case for
example of flat space (the boundary of the  AdS Poincar\'e patch) or the
Einstein Static Universe (the boundary of global AdS). Based on the
results for $O(D)$-symmetric spherical bubbles, one can infer nothing about vacuum
stability or decay rate in these other space-times. The existence of
lower-symmetry instantons which mediate vacuum decay on these geometry
remains an open question. What we can say  is that, if they exist,  these are {\em not}
coordinate transformations of spherical instantons.

As is well known, the interior of the bubble, in the real time
evolution after its nucleation,  is  an (open) FRW
universe. Here we have shown that, generically and unavoidably, it  ends in a big-crunch singularity.  One point to be stressed is that the singularity (and in fact the
whole FRW geometry) are hidden by a bulk horizon, which reaches the
boundary only in the infinite future. Therefore it is very
difficult to investigate the singularity from the QFT perspective.{
This is similar to the fact that it is hard  to obtain information on the interior
singularity of a black hole, although this topic is evolving.
In particular, certain quantum effects may be able to generate such transmissions, \cite{Gao:2016bin,Maldacena:2017axo,Maldacena:2018lmt}, and it is an interesting problem to study this
in the present context.}

It is appropriate here to comment on more general gravitational solutions describing RG flows dual to holographic QFTs defined on a sphere.
Such solutions generalize the solutions studied in this paper, and have been investigated, mostly in the Euclidean regime in \cite{Ghosh:2017big,F}.
There, their Lorentzian continuation has been studied and it was shown that the associated geometry involves dS slices that end in a bulk horizon.
This horizon is similar to the one described here, with the only difference that it extends to the infinite past. There is, however, the region behind the horizon, similar to the inside of the bubble in figure \ref{Penrose}. We portray the Penrose diagram in this case in figure \ref{fig13}.
 To find the fields of the orange region of \ref{fig13}, we must again solve the bulk equations with $V\to -V$ and initial conditions similar to the ones used here. Like it happened here, we expect for the same reasons that there will be singularities in the two orange quadrants the past and the future one.
It is an interesting problem to study the maximal extension of such geometries.

This work leaves some important questions open for future
investigation.  As the present analysis is limited to maximally
symmetric solutions, one is naturally  led to ask whether, in the
absence of such solutions, vacuum  decay may be mediated by
lower-symmetry instantons. This is particularly relevant for the
stability of global AdS (dual to field theories on $R\times S^{d-1}$),
where one would look naturally for instantons with $O(D-1)$
symmetry, such as those studied in \cite{Barbon:2010gn} in the
thin-wall approximation.

Another important question is the behavior of  the boundary field theory
correlators in the Lorentzian patch of the solution, and in particular
whether any effect of the crunching singularity can leave an imprint
on the boundary. A related question is to understand in detail what
the solution would look like in asymptotically global AdS coordinates,
and whether this signals a problem for the boundary QFT on  $R\times
S^{d-1}$.

Finally, in this work we have found that, in the explicit example we
considered,  the bubbles are generically not thin-walled bubbles, unless one
looks in very unnatural (from the CFT point of view) regions of
parameter space. While our example is not in any way special, it would
be interesting to investigate whether these facts apply to generic
theories. \\

This work is organized as follows. In Section \ref{sec:CDLanalytic} we present our setup, the general features of
$O(D)$ instanton solutions, their connection with holographic RG
flows of field theories on spheres, and the computation of the decay
rate.  We explain the generic (non)-existence of $O(D)$ instantons and connect
them to exotic holographic RG flows. In Section \ref{sec:flat} we show some results about flat domain walls, how
they can be constructed in holographic theories, and what the
thin-wall limit entails in this case. In Section \ref{sec:num} we return to spherical instantons and we present a
systematic procedure for obtaining these solutions from theories which
admit flat domain walls. We then turn to the numerical analysis of a
class of explicit examples, where in particular we discuss the
validity of the thin-wall approximation. In Section \ref{sec:analyticcont} we turn to the Lorentzian solution after bubble
nucleation, and we show that this generically result in a crunching
FRW space-time in the interior. In Section \ref{sec:spheretocylinder} we describe how to relate
the spherical instanton  to a solution with cylindrical
slicing, and what this implies for the dual field theory. Several technical details are left to the Appendix.


\section{Coleman-de Luccia processes and holographic RG flows} \label{sec:CDLanalytic}

\subsection{Setup: $O(D)$-instantons and flat domain-walls} \label{sec:CDLsetup}
Tunnelling processes of Coleman-De Lucia type involve spherical instantons of the gravity plus scalar field equations.
Such Euclidean solutions are relevant in a theory of Einstein gravity coupled to a scalar field theory with at least two vacua. In this work such vacua will be assumed to have both negative cosmological constant. The associated Lorentzian signature action in $D=d+1$ space-time dimensions is:
\begin{align}
\label{eq:setup1} S = M_P^{d-1} \int d^{d+1}x \, \sqrt{|g|} \bigg( R^{(g)} - \frac{1}{2} g^{\mu \nu} \partial_{\mu} \f \partial_{\nu} \f - V(\f) \bigg) + S_{\textsc{ghy}} \, ,
\end{align}
where we use the `mostly plus' convention and $R^{(g)}$ is the scalar curvature associated with the metric $g_{\mu \nu}$. As the solutions considered will exhibit boundaries, we also included the Gibbon-Hawking-York term $S_{\textsc{ghy}}$ to have a well-defined variational problem.

\begin{figure}[t]
\centering
\begin{overpic}[width=0.75\textwidth,tics=10]{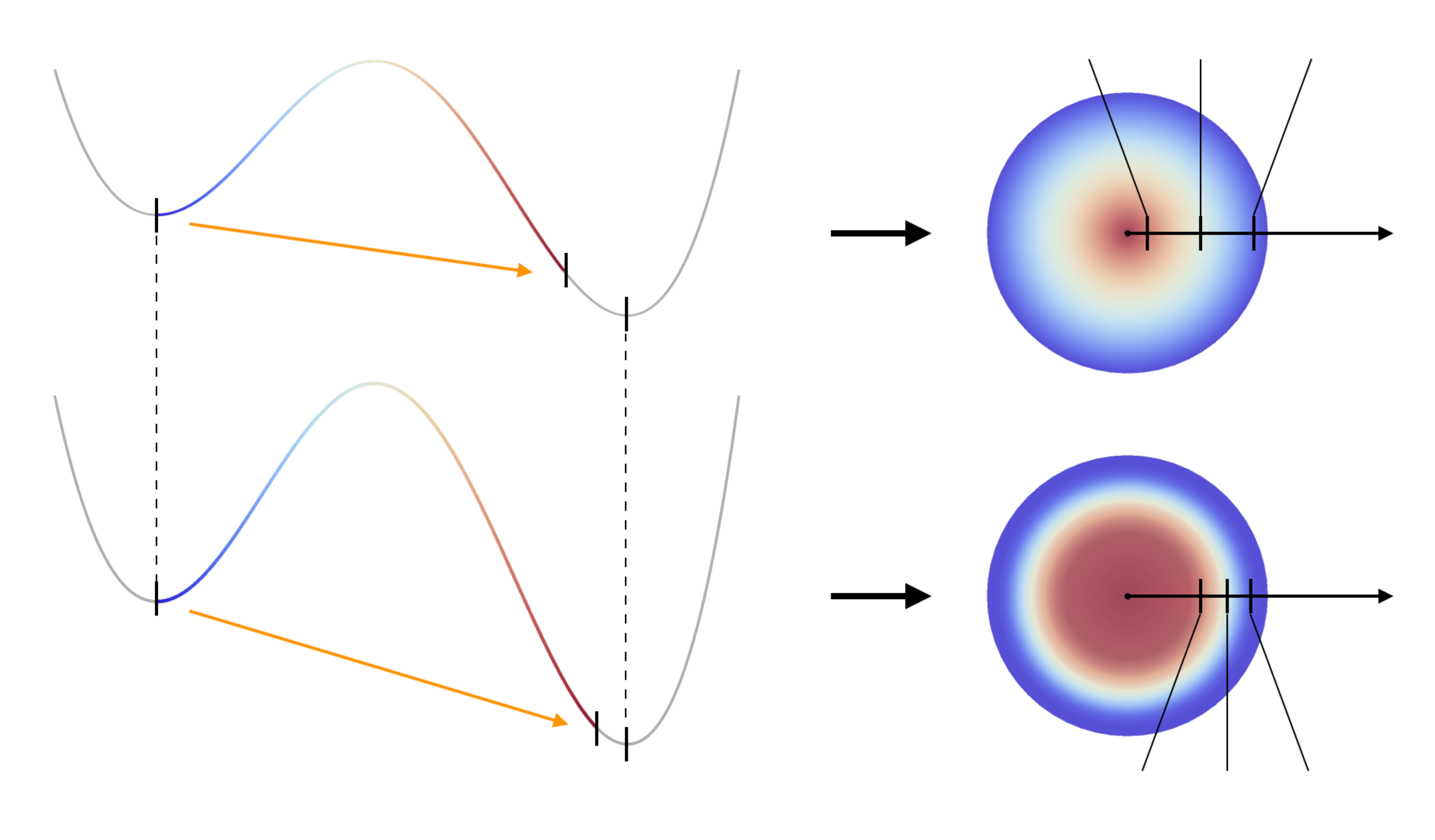}
\put(-5,51){$V(\f)$}
\put(-5,28.5){$V(\f)$}
\put(9.5,46.5){$\f_f$}
\put(42.5,40){$\f_t$}
\put(40,4){$\f_0$}
\put(37.5,35.25){$\f_0$}
\put(77.5,2.5){$r_{\textrm{in}}$}
\put(85,2.5){$\bar{r}$}
\put(91,2.5){$r_{\textrm{out}}$}
\put(73,56){$r_{\textrm{in}}$}
\put(83,56){$\bar{r}$}
\put(90,56){$r_{\textrm{out}}$}
\put(95,44){$r$}
\put(95,19){$r$}
\put(77,41){${}_{0}$}
\put(89.25,41){${}_{\infty}$}
\put(77,15.75){${}_{0}$}
\put(89.25,15.75){${}_{\infty}$}
\end{overpic}
\caption{
\textbf{LHS:} Potentials with minima at $\f_f$ (false vacuum) and $\f_t$ (true vacuum) permitting Coleman-de Luccia tunnelling solutions from $\f_f$ to a generic point $\f_0$. \textbf{RHS:} Cartoon of the corresponding $O(D)$-symmetric configuration in space-time. The coloring indicates the map of values of $\f$ in space-time. Here $r_{\textrm{in}}$ designates the inner limit of the wall, $r_{\textrm{out}}$ the outer limit of the wall and $\bar{r}$ is the center of the wall where the field $\f$ has interpolated half-way between $\f_f$ and $\f_0$. \textbf{Top:} Example with a thick wall, i.e.~a gradual transition from $\f_f$ to $\f_0$ along the radial direction $r$. \textbf{Bottom:} Example with a thin wall, i.e.~a sudden transition from $\f_f$ to $\f_0$ along the radial direction $r$.}
\label{fig:schem}
\end{figure}
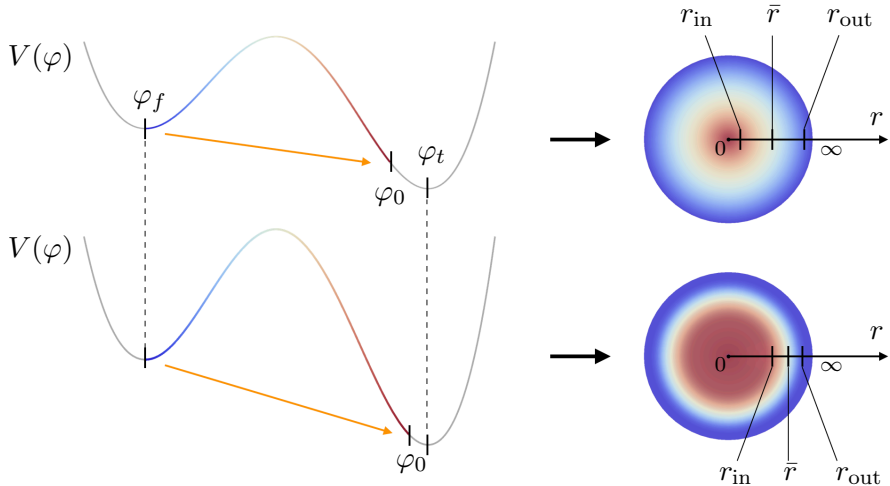

The corresponding Euclidean action is given by $S_E = -S$. The two AdS vacua in question correspond to two minima of $V(\f)$ which we choose to be located at $\f= \f_f$ (the `false' vacuum)\footnote{In a few cases that will be considered separately, $\f_f$ corresponds to an AdS maximum.} and $\f=\f_t$ (the `true' vacuum), respectively, with
\begin{align}
\label{eq:setup2} V(\f_f) \geq V(\f_t) \, , \quad \textrm{and} \quad V(\f_f), \, V(\f_t) < 0 \, .
\end{align}
See the LHS of figure \ref{fig:schem} for two example potentials. The solutions considered in this work shall be of two kinds:
\begin{itemize}
\item Coleman and de Luccia conjectured and have given sufficient arguments that the dominant instanton solutions must have $O(D)$ symmetry \cite{CdL}. Our primary focus therefore  will be on $O(D)$-symmetric solutions, which can be identified with CdL instantons describing tunnelling out of a false vacuum. These solutions also have an interpretation as holographic RG flows for a Euclidean field theory defined on a $d$-dimensional sphere.
\item In addition, we also consider flat AdS domain-wall solutions that interpolate between the two AdS minima. In holography these correspond to RG flow solutions dual to a QFT defined on $d$-dimensional flat space.
While these solutions do not describe decay processes, we still consider them here, as they turn out to be useful for understanding certain properties of the $O(D)$-symmetric solutions. The reason is that the flat domain walls can be understood as a particular limit of the $O(D)$-symmetric case, that at the same time allow for constructing explicit analytic expressions more easily.
\end{itemize}
In both cases, the metric and scalar field profile can be written as
\begin{align}
\label{eq:metricxirho} \f=\f(\xi), \quad ds^2=d \xi^2+\rho^2(\xi) \zeta_{\m\n}dx^\m dx^\n \, ,
\end{align}
with $\mu,\nu$ running over the remaining $d$ coordinates. For $O(D)$-symmetric solutions $\zeta_{\mu \nu}$ is the metric of a $d$-sphere with radius $\alpha$ and corresponding scalar curvature
\begin{align}
\label{eq:alphadef} R^{(\zeta)}=\frac{d(d-1)}{ \alpha^2} \, .
\end{align}
The coordinate $\xi$ describes a radial direction with range $\xi \in [0, \infty)$. For flat domain wall solutions $\zeta_{\mu \nu}$ is a metric of the Euclidean space $\mathbbm{R}^d$ and $\xi \in (-\infty, \infty)$ is unconstrained.

In this work, we primarily use a different set of space-time coordinates, that is more familiar from the study of holographic RG flows. These coordinates will be convenient in order to use insights from RG flows in the tunnelling context. In particular, we introduce
\begin{align}
u \equiv u_0 - \xi \, , \qquad e^{A(u)} \equiv \rho(\xi) \, ,
\end{align}
for some finite value of $u_0$. That is, the radial coordinate is now $u$ and for $O(D)$-symmetric solutions covers the range $u \in (-\infty, u_0]$. Using these coordinates the ansatz in \eqref{eq:metricxirho} becomes
\begin{align}
\f=\f(u), \quad ds^2=du^2+e^{2A(u)}\zeta_{\m\n}dx^\m dx^\n \ \label{metricansatz} \, .
\end{align}

We can obtain the (gravitational) equations of motion, by varying the action \eqref{eq:setup1} with respect to the metric and scalar field, finding
\begin{align}
\label{c3} 2(d-1) \ddot{A} + \dot{\f}^2 + \frac{2}{d} e^{-2A} R^{(\zeta)}  &=0  \, , \\
\label{c4} d(d-1) \dot{A}^2 - \frac{1}{2} \dot{\f}^2 + V - e^{-2A} R^{(\zeta)} &=0 \, , \\
\label{c5} \ddot{\f} +d \dot{A} \dot{\f} - V'  &= 0 \, ,
\end{align}
with
\be
\dot{x} \equiv dx / du \, , \qquad X' \equiv \partial X / \partial \f \, .
\ee
For the case of flat domain walls, we simply set $R^{(\zeta)}=0$ in (\ref{c3}),(\ref{c4}). Note that \eqref{c5} is identical to the (one-dimensional) equation of motion of a particle of unit mass and coordinate $\f$, performing damped motion in the inverted potential $-V$, with $u$ playing the role of time. We frequently employ this mechanical analogue for intuition in the behaviour of the system at hand.

In this work we shall be interested in $O(D)$-instantons corresponding to tunnelling from the false vacuum at $\f=\f_f$ (as defined above) to some point with scalar field value $\f=\f_0$.
These will correspond to solutions to \eqref{c3}--\eqref{c5} with the following asymptotic behavior:
\begin{align}
\label{eq:UVBC} O(D)\textrm{-instantons:} \quad & \f \rightarrow \f_f \, , \ \dot{\f}\rightarrow 0 \, , \ \rho=e^ A \rightarrow \infty \, , \quad \textrm{for} \quad u \rightarrow -\infty, \ \xi \rightarrow + \infty \, , \\
\nonumber &\f \rightarrow \f_0 \, , \ \dot{\f}\rightarrow 0 \, , \ \rho=e^ A \rightarrow 0 \, , \quad \ \,  \, \textrm{for} \quad u \rightarrow u_0, \ \, \xi \rightarrow 0 \, .
\end{align}
The flat domain-wall solutions studied here will interpolate between the extrema at $\f_f$ and $\f_t$. Therefore, the relevant {asymptotic behavior is:}
\begin{align}
\label{eq:IRBC} \textrm{Flat domain-walls:} \quad & \f \rightarrow \f_f \, , \ \dot{\f}\rightarrow 0 \, , \ \rho=e^ A \rightarrow \infty \, , \quad \textrm{for} \quad u \rightarrow -\infty, \ \xi \rightarrow + \infty \, , \\
\nonumber &\f \rightarrow \f_t \, , \ \, \dot{\f}\rightarrow 0 \, , \ \rho=e^ A \rightarrow 0 \, , \quad \ \, \textrm{for} \quad u \rightarrow +\infty, \ \xi \rightarrow - \infty \, .
\end{align}

In the following, we also examine under which circumstances $O(D)$-instantons are of the `thin-wall' type. By construction, $O(D)$-solutions are configurations corresponding to concentric circles of physical radius $r= \alpha \rho= \alpha e^ A$, with $\f$ constant on each spherical shell, but interpolating between its boundary values (here $\f_f$ and $\f_0$) along the radial direction. The wall is identified as the interval in the radial direction where the bulk of this interpolation happens. We refer to the wall as thin if the physical width of the wall is small compared to the radius at the center of the wall. See figure \ref{fig:schem} for two cartoons of $O(D)$-solutions with a thick and a thin wall, respectively. We shall find it useful to make this quantitative by introducing a parameter $\eta$ that measures the thickness (or rather the `thin-ness') of the wall.

First, we define the radius at the center of the wall $\bar{r}$ as the radius at the locus where $\f$ has interpolated half way between its boundary values $\f_f$ and $\f_0$, i.e.
\begin{align}
\bar{r} \equiv \alpha e^{A(\bar{u})} \, , \quad \textrm{with} \quad \f(\bar{u}) = \frac{\f_f + \f_0}{2} \, .
\end{align}
The wall itself denotes the region where the interpolation between $\f_f$ and $\f_0$ effectively occurs. There is no universal definition for this and here we make a choice. We define the wall as the interval $[u_{\textrm{in}}, u_{\textrm{out}}]$ with
\begin{align}
\label{eq:phiinoutdefOD} \f(u_{\textrm{in}}) = \frac{\f_f + \f_0}{2} - \gamma \frac{\f_f - \f_0}{2} \, , \quad \f(u_{\textrm{out}}) = \frac{\f_f + \f_0}{2} + \gamma \frac{\f_f - \f_0}{2} \, ,
\end{align}
with a parameter $\gamma <1$ that controls how close $\f(u_{\textrm{in}})$ is to $\f_0$ and $\f(u_{\textrm{out}})$ to $\f_f$.
In all practical examples we choose $\gamma=0.76$.\footnote{This implies that $\f(u_{\textrm{in}}) = \f_0 + 0.12 \, (\f_f - \f_0)$ and $\f(u_{\textrm{out}}) = \f_f + 0.12 \, (\f_0 - \f_f)$.} We can then define $r_{\textrm{out}} = \alpha e^{A(u_{\textrm{out}})}$ and $r_{\textrm{in}} = \alpha e^{A(u_{\textrm{in}})}$ as the radii corresponding to the outer and inner edge of the wall. Putting everything together, we define the `thin-ness' parameter $\eta$ as
\begin{align}
\label{eq:etadefOD} \eta \equiv \frac{r_{\textrm{out}} - r_{\textrm{out}}}{\bar{r}} = \frac{e^{A(u_{\textrm{out}})}-e^{A(u_{\textrm{in}})}}{e^{A(\bar{u})}} \, .
\end{align}
We  refer to a wall as `thin', if $\eta \ll 1$ with $\eta \rightarrow 0$ the limit of a vanishingly thin wall. For $\eta \gtrsim 1$ the interpolation happens over effectively the whole range of the radial direction, so that even the notion of a localised wall becomes ill-defined.

As mentioned above, both $O(D)$-instantons and flat domain walls permit an interpretation in terms of holographic RG flows. Hence, in the next section we briefly review the relevant holographic RG flow solutions.

\subsection{Holographic RG flow solutions in Einstein-scalar theories} \label{sec:CDLasHoloRG}
According to the holographic principle, a strongly-coupled large-$N$ quantum field theory (QFT) is dual to a weakly-coupled gravitational theory \cite{Maldacena:1997re,Aharony:1999ti}. Physical quantities on the QFT side can be obtained by computing analogous quantities on the gravity side \cite{Witten:1998qj, Gubser:1998bc}. In particular, the RG flow on the QFT side is geometrized on the gravity side as the   concept know as `holographic RG flow'.

A canonical setting (but not the most general one) for the study of holographic RG flows is Einstein-scalar theory with a non-trivial scalar potential \cite{Balasubramanian:1999jd,Freedman:1999gp, deBoer:1999tgo, Bianchi:2001de, Papadimitriou:2004ap, Ceresole:2007wx}, i.e.~the theoretical framework introduced in \eqref{eq:setup1} and \eqref{eq:setup2}, together with the ansatz \eqref{metricansatz}. We begin by reviewing some general aspects of holographic RG flows in this framework.
\begin{itemize}
\item AdS solutions on the gravitational side are dual to conformal field theories (CFTs). In the theory \eqref{eq:setup1}, such solutions correspond to   $\f =$constant and coinciding with an extremum $\f_{\textrm{ext}}$ of $V$ with $V(\f_{\textrm{ext}}) <0$. Here, the potential exhibits at least two such extrema at $\f=\f_f$ and $\f=\f_t$, which hence represent two different CFTs respectively.\footnote{If the maximum separating the two minima at $\f=\f_f$ and $\f=\f_t$ is an AdS maximum, one can also associate another CFT with this extremum.} The corresponding gravitational duals are two AdS$_{d+1}$ space-times, characterised by their AdS lengths $\ell_f$ and $\ell_t$, respectively, given by
\begin{align}
\label{eq:ellfelltdef}
\ell_f^2 \equiv - \frac{d(d-1)}{V(\f_f)} \, , \qquad \qquad \ell_t^2 \equiv - \frac{d(d-1)}{V(\f_t)} \, .
\end{align}
\item The scalar field $\f$ is dual to a scalar operator $\mathcal{O}$ perturbing the CFT associated with a given extremum and inducing a RG flow. The scaling dimension of this operator for a given extremum of $V$ is given by
\begin{align}
\label{eq:Deltageneraldef}
\Delta (\f_{\textrm{ext}}) \equiv \frac{d}{2} \left(1 + \sqrt{1 +\frac{4 (d-1)}{d} \frac{V''(\f_{\textrm{ext}})}{|V(\f_{\textrm{ext}})|}} \right) \, ,
\end{align}
which crucially depends on the curvature $V''$ of the potential at that locus. In particular, for a \emph{maximum} ($V''<0$) the scaling dimension is bounded as $\Delta < d$ and the operator $\mathcal{O}$ corresponds to a \emph{relevant} deformation of the CFT. In contrast, for a \emph{minimum} ($V''>0$) the scaling dimension takes values $\Delta > d$ and the perturbing operator $\mathcal{O}$ is \emph{irrelevant}. {In the following, unless specified differently, $\Delta$ will refer to the scaling dimension of the operator associated with the extremum at $\f=\f_f$, i.e.~the false vacuum.}

\item The coordinate $u$ (or, equivalently, $\xi$) can be used as a parameter along the flow induced by $\mathcal{O}$. Here we define flows such that the direction of increasing $u$ (i.e.~decreasing $\xi$) corresponds to the direction of flow towards the infrared. In particular, the locus $u \rightarrow -\infty$ (i.e.~$\xi \rightarrow + \infty$) will denote the deep UV, that corresponds to the UV fixed point from which the flow originates. In this work we shall be interested in solutions with boundary conditions \eqref{eq:UVBC} (i.e.~`tunnelling from $\f_f$'). In the language of holography, these solutions correspond to RG flows away from a UV fixed point governed by the CFT associated with $\f_f$. In all but a few special cases that will be considered separately, $\f_f$ will correspond to a minimum of $V$. Therefore, the flows considered here will be flows induced by the vev of an irrelevant operator. The bulk geometry exhibits a boundary at $u \rightarrow -\infty$ which can be identified as the boundary of a AdS$_{d+1}$ space-time with AdS length $\ell_f$.

\item The metric $\zeta_{\mu \nu}$ in \eqref{metricansatz} describes the background on which the dual $d$-dimensional field theory is defined.\footnote{More precisely, the metric of the background for the dual QFT is in the same conformal class as $\zeta_{\mu \nu}$. By choosing an integration constant appropriately when solving for $A(u)$, one can always ensure that the background for the field theory is described by $\zeta_{\mu \nu}$ exactly.} Here, we consider $\zeta_{\mu \nu}$ describing a $d$-sphere or flat $d$-dimensional space, corresponding to RG flows for field theories on the respective backgrounds.
\end{itemize}

In the context of holographic RG flows, it is often convenient not to work with the functions $A(u)$ and $\f(u)$, but with a different set of dynamical quantities $W(\f)$, $S(\f)$ and $T(\f)$ obeying first-order differential equations. These are defined as
\begin{align}
\label{eq:defWc} W(\f) \equiv -2 (d-1) \dot{A} \, , \qquad S(\f) \equiv \dot{\f} \, , \qquad T(\f) \equiv e^{-2A} R^{(\zeta)}  \, .
\end{align}
This can be done piecewise along a flow $\f(u)$, for intervals separated by loci where $\dot{\f} = 0$. In terms of these functions, the equations of motion \eqref{c3}--\eqref{c5} become
\begin{align}
\label{eq:EOM4} S^2 - SW' + \frac{2}{d} T &=0 \, , \\
\label{eq:EOM5} \frac{d}{2(d-1)} W^2 -S^2 -2 T +2V &=0 \, , \\
\label{eq:EOM6} SS' - \frac{d}{2(d-1)} SW - V' &= 0 \, ,
\end{align}
which are coordinate-independent, first-order non-linear differential equations.
In the following sections, when appropriate, we shall refer to RG flow solutions in terms of their solutions for $W$, $S$ and $T$.

The dual QFT is further specified by the UV value $j$ for the source of the operator $\mathcal{O}$ and its vev $\langle \mathcal{O} \rangle$. For the ansatz \eqref{metricansatz}, the QFT is additionally characterised by the value of the scalar curvature $R^{(\zeta)}$ of the QFT background space-time.\footnote{Strictly speaking, one needs to specify the UV value of the scalar curvature of the QFT background space-time $R^{\textsc{uv}} = \lim_{u \rightarrow -\infty} e^{-2u/\ell_f} R^{(\textrm{ind})}$, where $R^{(\textrm{ind})} = e^{-2A(u)} R^{(\zeta)}$ is the induced curvature on a fixed-$u$-slice. As we describe in \cite{Ghosh:2017big}, we can always set  $R^{\textsc{uv}} = R^{(\zeta)}$ without loss of generality, so that a choice of $R^{(\zeta)}$ is equivalent to fixing $R^{\textsc{uv}}$.} The parameters $j$ and $R^{(\zeta)}$ act as boundary conditions in the UV when solving for $A(u)$ and $\f(u)$ and hence will be referred to as boundary data. The vev $\langle \mathcal{O} \rangle$ is then determined by the solution for a given set of boundary data.

For a given holographic RG flow, the QFT data $j$, $R^{(\zeta)}$ and $\langle \mathcal{O} \rangle$ can be read off from the asymptotic behavior of the corresponding solutions, in the vicinity of a UV fixed point. This can be done using the expressions for $A(u)$ and $\f(u)$, but here we focus on $W(\f)$. The UV fixed point is identified with an extremum of the potential, so we need to consider the solutions for $W$ in the vicinity of an extremum. Here we just present the results. More details and the derivation can be found in \cite{Kiritsis:2016kog,Ghosh:2017big}.

In particular, one has to distinguish between maxima and minima of $V$. Near a maximum of $V$, there exist two branches of solutions for $W$, which we denote by $(-)$ and $(+)$. In contrast, near a minimum of $V$ only the $(+)$-branch solution exists. Expanding about an extremum at $\f=\f_f=0$ the solutions on the two branches are given by \cite{Ghosh:2017big}:\footnote{In \cite{Ghosh:2017big} the parameters $\mathfrak{R}$ and $\mathcal{R}$ appearing in \eqref{eq:Wminus} and \eqref{eq:Wplus} were denoted by the same symbol $\mathcal{R}$. In the more detailed analysis of \cite{Ghosh:2017big} this could be done without creating confusion, but here we choose different symbols for clarity.}
\begin{align}
\label{eq:Wminus} W_-(\f) &= W_-^{\textrm{reg}}(\f) + \frac{\mathfrak{R}}{d \ell_f} \, |\f|^{2/(d-\Delta)} \big( 1 + \ldots \big) + \frac{C}{\ell_f} \, |\f|^{d/(d-\Delta)}  \big( 1 + \ldots \big) \, , \\
\label{eq:Wplus} W_+(\f) &= W_+^{\textrm{reg}}(\f) + \frac{\mathcal{R}}{d \ell_f} \, |\f|^{2/(\Delta)} \big( 1 + \ldots \big) \, ,
\end{align}
where $\mathfrak{R}$, $C$ and $\mathcal{R}$ are dimensionless parameters and the ellipses denote subleading terms. The functions $W_\pm^{\textrm{reg}}(\f)$ have a regular expansion in powers of $\f$ and are given by:
\begin{align}
\label{eq:Wmreg} W_-^{\textrm{reg}}(\f) &= \frac{2(d-1)}{\ell_f} + \frac{d-\Delta}{2 \ell_f} \f^2 + \mathcal{O}(\f^3) \, , \\
\label{eq:Wpreg} W_+^{\textrm{reg}}(\f) &= \frac{2(d-1)}{\ell_f} + \frac{\Delta}{2 \ell_f} \f^2 + \mathcal{O}(\f^3) \, .
\end{align}
The superpotentials (\ref{eq:Wminus}) and (\ref{eq:Wplus}  give rise
to solutions with   scalar field near-boundary asymptotics:
\be  \label{phisolution1}
\f(\xi) \simeq \f_f + \f_- \ell_f^{d-\Delta}e^{(d-\Delta) u/ \ell_f} +  \f_+
\ell_f^{\Delta}e^{\Delta u / \ell_f} +\ldots \qquad u\to -\infty
\ee
where $\f_- = j$ is the source of the dual operator $\mathcal{O}$, and $\f_+$ is
related to its vev by:
\be \label{vvev}
\< \mathcal{O}\> = (2\Delta -d) (M \ell_f)^{d-1}\f_+ .
\ee
In the case of a $W_+$-type solution, the source $\f_-=0$ and the scalar
field has only subleading boundary asymptotics. The solution is
controlled by a single parameter, $\f_+$.

The numerical parameters $\mathfrak{R}$, $C$ and $\mathcal{R}$  encode the QFT data. In particular, they are dimensionless combinations of the QFT data $j$, $R^{(\zeta)}$ and $\langle \mathcal{O} \rangle$ \cite{Klebanov:1999tb, Ghosh:2017big}:
\begin{align}
\mathfrak{R} &= R^{(\zeta)} \, |j|^{-2/ (d-\Delta)} \, , \\
C &= \frac{d-\Delta}{d} \frac{\langle \mathcal{O} \rangle}{(M\ell_f)^{d-1}} \, |j|^{-\Delta/(d-\Delta)} \, ,\\
\label{eq:curlyRdef} \mathcal{R} &= (2 \Delta -d)^{2/\Delta} (M \ell_f)^{2(d-1)/ \Delta} R^{(\zeta)} {\langle \mathcal{O} \rangle}^{-2 / \Delta} \, .
\end{align}
The above expressions also hold for RG flows for theories on flat
space, in which case one just sets $R^{(\zeta)}=0$ and hence
$\mathfrak{R}=0=\mathcal{R}$ in the above.

The fact that the solution on the $(+)$-branch only contains the parameter $\mathcal{R}$ but not $\mathfrak{R}$ or $C$ can then be understood as follows. The $(+)$-branch describes a RG flow driven purely by a non-zero vev $\langle \mathcal{O} \rangle$, i.e.~the source vanishes, $j=0$. The solution $W_+$ hence only contains $\mathcal{R}$ as this is the only dimensionless combination of QFT data that does not contain $j$. For RG flows from a maximum of $V$, as originally observed in \cite{Papadimitriou:2007sj}, it can be shown that the $(+)$-branch solution can be understood as the limit $j \rightarrow 0$ with $\langle \mathcal{O} \rangle = \textrm{const.}$ and $R^{(\zeta)}= \textrm{const.}$ of a $(-)$-branch solutions. Therefore, for a maximum as a UV fixed point, the $(+)$-branch solution can be seen as a special case from the family of $(-)$-branch solutions.

 One can also show, see e.g.~\cite{Ghosh:2017big}, that if a potential admits RG flow solutions from the $(-)$-branch, these solutions come as a continuous one-parameter family. The reason is as follows. As reviewed above, $(-)$-branch solutions depend on the two numerical parameters $\mathfrak{R}$ and $C$. The constraint that the holographic RG flow is regular at the IR end point, denoted by $\f_0$, fixes one combination of $\mathfrak{R}$ and $C$. This leaves one free parameter, which can be shown to be equivalent to a choice of the value of the IR end point $\f_0$, \cite{Ghosh:2017big}. In contrast, a $(+)$-branch solution only contains one free parameter $\mathcal{R}$. As a result, there is not enough freedom to ensure regularity of a $(+)$-type flow at any potential IR end point $\f_0$ and hence $(+)$-branch solutions do not come in continuous families. Instead, $(+)$-type solutions, if they exist at all in a given potential, exist as isolated solutions for selected values of the IR end point $\f_0$. Each such solution comes with a fixed value of the parameter $\mathcal{R}$.

Having discussed holographic RG flow solutions in some detail, we now turn our attention to their role in describing Coleman-de Luccia tunnelling processes. An important question at this stage is whether all holographic RG solutions satisfying the metric ansatz \eqref{metricansatz} with $R^{(\zeta)} \neq 0$ automatically also possess an interpretation as an $O(D)$-instanton. The answer to this is `no', as we explain in the following section. In particular, we shall argue that only $(+)$-branch solutions admit an interpretation as $O(D)$-instantons while $(-)$-branch solutions cannot be understood as mediating tunnelling.

\subsection{$O(D)$-instantons as holographic RG flows} \label{sec:linkCDLandHoloRG}
$O(D)$-instantons are solutions to \eqref{c3}--\eqref{c5} with $R^{(\zeta)} \neq 0$. From the discussion in the previous section it follows that such solutions, when describing tunnelling from AdS vacua, also describe holographic RG flows with the following properties:
\begin{itemize}
\item The holographic RG flow corresponding to an $O(D)$-instanton is a flow for a field theory defined on a $d$-sphere, since in this case the metric $\zeta_{\mu \nu}$ in \eqref{metricansatz} is a metric on a $d$-sphere. This in turn implies that holographic RG flows for field theories on flat space-time can never have an interpretation in terms of an $O(D)$-instanton.
\item An $O(D)$-instanton describing tunnelling away from an AdS extremum $\f_f$ will correspond to an RG flow away from a UV fixed point at $\f_f$. The dimension $\Delta$ of the operator $\mathcal{O}$ perturbing the UV CFT is related to the potential and its curvature at $\f_f$ via \eqref{eq:Deltageneraldef}.
\end{itemize}

The fact that the $O(D)$-instantons permit an interpretation in terms of holographic RG flows implies that there is a way of describing tunnelling in AdS from the dual QFT point of view. In particular, an important set of quantities that characterise a holographic RG flow are the boundary QFT data. These are given by  $j$, the UV value of the source of the scalar operator $\mathcal{O}$ dual to $\f$, and, for a field theory on a $d$-sphere, the scalar curvature $R^{(\zeta)}$ of the $d$-sphere. To describe an $O(D)$-instanton from the QFT perspective therefore corresponds to specifying the relevant QFT data that will reproduce the holographic RG flow corresponding to the desired $O(D)$-instanton.

In the following we argue that, from the QFT point of view, a given $O(D)$-instanton solution will correspond to not just one QFT, but a family of QFTs with boundary data $j=0$, $R^{(\zeta)} \neq 0$ with $R^{(\zeta)}$ otherwise arbitrary. The fact that $O(D)$-instantons correspond to QFTs with $j=0$ can be understood as following from the requirement that the tunnelling rate is well-defined. According to Coleman, the tunnelling probability per unit time and per unit volume will be denoted by $\Gamma / V$ and can be computed in the semiclassical limit as \cite{Coleman:1977py, Callan:1977pt, CdL}:
\begin{equation}
\label{eq:TunnellingRateGamma}
\frac{\Gamma}{V}=A ~e^{-{\frac{B}{\hbar}}}\left[ 1+\mathcal{O}(\hbar) \right] \ ,
\end{equation}
where $\hbar$ is shown explicitly here, but will be suppressed in the following.
The prefactor $A$ depends on the one-loop determinant of the theory under consideration and we leave it undetermined here. The exponent $B$ is given by \cite{Coleman:1977py, Callan:1977pt, CdL}:
\begin{equation}
B=S_{E, \, \textrm{inter}}-S_{E, \, \textrm{false}} \ \label{eq:instactB}
\end{equation}
where $S_{E, \, \textrm{inter}}$ is the Euclidean on-shell action for the interpolating $O(D)$-tunnelling solution and $S_{E, \, \textrm{false}}$ is the background Euclidean action, i.e.~the action for the solution with the false vacuum throughout space-time. Here, for tunnelling from an AdS vacuum, both $S_{E, \, \textrm{inter}}$ and $S_{E, \, \textrm{false}}$ exhibit divergences due to the infinite volume of the AdS boundary in the UV, with the divergent terms depending explicitly on the UV data. These divergences only cancel (and therefore $B$ is well-defined) if the boundary data of the interpolating solution and the background solution are the same.\footnote{The definition of the on-shell action typically includes a set of counterterms that regulate the UV divergences and render the action finite. These counterterms depend explicitly on the boundary data and also contain free parameters whose choice corresponds to picking a renormalization scheme. For the Coleman instanton action $B$ in \eqref{eq:instactB} to be scheme-independent the QFT data of the two theories compared needs to coincide, which again enforces that the interpolating solution must have a vanishing source, $j=0$.} This is shown in detail for $D=d+1=4$ in appendix \ref{App:Pert}. The background solution corresponds to the UV CFT associated with the false vacuum and therefore has $j=0$ by definition. This in turn implies that the $O(D)$-tunnelling solution must also have $j=0$. In the terminology introduced in the previous section, this implies that $O(D)$-instantons correspond to holographic RG flows on the $(+)$-branch, as it is these solutions that have $j=0$, while $(-)$-branch solutions have $j \neq 0$.

We turn to the remaining QFT data. As stated before, the $O(D)$ symmetry of the tunnelling solution implies that the corresponding field theory is defined on a $d$-sphere, and hence we require $R^{(\zeta)} \neq 0$. But what about the value of $R^{(\zeta)}$? How is this related to properties of the $O(D)$-tunnelling solution?

To answer this question, we return to our result that $O(D)$-instantons correspond to $(+)$-branch type RG flows. In \eqref{eq:Wplus} we recorded the near-boundary expansion of $W$ on a $(+)$-branch solution. Note that this depends on $R^{(\zeta)}$ only through the dimensionless combination $\mathcal{R}$, which has the interpretation of the UV curvature $R^{(\zeta)}$ in units of the vev $\langle \mathcal{O} \rangle$. That is, a given $O(D)$-instanton corresponds to a holographic RG flow with a fixed value of $\mathcal{R}$, not $R^{(\zeta)}$. This implies that we can pick the value of the boundary curvature $R^{(\zeta)}$ freely, and the vev $\langle \mathcal{O} \rangle$ will adjust accordingly in the solution so that the dimensionless combination $\mathcal{R}$ remains fixed. The only constraint on $R^{(\zeta)}$ is that we have to choose the same value for the $O(D)$-tunnelling solution and for the background solution, so that the instanton action $B$ in \eqref{eq:instactB} is again well-defined.

To conclude: Holographic RG flow solutions that correspond to $O(D)$-instantons describe QFTs with vanishing UV source, $j=0$, defined on a $d$-sphere with curvature $R^{(\zeta)} \neq 0$, but otherwise unconstrained.

\subsection{$O(D)$-instanton action}
\label{sec:instaction}
For a tunnelling event mediated by an $O(D)$-instanton, the tunnelling probability per unit time and per unit volume  was recorded above in \eqref{eq:TunnellingRateGamma}, and depends on the one-loop determinant $A$ of the theory under consideration and the so-called instanton action\footnote{There
  is a caveat though, if $A$ turns out to be imaginary (this happens if the
  determinant has an odd number of negative modes) then
  $\Gamma$ is imaginary and the decay is unsuppressed, \cite{Callan:1977pt}.} denoted by $B$  \cite{Coleman:1977py, Callan:1977pt, CdL}.
In the following, we describe how the instanton action $B$ is computed explicitly for the tunnelling processes considered in this work.

Following \cite{Coleman:1977py, Callan:1977pt, CdL}, the instanton action given in \eqref{eq:instactB} is defined as the difference between the on-shell action for the tunnelling solution $S_{E, \, \textrm{inter}}$ and the on-shell action for the background solution $S_{E, \, \textrm{false}}$ with the false vacuum filling all of space-time. For AdS false vacua, both of these diverge due to the infinite volume of the AdS boundary and \eqref{eq:instactB} should be modified as
\begin{equation}
B=\lim_{u_{\textsc{uv}} \rightarrow -\infty} S_{E, \, \textrm{inter}}-S_{E, \, \textrm{false}} \ \label{eq:acdiff} \, ,
\end{equation}
where $u_{\textsc{uv}}$ is a UV-cutoff that regulates the divergences.

Then, both $S_{E, \, \textrm{inter}}$ and  $S_{E, \, \textrm{false}}$ can be computed by evaluating the action in \eqref{eq:setup1} on the corresponding solutions. Both the instanton and the background solutions obey the ansatz \eqref{metricansatz}, in which case the action in \eqref{eq:setup1} can be written as:\footnote{For the ansatz \eqref{metricansatz} one can re-express $R^{(g)}$ as a function of $A$, $\dot{A}$ and $\ddot{A}$. Using the equations of motion \eqref{c3}--\eqref{c5} it is then possible to eliminate all explicit appearances of $\ddot{A}$, $\dot{\f}$ and $V$ to arrive at the expression in \eqref{eq:Son} (see e.g.~\cite{Ghosh:2017big,F} for more details).}
\begin{align}
\label{eq:Son} S_{E, \, O(D)} = \ & 2(d-1) M^{d-1} V_d \, {\big[e^{dA} \dot{A} \big]}_{u_{\textsc{uv}}} - \frac{2}{d} M^{d-1} R^{(\zeta)} V_d \int_{u_{\textrm{uv}}}^{u_0 } du \, e^{(d-2)A} \, , \\
\nonumber &\textrm{with} \quad V_d \equiv \int \textrm{d}^dx \sqrt{|\zeta |} = \textrm{Vol}(S^d) = \tilde{\Omega}_d \big( R^{(\zeta)} \big)^{-d/2} \, .
\end{align}
Then $S_{E, \, \textrm{inter}}$ can be computed by evaluating $S_{E, \, O(D)}$ in \eqref{eq:Son} on a solution $A(u)$ for the interpolating case. Similarly, $S_{E, \, \textrm{false}}$ is given by \eqref{eq:Son} evaluated on the solution $A(u)$ corresponding to the false vacuum. For the latter, we can give an explicit expression for $A(u)$. The false vacuum corresponds to a solution to \eqref{c3}--\eqref{c5} with $\f=\f_f$ and $\dot{\f}=0$. One can check that for $R^{(\zeta)} > 0$ a corresponding solution for $A(u)$ is given by
\begin{align}
\label{eq:AofuCFTfalse}
A(u)=\ln\left[-\frac{\ell_f}{\a} \sinh \left(\frac{u-u_{f,0}}{\ell_f} \right) \right] \, ,
\end{align}
where $\ell_f$ was defined in \eqref{eq:ellfelltdef} and $\alpha$ in \eqref{eq:alphadef}.
The parameter $u_{f,0}$ is an integration constant and we can choose it freely. Here we set it to $u_{f,0} = (\ell_f /2) \log (4 \alpha^2 / \ell_f^2)$.

We can then calculate $B$ in \eqref{eq:acdiff} by inserting the
regulated expressions for $S_{E, \, \textrm{inter}}$ and $S_{E, \,
  \textrm{false}}$ obtained from \eqref{eq:Son} and removing the
regulator by setting $u_{\textsc{uv}} \rightarrow -\infty$.

The calculation parallels the one presented in \cite{F} for
source-driven RG flows. We observed that:
\begin{itemize}
\item The first term in equation (\ref{eq:Son}) cancels exactly when
  taking the difference (\ref{eq:acdiff}). The reason is that the
  finite contribution to this term comes purely  from the term
  proportional to $C$ in the expansion of $W$, equation
  (\ref{eq:Wminus}-\ref{eq:Wplus}), and this term is absent for
  $W_+$-type solution. This is unlike the case analyzed in \cite{F}
  where the $C$ terms gives the leading contribution to the free
  energy.
\item The second term  in (\ref{eq:Son}) gives a finite
  result (after subtracting the AdS contribution) and can be written as:
\be \label{eq:B}
S_{E, \, \textrm{inter}} - S_{E,  \, \textrm{false}} \to  - (M \ell_f)^{d-1} {\cal B} \int d^dx
\sqrt{|\zeta|} \left(R^{(\zeta)}\right)^{d/2} \, , \quad \textrm{for} \quad u_{\textsc{uv}} \to -\infty \, ,
\ee
where we introduced a dimensionless constant ${\cal B}$ which does not depend on
curvature,\footnote{To make contact
  with \cite{F}, ${\cal B}$ is related to the
  vev-parameter appearing in the  the $U$-function.} which is uniquely determined for a given interpolating solution, like the value of the
parameter ${\cal R}$ appearing in equations
(\ref{eq:Wplus}-\ref{eq:curlyRdef}).
\end{itemize}
The integral in equation (\ref{eq:B}) gives a curvature-independent
result, since the volume of $S^d$ scales as  $\alpha^d$  where
$\alpha$ is the sphere radius given by equation
(\ref{eq:alphadef}). This leads to the final result:
\be \label{eq:B2}
B  =  - (M \ell_f)^{d-1} \tilde{\Omega}_d \, {\cal B} \, ,
\ee
 where $\tilde{\Omega}_d $ is a geometric factor related to the volume
 of the unit $d$-dimensional sphere, defined in \eqref{eq:Son}.

Although $B$ can only be computed numerically, we can determine its
sign  without doing any extra computation. The sign of $B$ is very
important, since by equation (\ref{eq:TunnellingRateGamma}) it governs
the decay rate of the false vacuum: a negative $B$ would imply that
the decay is unsuppressed and the false vacuum is actually unstable,
rather than metastable.

To determined the sign of equation (\ref{eq:B2}), we can go back to
the expression (\ref{eq:Son}), and rewrite (recall the first term
gives no contribution for $W_+$-flows):
\be \label{eq:B3}
B = \left( -{\frac{2}{d}} M^{d-1} R^{(\zeta)} V_d \right)\lim_{u_{\textsc{uv}} \to -\infty}
\left[ \int_{u_{\textsc{uv}}}^{u_0} du\, e^{(d-2)A_{\textrm{inter}}(u)} -
  \int_{u_{\textsc{uv}}}^{u_{f,0}} du\, e^{(d-2)A_{\textrm{false}}(u)} \right]
\ee
where $A_{\textrm{inter}}(u)$ and $A_{\textrm{false}}(u)$ are the warp factors of the CdL
instanton and of the AdS false-vacuum solution, respectively. It is
easy to see that the quantity in the square brackets is negative, because the
scale factor $e^A$ decreases faster in the CdL solution than in
AdS. One can see this using the definition (\ref{eq:defWc}) and the
relation  (cfr.~eq.~(\ref{eq:Wplus}), (\ref{eq:Wpreg})):
\be \label{eq:B4}
W_+ > W_{f} = {\frac{2(d-1)}{\ell_f}}.
\ee
The above argument shows that, regardless of its numerical value,
$B > 0$.

From the point of view of the boundary field theory, $B$ computes the
difference between the free energies on $S^d$  of the two semiclassical states
corresponding to the CFT stuck at the UV fixed point  $\f_f$ and the
RG-flow driven away from the fixed point by the non-zero vev $\f_+$:
\be
B = {\cal F}_{\textrm{RG-flow}} - {\cal F}_{\textrm{fixed-point}}  >0 \, .
\ee
The positivity of $B$ implies the semiclassical state of lower free
energy is the pure AdS solution with no flow.

\subsection{No-go result for tunnelling to AdS minima or maxima} \label{sec:CDLnogo}
In this section we show that a process described by an $O(D)$-instanton can never result in $\f$ arriving at \emph{exactly} an AdS extremum of the potential. One consequence of this is that tunnelling from a false vacuum to \emph{exactly} the true vacuum is forbidden. This result is not new and has already been observed in the foundational work by Coleman and de Luccia \cite{CdL}. However, in the approximation considered in \cite{CdL} the end point of tunnelling can be made arbitrarily close to the true vacuum, so that this observation was not emphasised there. This may leave casual readers of the topic with the false impression that tunnelling exactly to a true vacuum is possible, while in fact is is not. The purpose of this section is to emphasize this fact about tunnelling.

\subsubsection*{No tunnelling to exactly an AdS extremum}

We demonstrate this by assuming that a solution exists that arrives in the vicinity of an AdS extremum of $V$ and then show that there is no solution of $O(D)$ type that can reach the extremum exactly. To start, note that in the vicinity of an AdS extremum at $\f=\f_{\textrm{ext}}$, the potential can always be written as
\begin{align}
\label{eq:VexpIR} V(\f) = -\frac{d(d-1)}{\ell^2} - \frac{\Delta(d-\Delta)}{2 \ell^2} (\f-\f_{\textrm{ext}})^2 + \mathcal{O} \big( (\f-\f_{\textrm{ext}})^3 \big) \, ,
\end{align}
where we have used \eqref{eq:Deltageneraldef} to re-express $V''(\f_{\textrm{ext}})$ in terms of the dual scaling dimension $\Delta$.

We now construct the solution for an $O(D)$-instanton in the vicinity of the extremum by perturbing about the solution at the extremum. At the extremum, i.e.~for $\f=\f_{\textrm{ext}}=\textrm{const}$, the solution for the space-time is just $D$-dimensional AdS space with AdS length $\ell$. We encountered the corresponding solution for $A(u)$ before in \eqref{eq:AofuCFTfalse} for the case $f_{\textrm{ext}}=\f_f$. Denoting this solution by $A_0(u)$, we have
\begin{align}
\label{eq:A0def} A_0(u)=\ln\left[-\frac{\ell}{\a} \sinh \left(\frac{u-u_0}{\ell} \right) \right] \, ,
\end{align}
with $\alpha$ defined in \eqref{eq:alphadef}. To find solutions in the vicinity of $\f_{\textrm{ext}}$, we then expand about the solution at $\f=\f_{\textrm{ext}}$. We choose the ansatz
\begin{align}
\nonumber A(u) &= A_0(u) + \epsilon^2 A_2(u) + \mathcal{O}(\epsilon^3) \, , \\
\label{eq:IRexpphi} \f(u) &= \f_{\textrm{ext}} + \epsilon \f_1(u) + \mathcal{O}(\epsilon^2) \, ,
\end{align}
where $\epsilon \sim |\f-\f_{\textrm{ext}}|$ is a small parameter measuring the departure from the extremum. This will allow for the construction of solutions for $\f(u)$ and $A(u)$ iteratively, by solving order by order in $\epsilon$ upon inserting this ansatz into the equations of motion \eqref{c3}--\eqref{c5}.

Here, we are mainly interested in whether interpolating solutions exist that smoothly arrive at $\f=\f_{\textrm{ext}}$. Therefore, we focus on the correction $\f_1(u)$ to the constant solution $\f=\f_{\textrm{ext}}$. The corresponding equation of motion is given by:
\begin{align}
\label{eq:minexp1}
\ddot{\f}_1(u) + \frac{d}{\ell} \coth \left( \frac{u-u_0}{\ell} \right) \dot{\f}_1 + \frac{\Delta (d-\Delta)}{\ell^2} \f_1(u) =0 \, .
\end{align}
We need to complement this with the appropriate boundary conditions. We are interested in $O(D)$-instantons that correspond to tunnelling to $\f=\f_{\textrm{ext}}$. The end point of a tunnelling process by an $O(D)$-instanton is the centre of the solution, where the radius $e^A \rightarrow 0$ shrinks to zero. From \eqref{eq:A0def} it follows that at leading order this corresponds to $u \rightarrow u_0$. Therefore, a solution corresponding to tunnelling to $\f_{\textrm{ext}}$ would have to satisfy the boundary conditions, see \eqref{eq:UVBC}:
\begin{align}
\f_1(u_0) = 0 \, , \qquad \dot{\f}_1(u_0) = 0 \, .
\end{align}
The task now simply is to solve \eqref{eq:minexp1} and implement the boundary conditions. While this can be done in all generality (see e.g.~\cite{F}), a simplified analysis will be sufficient to prove the point without sacrificing rigour. In particular, by restricting attention to the leading terms for $u \rightarrow u_0$ we can expand the $\coth$-term in \eqref{eq:minexp1} for $u \rightarrow u_0$ and solve \eqref{eq:minexp1} as\footnote{For $d=3$, which is the case of most relevance for our $4$-dimensional universe, the first subleading correction in the term with integration constant $c_1$ should be $\mathcal{O} \big( \log ( \tfrac{u-u_0}{\ell} ) \big)$. However, this does not affect anything that follows.}
\begin{align}
\f_1(u) \underset{u \rightarrow u_0}{=} & \ \hphantom{+} c_1 \bigg[ \Big(\frac{u-u_0}{\ell}\Big)^{-(d-1)} + \mathcal{O}\bigg(\Big(\frac{u-u_0}{\ell}\Big)^{-(d-3)} \bigg) \bigg] \\
\nonumber & + c_2 \bigg[ 1- \frac{\Delta(d-\Delta)}{2(d+1)} \Big(\frac{u-u_0}{\ell}\Big)^2 + \mathcal{O}\bigg(\Big(\frac{u-u_0}{\ell}\Big)^4 \bigg) \bigg] \, ,
\end{align}
with $c_1$ and $c_2$ integration constants. It is now easy to verify that $O(D)$-solutions arriving at the extremum are not allowed. The boundary condition $\dot{\f}_1(u_0) = 0$ can only be satisfied by setting $c_1=0$ and similarly the boundary condition $\f_1(u_0) = 0$ requires $c_2=0$. As a result the only possible solution is that $\f_1$ vanishes identically. For the ansatz in \eqref{eq:IRexpphi}, if $\f_1$ vanishes then all higher corrections, i.e.~$A_2$ etc., will also vanish, as one can verify from the equations of motion. Therefore,  we arrive at the result that a solution that arrives at $\f_{\textrm{ext}}$ does not exist. This confirms what we have set out to prove: \textit{there is no solution of $O(D)$ type can describe tunnelling to exactly an extremum of the potential}.

Instead, tunnelling processes via $O(D)$-instantons deposit $\f$ at generic points of the potential $V$ that are not extrema. For completeness, we record the expressions for $\f(u)$ and $A(u)$ in the vicinity of such a generic point, denoted by $\f=\f_0$. More details regarding the derivation can be found in \cite{Ghosh:2017big}. In particular, near  $\f=\f_0$ we can expand $\f(u)$ and $A(u)$ for $u \rightarrow u_0$ as
\begin{align}
\label{eq:AIR} A(u) & = \ln \Big(- \frac{u-u_0}{\alpha} \Big) + \mathcal{O} \big( (u-u_0) \big) \, , \\
\label{eq:phiIR} \f(u) &= \f_0 + \frac{V'(\f_0)}{2(d+1)} (u-u_0)^2 + \mathcal{O} \big( (u-u_0)^3 \big)  \, ,
\end{align}
with $\alpha$ defined as in \eqref{eq:alphadef}. {\em Note that an end point $\f_0$ with $V'(\f_0) <0$ can only be approached from the left, while an end point with $V'(\f_0) >0$ can only be arrived at coming from the right.}

\subsubsection*{A quick look at flat domain walls}
In contrast to $O(D)$-instantons, flat domain walls interpolate between two extrema of the potential, as we briefly review here. Consider again an extremum at $\f=\f_{\textrm{ext}}$ with $V$ in its vicinity given by \eqref{eq:VexpIR}. Approaching an extremum, the bulk space-time asymptotes to AdS$_{d+1}$, which for $R^{(\zeta)}=0$ implies that
\begin{align}
\label{eq:Anearextflat}
A(u) = -\frac{u}{\ell} \, .
\end{align}
Inserting this into \eqref{c5} one finds
\begin{align}
\label{eq:minexp1a}
\ddot{\f}(u) + \frac{d}{\ell} \dot{\f} + \frac{\Delta (d-\Delta)}{\ell^2} \f(u) =0 \, ,
\end{align}
which can be solved as
\begin{align}
\label{eq:phinearextflat} \f(u) = \f_{\textrm{ext}} + \f_- \, \ell^{(d-\Delta)} \, e^{(d-\Delta)u / \ell} + \f_+ \, \ell^{\Delta} \, e^{\Delta u / \ell}  \, ,
\end{align}
with $\f_-$ and $\f_+$ integration constants. The UV and IR ends of the domain wall are given by the loci $e^A \rightarrow +\infty$ and $e^A \rightarrow 0$, respectively, corresponding to $u \rightarrow -\infty$ and $u \rightarrow +\infty$. At both the UV and IR end points we need to implement the boundary conditions
\begin{align}
\f(u \rightarrow \pm \infty) = \f_{\textrm{ext}} \, , \qquad \dot{\f}(u \rightarrow \pm \infty) = 0 \, .
\end{align}
We can then make the following observations:
\begin{itemize}
\item Both maxima and minima of $V$ can be UV ends of flat domain walls. In the case of a minimum of $V$, $\Delta > d$, the boundary conditions require setting $\f_-=0$, but $\f_+$ can be non-zero.
\item Maxima of $V$, $\Delta < d$, cannot be IR end points of flat domain walls. The boundary conditions can only be satisfied by setting both $\f_-=0=\f_+$ and hence the wall cannot arrive there.
\item Instead, flat domain walls have their IR end points at minima of $V$ ($\Delta > d$). The boundary conditions require setting $\f_+=0$, but $\f_-$ can be non-zero.
\end{itemize}

\subsection{Tunnelling from AdS minima is non-generic} \label{sec:CDLnongeneric}
Without gravity, a physical system with two non-degenerate ground states will generically permit tunnelling from the false to the true vacuum \cite{Coleman:1977py}. However, in the presence of gravity this is no longer the case. The existence of two non-degenerate ground states does not automatically imply the existence of an $O(D)$-instanton solution. This has already been observed in the foundational work by Coleman and de Luccia  \cite{CdL} in the context of thin-wall solutions. Here, we  present a general argument for this observation valid beyond the thin-wall approximation, and link it to known features of holographic RG flows. In addition, we derive a sufficient condition on a potential for possessing a $O(D)$-instanton describing tunnelling out of an AdS minimum.

\noindent \textbf{A mechanical analogue --} A useful tool for constructing the argument of this section will be a mechanical analogue to the problem at hand, originally used by Coleman in \cite{Coleman:1977py}. The observation is that the equation of motion \eqref{c5} is identical to that for the Lorentzian problem of a particle of unit mass performing one-dimensional motion in the inverted potential $-V$. The field $\f$ plays the role of the coordinate of the particle and the term $\sim \dot{A} \dot{\f}$ implements friction. For positive friction the time direction for the motion of the particle needs to be in the negative $u$-direction. This follows from the equation of motion \eqref{c4} together with the boundary conditions \eqref{eq:UVBC} which imply that $\dot{A}$ is strictly negative, $\dot{A}<0$. Hence, for positive friction the time variable $t$ for the particle has be taken as $t=-u$. As a result \eqref{c5} can be written as
\begin{align}
\label{eq:EOMparticle} \frac{d^2 \f}{dt^2} + b(t) \frac{d\f}{dt} - V' =0 \, , \quad \textrm{with} \quad b(t) \equiv - d \dot{A}(u) > 0 \, ,
\end{align}
which can be identified as the equation of motion for damped motion in the inverted potential $-V$.

Due to the identification $t=-u$, the initial conditions for the particle are set by the boundary conditions at the end point of the tunnelling process. An $O(D)$-solution describing tunnelling from $\f_f$ to some locus $\f_0$ exhibits the behaviour\footnote{See \eqref{eq:phiIR} for the precise behaviour near $u=u_0$.}
\begin{align}
\nonumber \f \rightarrow \f_f \, , \ \dot{\f} \rightarrow 0 \, , \textrm{ for } u \rightarrow -\infty \, , \qquad \quad \f \rightarrow \f_0 \, , \ \dot{\f} \rightarrow 0 \, , \textrm{ for } u \rightarrow u_0 \, ,
\end{align}
This implies that in the mechanical analogue the particle is initially released from $\f_0$ \emph{at rest} ($\tfrac{d \f}{dt} =0$). To describe tunnelling from an extremum at $\f=\f_f$, the particle then has to come to rest again at $\f_f$. As a result, an $O(D)$-solution describing tunnelling from an AdS minimum (maximum) at $\f_f$ to some generic point $\f_0$ has the following mechanical analogue: It corresponds to a solution, where the particle starts rolling from rest from $\f_0$ and after performing damped motion comes to rest again at the maximum (minimum) $\f_f$ of the inverted potential.

We can also consider the mechanical analogue of a flat domain-walls. For flat domain-walls interpolating between $\f_f$ and $\f_t$ one has
\begin{align}
\nonumber \f \rightarrow \f_f \, , \ \dot{\f} \rightarrow 0 \, , \textrm{ for } u \rightarrow -\infty \, , \qquad \quad \f \rightarrow \f_t \, , \ \dot{\f} \rightarrow 0 \, , \textrm{ for } u \rightarrow +\infty \, .
\end{align}
The mechanical analogue now corresponds to a particle being released \emph{at rest} from $\f_t$ and coming to rest again at $\f_f$.

In the following, we test whether a potential admits an $O(D)$-instanton solution or flat domain-wall solution by using the mechanical framework and releasing a test particle from various points $\f_{\textrm{start}}$ with the appropriate initial conditions, and follow its subsequent trajectory. In fact, we can discuss both $O(D)$-instanton solutions and flat domain-wall solutions in a unified way as in both cases the particle has to be released from rest.

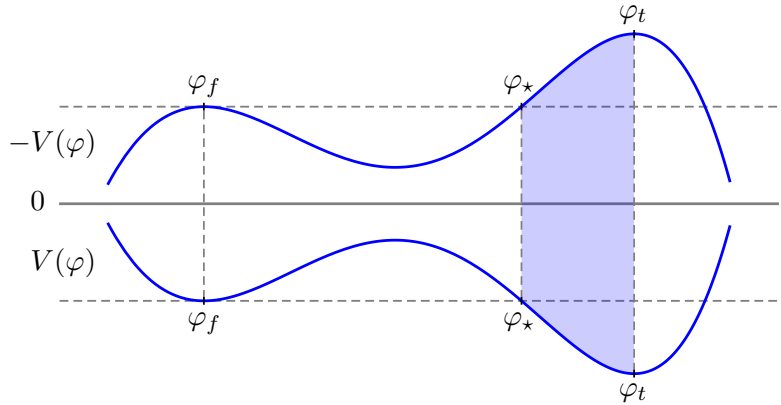
\begin{figure}[t]
\centering
\begin{overpic}[width=0.65\textwidth,tics=10]{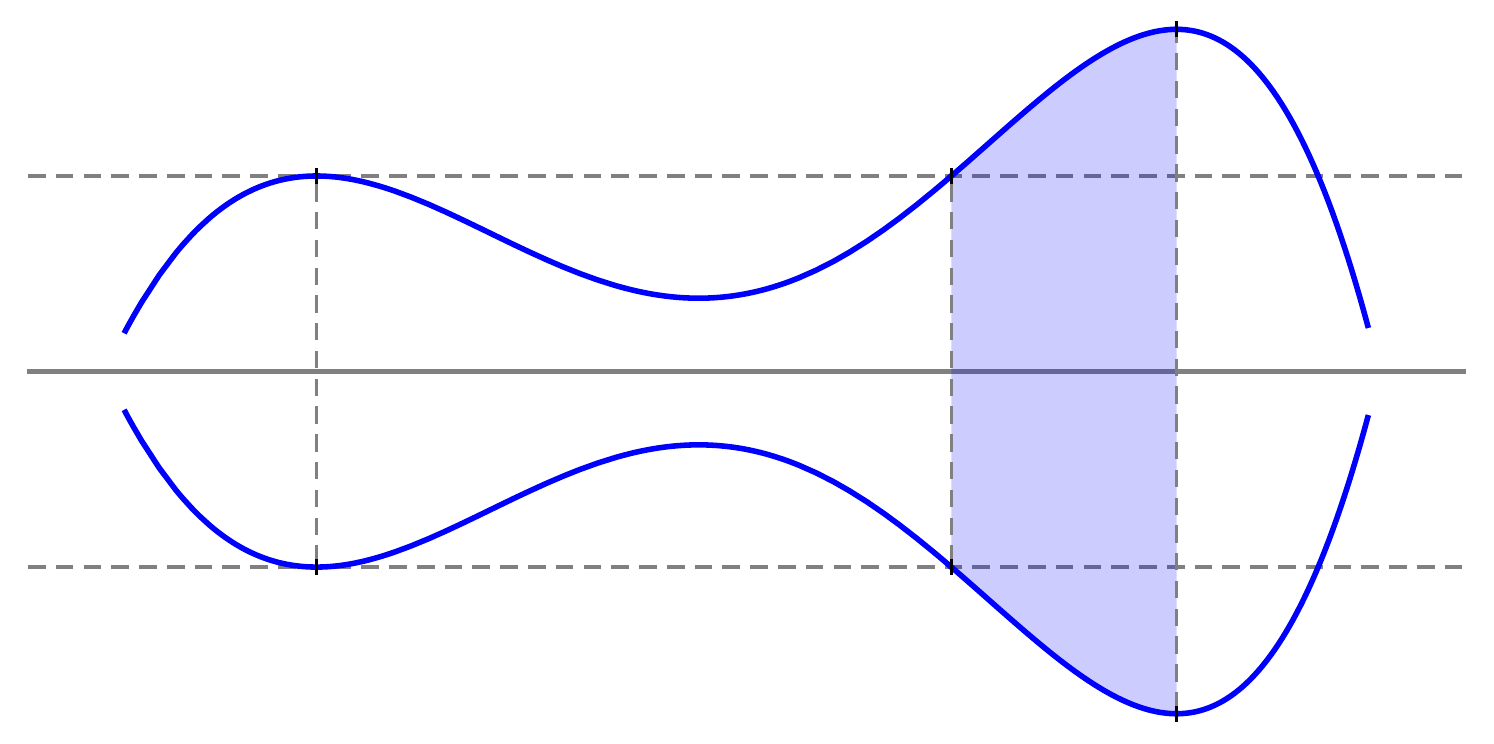}
\put(-5,32){$-V(\f)$}
\put(-2,24){$0$}
\put(-2,16){$V(\f)$}
\put(19,8.5){$\f_f$}
\put(61.25,8.5){$\f_\star$}
\put(76.75,-1){$\f_t$}
\put(19,40.5){$\f_f$}
\put(61.25,40.5){$\f_\star$}
\put(76.75,50){$\f_t$}
\end{overpic}
\caption{Potential $V(\f)$ with two AdS minima at $\f=\f_f$ (`false vacuum') and $\f=\f_t$ (`true vacuum') with $V(\f_f) > V(\f_t)$, separated by a barrier. The inverted potential $-V(\f)$ exhibits two maxima at $\f_f$ and $\f_t$ separated by a valley. The locus $\f_\star$ is defined as the point with $V(\f_\star) = V(\f_f)$, but on the other side of the barrier/valley from $\f_f$. A particle released from rest and performing damped motion in the inverted potential $-V$ cannot reach $\f_f$ if it is initially released outside the shaded region.}
\label{fig:undersec2}
\end{figure}

\noindent \textbf{Tunnelling from AdS minima is non-generic --} Here we present the main argument of this section, starting with the mechanical picture, before turning to a holographic point of view. As explained above, solutions describing tunnelling from a \emph{minimum} of $V$ correspond to mechanical solutions describing a particle coming to rest exactly at a \emph{maximum} of $-V$. In the mechanical framework it is then easy to see that minima of $-V$ are attractors while maxima are not. That is, if the particle is released from some point in the inverted potential, the particle will roll down the inverted potential and, as a result of friction, generically come to rest again in a minimum of $-V$. In contrast, to come to rest at a maximum of $-V$ the trajectory of the particle has to be tuned carefully to avoid any over- or undershooting.

The observation then is that in some potentials this can be done by adjusting the starting point $\f_{\textrm{start}}$ for the particle, but in other potentials no such starting point can be found. This is best illustrated by an example. Consider a potential with exactly two AdS minima at $\f=\f_f$ and $\f=\f_t$ separated by a barrier by default, with $V(\f_f) > V(\f_t)$ as shown in fig.~\ref{fig:undersec2}. As described before, solutions describing tunnelling from $\f_f$ will correspond to mechanical solutions in the inverted potential $-V$ where the particle comes to rest at $\f_f$. The inverted potential exhibits a local maximum at $\f_f$ and a global maximum $\f_t$ separated by an intermediate minimum, and is also shown in fig.~\ref{fig:undersec2} Due to the existence of friction, a necessary condition for the particle to reach $\f_f$ is that its initial potential energy is at least as large as the potential energy at $\f_f$, i.e.~$-V(\f_{\textrm{start}}) > - V(\f_f)$. This implies that the particle has to start in the interval $\f_{\textrm{start}} \in [\f_\star, \f_t]$ (the shaded region in fig.~\ref{fig:undersec2}) where at $\f_\star$ one has $V(\f_{\star}) =V(\f_f)$. However, this is not a sufficient condition for a solution to exist. In fact, depending on the precise shape of the potential, there may not exist a choice of starting point so that the particle successfully reaches $\f_f$. In particular, friction may be sufficiently strong that for any starting point $\f_{\textrm{start}} \in [\f_\star, \f_t]$ the particle will not reach $\f_f$, but undershoot and subsequently roll down to the bottom of the valley. In the Euclidean picture this implies that no solutions describing tunnelling from $\f_f$ exist in this potential. This shows that trajectories that will allow the particle to come to rest at a maximum of $-V$ only exist for certain potentials, but not generically. In the Euclidean picture the corresponding finding is that solutions describing tunnelling from a minimum of $V$ are non-generic.

This result can also be deduced from a holographic point of view, as we proceed to show in the following. To begin, we use the formulation in terms of holographic RG flows to re-derive the statement from above that maxima of $V$ (i.e.~minima of $-V$) are attractors while minima of $V$ (i.e.~maxima of $-V$) are not. To this end we recall the asymptotic behaviour of holographic RG flow solutions near an extremum. For example, consider the solutions for $W(\f)$ near an extremum of $V$ that represents the UV fixed point for a holographic RG flow. We encountered these before in \eqref{eq:Wminus} for an expansion around a maximum of $V$, and in \eqref{eq:Wplus} for an expansion around a minimum of $V$. The observation was that in the vicinity of a maximum, $W$ depends on two free parameters $\mathfrak{R}$ and $C$, while in the vicinity of a minimum there is only one free parameter $\mathcal{R}$. That is, solving near a maximum of $V$, solutions for $W$ come as a two-parameter family, while near a minimum of $V$, solutions for $W$ represent a one-parameter family:
\begin{align}
\textrm{Expansion near max.~of } V\textrm{:} \quad & W(\f) = W_{\mathfrak{R}, C} (\f) \, , \\
\textrm{Expansion near min.~of } V\textrm{:} \quad  & W(\f) = W_{\mathcal{R}} (\f) \, .
\end{align}
In a complete RG flow, these parameters are fixed by the boundary/ regularity conditions on the other end of the flow. In particular, setting the value of $\f$ where the flow ends on the other side fixes one combination of the parameters. This does however not yet guarantee that the solution is regular, i.e.~regularity is an additional requirement to be satisfied.

\begin{figure}[t]
\centering
\begin{overpic}[width=0.45\textwidth,tics=10]{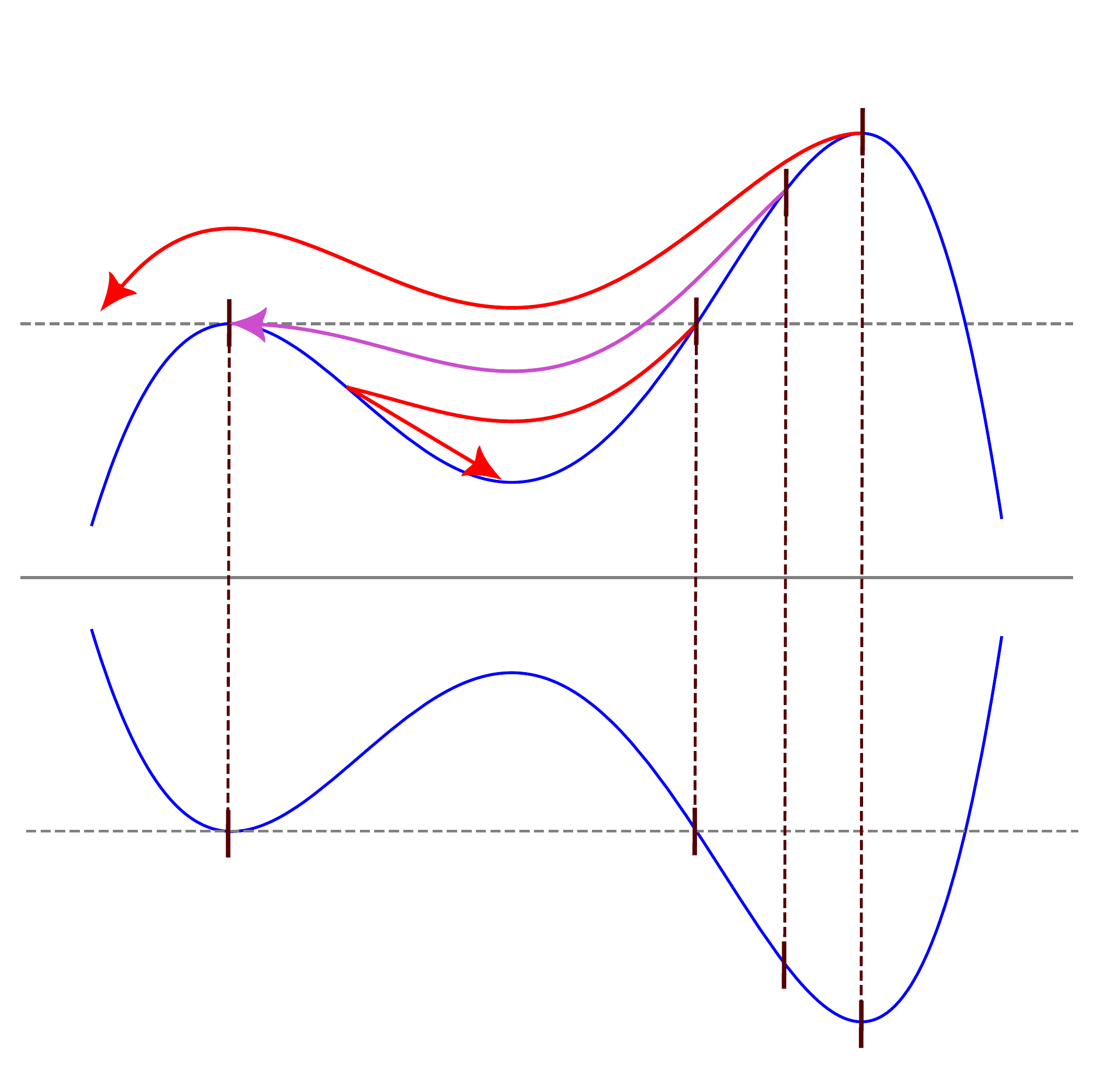}
\put(-8,60){$-V(\f)$}
\put(-2,45.5){$0$}
\put(-4,31){$V(\f)$}
\put(19,18){$\f_f$}
\put(61.25,17.25){$\f_\star$}
\put(69,5.5){$\f_0$}
\put(76.75,0.75){$\f_t$}
\end{overpic}
\caption{Potential $V(\f)$ with two minima at $\f=\f_f$ (`false vacuum') and $\f=\f_t$ (`true vacuum') with $V(\f_f) > V(\f_t)$, separated by a barrier. The inverted potential $-V(\f)$ exhibits two maxima at $\f_f$ and $\f_t$ separated by a valley. For the depicted potential, using the mechanical analogue of tunnelling solutions, a particle released at rest from $\f_t$ towards the left will overshoot $\f_f$, as depicted the uppermost red trajectory. A particle released at rest from $\f_\star$ with $V(\f_\star)=V(\f_f)$ will undershoot $\f_f$ and eventually settle at the bottom at the valley as depicted by the lower red trajectory. By continuity, there exists a locus $\f_0$, so that if the particle is released at rest from there, it will neither over- nor undershoot, but come to rest again at exactly $\f_f$. This trajectory is shown in purple.}
\label{fig:overundersec2}
\end{figure}
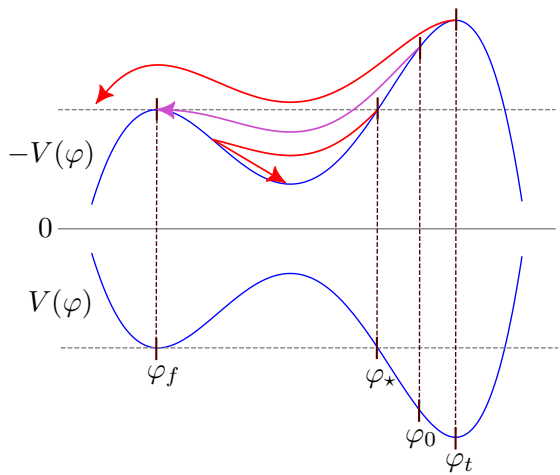

The observation is that in the case of a RG flow from a UV fixed point at a maximum of $V$, we can generically ensure regularity by adjusting the remaining free parameter. As a result, in a generic potential $V$ with a maximum, there generically exist holographic RG flows that are regular away from this maximum. In other words, maxima of $V$ are attractors for holographic RG flows.

For RG flow emanating from a UV fixed point at a minimum of $V$, there is no further free parameter that can be adjusted to ensure regularity and typically no regular solution can be found. Therefore, we confirm once more what we endeavoured to show in this section: In a generic potential, \emph{regular} holographic RG flows emanating from an AdS minimum of $V$ generically do not exist. Using the identification between $O(D)$-instantons and holographic RG flows, this confirms that tunnelling from AdS minima is non-generic, as we have concluded previously using the mechanical picture. The exceptions are potentials which are to some extent tuned, so that the boundary and regularity condition can be satisfied simultaneously by adjusting just one parameter.

An important question then is: how can one identify potentials that admit $O(D)$-instantons for tunnelling from an AdS minimum? In the following, we derive a sufficient condition on $V$ for such solutions to exist.

\noindent \textbf{Tunnelling from AdS minima: a sufficient condition on $V$ --}
Here we again consider a potential $V$ with two AdS minima at $\f=\f_f$ and $\f=\f_t$ with $V(\f_f) > V(\f_t)$, separated by a barrier.
To employ the mechanical analogue of $O(D)$-instantons, we now turn to the inverted potential $-V$. This has two maxima $\f_f$ and $\f_t$ with $-V(\f_f) < -V(\f_t)$, with a valley between them. Both the potential and inverted potential are shown in blue in fig.~\ref{fig:overundersec2}. To determine whether the potential $V$ permits a $O(D)$-instanton solution we shall study whether the corresponding mechanical solution in $-V$ exists.

In the following, we employ a form of the overshoot/undershoot argument originally introduced in \cite{Coleman:1977py}.
Consider starting at rest at $\f_t$ and letting the particle roll towards $\f_f$.\footnote{In practice, when solving numerically, for the particle to roll away in finite time, one has to start some finite distance away from $\f_t$.} There are three possibilities for the subsequent behaviour of the particle:
\begin{enumerate}
\item \textbf{Overshooting case:} As the initial potential energy at $\f_t$ is higher than the potential energy at $\f_f$, the particle can pass the trough and overshoot the maximum of $-V$ at $\f_f$.
\item \textbf{Undershooting case:} If the friction is sufficiently large, enough kinetic energy can be dissipated so that the particle undershoots the maximum of $-V$ at $\f_f$. Instead, the particle eventually settles at the bottom of the valley, after performing damped oscillations about this minimum of $-V$.
\item \textbf{Interpolating case:} The particle can cross the valley and come to rest exactly at $\f_f$. In the original Euclidean interpretation, this corresponds to a flat domain-wall solution. This is not expected to occur in a generic potential, which will either lead to the particle overshooting or undershooting.
\end{enumerate}

To continue, we first examine the overshooting case from above, i.e.~we assume that the potential is such that the particle overshoots $\f_f$ when released from rest at $\f_t$. This is shown by the upper red trajectory in fig.~\ref{fig:overundersec2}. Now consider releasing the particle not from the top of the hill of $-V$ at $\f_t$, but further down the slope. In particular, as a starting point consider the locus $\f=\f_\star$ where the particle has the same potential energy as at $\f=\f_f$, i.e.~$-V(\f_\star) = -V(\f_f)$, but on the other side of the trough from $\f_f$, as shown in fig.~\ref{fig:overundersec2}. Even though the particle initially has just sufficient potential energy to exactly reach $\f_f$ on the other side of the valley, it will undershoot and become trapped in the valley as a consequence of friction (see the lower red trajectory in fig.~\ref{fig:overundersec2}). This also applies to all other starting points that lie further down the slope than $\f_\star$. To summarise, here, when released from the top of the hill at $\f_t$, the particle will overshoot $\f_f$, but when released from $\f_\star$ or lower down the slope it will undershoot $\f_f$. By continuity it then follows that there exists some locus $\f = \f_0 \in [\f_{\star}, \f_t]$ for which the particle will neither over- nor undershoot, but exactly come to rest at $\f_f$. This is the mechanical version of the solution we seek: the Euclidean analogue will be an $O(D)$-solution describing tunnelling from $\f_f$. In fig.~\ref{fig:overundersec2} this is indicated by the purple trajectory.

We now turn to the undershooting and interpolating cases from above, i.e.~we assume that the inverted potential $-V$ is such that the particle either undershoots or arrives at rest at $\f_f$ when starting from $\f_t$. Now consider releasing the particle further down the hill. In this case the initial potential energy is lower, so it will be more difficult for the particle to reach $\f_f$ on the other side of the valley. At the same time, the velocity of the particle, when passing the valley will be lower, which lowers friction, so that less kinetic energy is dissipated. While at this stage we cannot exclude that these two effects may cancel one another, so that one may find a solution that comes to rest at $\f_f$ exactly, this is not expected to happen generically.\footnote{Correspondingly, we did not find any examples for this in any of the numerical examples studied.}  As before, once the starting point is lowered beyond $\f_{\star}$, i.e.~$-V(\f_{\textrm{start}}) < -V(\f_\star) = -V(\f_f)$,  the particle again must undershoot. All things considered, we do not expect potentials from the undershooting and interpolating cases above to permit $O(D)$-instanton solutions.

Based on these observations, we can identify a sufficient condition on the potential $V$ to permit a solution describing tunnelling from $\f_f$: \newline
\noindent \emph{A potential $V$ with two AdS minima at $\f=\f_f$ and $\f=\f_t$ with $V(\f_f) > V(\f_t)$ will permit an $O(D)$-instanton describing tunnelling from $\f_f$, if in the mechanical formulation, the particle overshoots $\f_f$ when released from rest at $\f_t$.}

Given a potential $V$ with two non-degenerate minima $\f_f$ and $\f_t$, the above condition allows for an easy  test to determine whether the false minimum $\f_f$ is prone to decay via $O(D)$-instantons. One simply has to solve \eqref{c3}--\eqref{c5} with the boundary conditions corresponding to the particle released from rest at $\f_t$. In the Euclidean picture, these are the boundary conditions for a flat domain-wall solution ending at $\f_t$ i.e.\footnote{In practice, the boundary conditions are implemented at some large but finite value of $u$. The appropriate boundary conditions can be read-off from the near-extremum behaviour of $A(u)$ and $\f(u)$ shown in \eqref{eq:Anearextflat} and \eqref{eq:phinearextflat}.}
\begin{align}
\f(u \rightarrow + \infty) \rightarrow \f_t \, , \qquad e^A(u \rightarrow + \infty) \rightarrow 0 \, .
\end{align}
In section \ref{sec:num} we shall use this criterion to design potentials that admit $O(D)$-tunnelling solutions.

\subsection{$O(D)$-instantons and exotic holographic RG flows}
\label{sec:exotic}
We now show that the existence of an $O(D)$-instanton describing
tunnelling from an AdS minimum is tied to the existence of a
holographic RG flow solution that exhibits `exotic' phenomena, such as
reversals of the flow in $\f$-space and the skipping of fixed
points. Such exotic RG flows have been studied in detail in
\cite{Kiritsis:2016kog} for field theories on flat space-time and in
\cite{Ghosh:2017big} for field theories on constant curvature
backgrounds. Here, we  argue that a potential that admits an
$O(D)$-instanton, describing tunnelling from an AdS minimum, will also
admit holographic RG flows with these exotic
features. Conversely, the existence of exotic RG flows will imply the
  existence of $O(D)$-instanton solutions.

\begin{figure}[t]
\centering
\begin{overpic}[width=0.75\textwidth,tics=10]{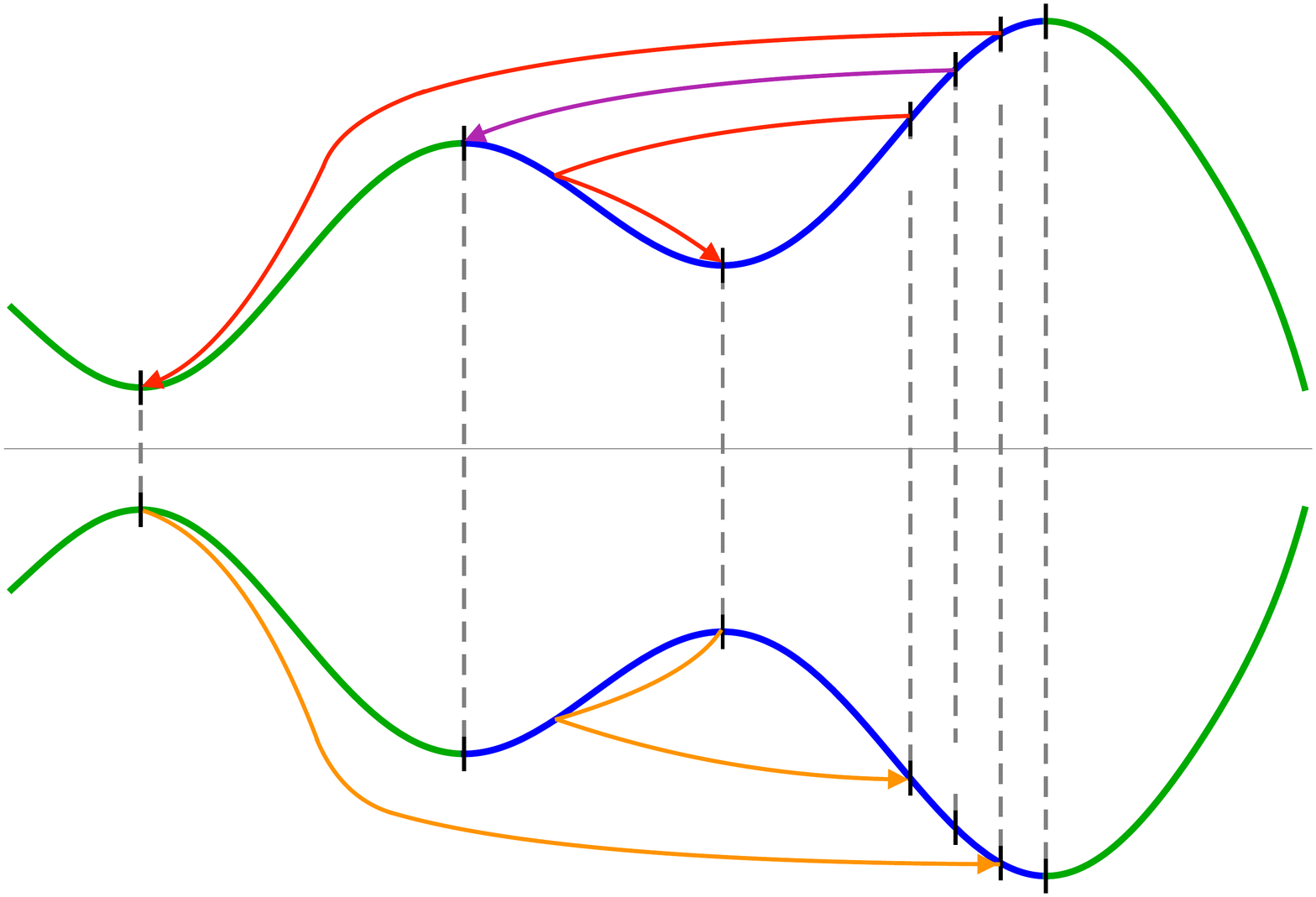}
\put(-2,48){$-V(\f)$}
\put(-2,34.5){$0$}
\put(-2,21){$V(\f)$}
\put(33.5,9.5){$\f_f$}
\put(67.1,55.75){$\f_b$}
\put(70.25,11.2){$\f_0$}
\put(74,62.5){$\f_a$}
\put(77.4,0){$\f_t$}
\put(9,27){$\f_{\textsc{uv}_1}$}
\put(53.5,18.25){$\f_{\textsc{uv}_2}$}
\end{overpic}
\caption{Potential $V(\f)$ with two AdS minima at $\f_f$ and $\f_t$ with $V(\f_f)>V(\f_t)$, separated by a maximum at $\f_{\textsc{uv}_2}$ (blue region). The potential outside the region $[\f_f,\f_t]$ is shown in green. Beyond the minimum at $\f_f$ the potential exhibits another maximum at $\f_{\textsc{uv}_1}$. The inverted potential exhibits two maxima at $\f_f$, $\f_t$ and two minima at $\f_{\textsc{uv}_1}$, $\f_{\textsc{uv}_2}$. The potential admits an $O(D)$-instanton describing tunnelling from the false minimum $\f_f$ to the locus $\f_0$. The mechanical analogue of this is shown in purple denoting the trajectory of a particle released at rest from $\f_0$ and rolling to $\f_f$. If the particle is released from $\f_b$ somewhat below $\f_0$ it undershoots $\f_f$, reverses direction and eventually settles at $\f_{\textsc{uv}_2}$ (lower red trajectory). In the Euclidean framework this corresponds to a holographic RG flow from a UV fixed point at $\f_{\textsc{uv}_2}$ to an IR end point at $\f_b$ that `bounces', i.e.~reverses direction in $\f$ (upper orange flow). If the particle is released from $\f_a$ somewhat above $\f_0$ it overshoots $\f_f$ and comes to rest in $\f_{\textsc{uv}_1}$ (upper red trajectory). In the Euclidean framework this corresponds to a holographic RG flow from a UV fixed point at $\f_{\textsc{uv}_1}$ to $\f_a$ (lower orange trajectory). This flow `skips' over the other UV fixed point at $\f_{\textsc{uv}_2}$.}
\label{fig:bounceskip}
\end{figure}

To be specific, consider once more a potential with two minima at $\f=\f_f$ and $\f=\f_t$ with $V(\f_f) > V(\f_t)$. Now we also assume that the potential admits an $O(D)$-instanton solution describing tunnelling from $\f_f$ to some locus $\f=\f_0$ that is not an extremum ($V'(\f_0)\not=0$). We shall argue that this potential will also admit  holographic RG flows with the exotic features mentioned above.

We turn once more to the mechanical picture introduced in the previous section. There, we found that the $O(D)$-instanton can be identified as the solution at the interface of solutions where the particle overshoots or undershoots $\f_f$. That is, if the particle is released not at $\f_0$, but slightly further up the slope of $-V$, it will overshoot. If instead it is released just below $\f_0$ it will undershoot. In the following, we focus on the solutions in the over- and undershooting case. These are solutions to the equations of motion \eqref{c3}--\eqref{c5} and hence also possess an interpretation in terms of holographic RG flows. We find that these solutions can be matched with the so-called exotic holographic RG flows.

\vspace{0.1cm}

\noindent \textbf{Undershooting solutions as bouncing flows --} Consider releasing the particle not from $\f_0$ but further down the slope of $-V$ from $\f_0$ so that the particle undershoots $\f_f$. Once it undershoots, the particle is trapped in the valley of $-V$ between $\f_f$ and $\f_t$ and will eventually settle at the bottom. 
In the Euclidean framework, solutions that end at the bottom of the valley of $-V$ between $\f_f$ and $\f_t$ correspond to holographic RG flows from a UV fixed point at the maximum that is located at the top of the barrier separating $\f_f$ and $\f_t$. This UV fixed point is denoted as $\f_{\textsc{uv}_2}$ in fig.~\ref{fig:bounceskip}.

The observation is that if the particle is released sufficiently close to $\f_0$, it will not just roll to the bottom of the valley directly, but it will oscillate about the bottom of the valley (at least once). The reason is that if the particle is released sufficiently close to $\f_0$, it should just fail to reach $\f_f$ on the other side of the valley and hence passes the bottom of the valley at least once before eventually settling there. This is shown in fig.~\ref{fig:bounceskip} as the red trajectory starting at $\f_b$. In the Euclidean picture this will correspond to an RG flow that leaves the UV fixed point at $\f_{\textsc{uv}_2}$ towards $\f_f$, reverses direction and passes the maximum at least once, before finally ending at $\f_b$. This is shown as the orange flow from $\f_{\textsc{uv}_2}$ in fig.~\ref{fig:bounceskip}. The reversal of flow direction in $\f$-space was termed a `bounce' in \cite{Kiritsis:2016kog,Ghosh:2017big}. The observation is that a potential that admits $O(D)$-tunnelling from an AdS minimum will typically also permit bouncing flows from the UV fixed point $\f_{\textsc{uv}_2}$, i.e.~the top of the barrier separating the false and true minimum.

\vspace{0.1cm}

\noindent \textbf{Overshooting solutions as skipping flows --} Now consider releasing the particle not from $\f_0$ but further up the slope of $-V$ from $\f_0$ so that the particle overshoots $\f_f$. To discuss the subsequent evolution of the particle, we need to extend the trajectory of the particle beyond $\f_f$. As $\f_f$ is a minimum, the potential beyond it can either rise monotonically and eventually diverge, or it could reach another maximum.\footnote{We ignore the possibility that $V$ remains constant beyond $\f_f$ as this seems highly artificial and we have no such examples in string theory.} For the divergent case, no regular solution for the subsequent evolution of the particle can be found and we ignore this possibility.

We therefore consider the case that $V$ possesses another maximum beyond $\f_f$, which in fig.~\ref{fig:bounceskip} we denoted by $\f_{\textsc{uv}_1}$. In the inverted potential $-V$ the maximum at $\f_{\textsc{uv}_1}$ translates into a minimum. As seen before, minima of $-V$ are attractors for the particle. Therefore, once the particle overshoots $\f_f$ and enters the valley of $\f_{\textsc{uv}_1}$ it will typically settle at the bottom of this minimum (or another minimum of $-V$ if the particle overshoots the valley of $\f_{\textsc{uv}_1}$ as well). This is shown as the red trajectory originating from $\f_a$ in fig.~\ref{fig:bounceskip}

In the Euclidean picture this will correspond to a RG flow from a UV fixed-point at $\f_{\textsc{uv}_1}$ to $\f_a$. Most importantly, this RG flow skips the other potential UV fixed point at $\f_{\textsc{uv}_2}$. Flows that bypass viable fixed points at extrema of the potential were denoted as `skipping' flows in \cite{Kiritsis:2016kog,Ghosh:2017big}. From the above reasoning we expect that  a potential that admits $O(D)$-tunnelling from an AdS minimum will also permit skipping flows. For the skipping flow the relevant UV fixed point will be another maximum outside the range $[\f_f, \f_t]$, here denoted by $\f_{\textsc{uv}_1}$.

The same argument can also be run in reverse: the existence of
  curved skipping RG flows which overshoot $\f_f$ and of curved
  bouncing RG flows which undershoot it imply the existence of a single flow as  the limiting case which
  separates the two classes of solutions, and which connects $\f_f$
  with  point $\f_0$ on the slope of the potential. This flow is the
  $O(D)$-instanton which describes decay of the false AdS$_D$ vacuum.

In the discussion above, the RG flows were for QFTs defined on $S^d$, but interesting statements can also be made for flat-sliced flows that connect extrema of the potential. Consider for example a potential that admits a flat-sliced skipping flow from $\f_{\textsc{uv}_1}$ to $\f_t$. This is the RG flow equivalent of the overshooting mechanical solution for a particle starting at rest from $\f_t$, which we identified as a sufficient condition for a potential to admit an $O(D)$-instanton.

An explicit example for a potential that exhibits the features discussed here can be found in \cite{Kiritsis:2016kog,Ghosh:2017big}. In particular in \cite{Ghosh:2017big} the various RG flow solutions for all possible flow end points in this potential were derived and their holographic interpretation was discussed systematically. We refer readers to this work for more details.


\section{Flat AdS domain-walls and a thin-wall limit}  \label{sec:flat}
In this section we shall construct explicit solutions for flat AdS domain-walls and the corresponding potential permitting these solutions. While these solutions do not describe tunnelling processes, we find it nevertheless convenient to discuss them for the following reasons. The potentials that we derive here, will be good starting points for constructing potentials that will admit $O(D)$-instanton solutions, our primary target. This is valuable, as generic potentials typically do not permit these solutions as discussed above. Furthermore, flat domain-wall solutions can be understood as a limit of $O(D)$-instanton solutions where the curvature of the bubble wall is taken to zero. As a result, the analytic expressions we shall compute for flat domain-walls, will be helpful for making analytic statements about $O(D)$-instanton solutions in a certain limit, at least approximately.

\subsection{Analytic flat domain-wall solutions with a thin-wall limit} \label{sec:flatthinwall}
Here we  construct families of flat AdS domain-walls that will explicitly admit a limit in which the domain wall becomes infinitely thin. It behoves to point out that this `thin-wall limit' for flat domain walls is a priori completely unrelated to the more familiar thin-wall limit of $O(D)$-instantons. For $O(D)$-symmetric solutions, the bubble wall is deemed thin, if the interval in the physical radius $r$ over which the interpolation occurs, is small compared to the bubble radius $\bar{r}$, as described in section \ref{sec:CDLsetup}. For flat domain walls there is no notion of a radial coordinate and a different definition is needed. In this case we define the wall as the interval in the coordinate $u$ over which the interpolation between the two AdS backgrounds occurs. To decide whether this wall is thin, we need to compare this to another length scale. For AdS domain-walls, there exist two further length scales in the form of the AdS lengths of the two AdS backgrounds connected by the wall. Therefore, {\em we define a flat AdS domain-wall to be thin if the length of the interval in $u$ where interpolation occurs is small compared to either AdS length scale}.
This  gives an invariant definition in the gravitational theory, but has no clear meaning in the dual QFT.

We now proceed to constructing such a family of solutions explicitly. The relevant ansatz is given by \eqref{metricansatz} with $\zeta_{\mu \nu}$ a metric on flat Euclidean space. The corresponding equations of motion are given in \eqref{c3}-\eqref{c5} with $R^{(\z)}=0$. We seek solutions that:
\begin{itemize}
\item interpolate between two AdS space-times with length scales $\ell_f$ and $\ell_t$, respectively, with $\ell_t < \ell_f$;
\item interpolate at the same time between the values $\f=\f_f$ and $\f=\f_t$.
\end{itemize}
AdS solutions for the ansatz \eqref{metricansatz} with $\zeta_{\mu \nu}$ describing Euclidean space correspond to a scale factor $A(u) = - u / \ell$, $\dot{A}(u)= - 1 / \ell$, with $\ell$ the relevant AdS length. The two demands in the list above are hence equivalent to the following boundary conditions:
\be
\dot A=\left\{ \begin{array}{lll}
\displaystyle - \frac{1}{\ell_f} ,&\phantom{aa} & u \rightarrow - \infty     ,\\ \\
\displaystyle - \frac{1}{\ell_t} ,&\phantom{aa}& u \rightarrow \infty    .
\end{array}\right.\sp
\f=\left\{ \begin{array}{lll}
\displaystyle \f_f,&\phantom{aa} & u \rightarrow - \infty    ,\\ \\
\displaystyle \f_t,&\phantom{aa}& u \rightarrow \infty  .
\end{array}\right.
\label{b5}
\ee

We now turn to another demand on our solution, the existence of a thin-wall limit in the sense described above. The limit of a vanishingly thin wall corresponds to the case where the interpolation happens instantaneously, i.e.~$\dot{A}$ and $\f$ jump discontinuously between the two boundary values at some locus $u =\bar{u}$. To implement this, we try the ansatz $\dot{A} \sim \Theta(u-\bar{u})$, $\f \sim \Theta(u-\bar{u})$, with $\Theta$ referring to the Heaviside theta function, which in turn implies $\ddot A\sim \delta(u-\bar{u})$ and $\dot \f\sim \delta(u-\bar{u})$. However, one can easily confirm that in this case, none of the equations of motion \eqref{c3}--\eqref{c5} can be solved, i.e.~the ansatz is invalid. This does not imply that an extreme thin-wall limit is impossible, it just shows that \emph{not both} of $\ddot{A}$ and $\dot{\f}$ can simultaneously be a $\delta$-function. One way of overcoming the problem is to demand that only one of $\ddot{A}$ or $\dot{\f}$ becomes a $\delta$-function in the extreme thin-wall limit. This is what we do here, choosing that $\dot{\f} \sim \delta(u-\bar{u})$ in the extreme limit. The alternative analysis for $\ddot{A} \sim \delta(u-\bar{u})$ can also be done and is shown in appendix \ref{app:altthinwall}.

One way of ensuring that $\dot{\f}$ permits an extreme thin-wall limit with $\dot{\f} \sim \delta(u-\bar{u})$ is to propose a solution for $\dot{\f}$ in terms of a resolved $\delta$-function. There are many ways of resolving $\delta$-functions\footnote{We can  use any smooth interpolating function for this argument.} , but inspired by Coleman's analysis in \cite{Coleman:1977py, Callan:1977pt} we  write
\begin{align}
\label{eq:phidotansatzflat} \dot{\f}(u) = \frac{\Delta}{4 \ell_f} \, \frac{\varrho}{\cosh^2 \big( \frac{\Delta}{2} \frac{u-\bar{u}}{\ell_f} \big)}\, ,
\end{align}
with $\Delta$ a numerical constant and $\varrho$ a free parameter to be ultimately fixed by the boundary conditions. We also included $\ell_f$ for dimensional reasons. The extreme thin-wall limit corresponds to $\Delta \rightarrow +\infty$, for which $\dot{\f} \rightarrow \varrho \, \delta(u-\bar{u})$ as required. {The choice of the symbol $\Delta$ here is no accident. So far we have used $\Delta$ to refer to the scaling dimension of the operator perturbing the CFT associated with an AdS extremum of $V$. As will be confirmed later, for $\f_f$ a minimum of $V$, the parameter $\Delta$ introduced above will indeed correspond to the scaling dimension, consistent with our previous use of $\Delta$. If $\f_f$ is a maximum of $V$, this identification does not always hold, as we shall explain later. This will not be a problem, as in this paper $\f_f$ always describes a minimum of $V$ unless explicitly stated otherwise.}

Using \eqref{eq:phidotansatzflat}, the equations of motion \eqref{c3}--\eqref{c5} and the boundary conditions \eqref{b5} we can construct the desired solutions for $\f(u)$, $A(u)$ and even the potential $V(\f)$. Integrating \eqref{eq:phidotansatzflat} and implementing the boundary conditions for $\f$ we find
\begin{align}
\label{eq:phisolflat} \f(u) = \bar{\f} + \frac{\varrho}{2} \tanh \left( \frac{\Delta}{2} \frac{u-\bar{u}}{\ell_f} \right) \, , \quad \textrm{with} \quad \bar{\f} \equiv \frac{\f_f+\f_t}{2} \, , \quad \varrho \equiv \f_t-\f_f \, ,
\end{align}
i.e.~the boundary conditions for $\f$ fix the value of $\varrho$. Inserting this into \eqref{c3} we can integrate to obtain the expression for $\dot{A}$. After implementing the boundary conditions in \eqref{b5} one finds
\begin{align}
\label{eq:Adotsolflat} \dot{A}(u) = - \frac{1}{\ell_f} - \frac{\Delta \, \varrho^2}{48(d-1) \ell_f} \left[ 2 + 2 \tanh \left( \frac{\Delta}{2} \frac{u-\bar{u}}{\ell_f} \right) + \frac{\tanh \left( \frac{\Delta}{2} \frac{u-\bar{u}}{\ell_f} \right)}{\cosh^2 \left( \frac{\Delta}{2} \frac{u-\bar{u}}{\ell_f} \right)}\right] \, ,
\end{align}
together with the relation
\begin{align}
\label{eq:Vparameterconstraint} \frac{1}{\ell_t} = \frac{1}{\ell_f} + \frac{\Delta \, \varrho^2}{12(d-1) \ell_f} \, ,
\end{align}
i.e.~out of the four parameters $\ell_f$, $\ell_t$, $\Delta$ and $\varrho$ only three can be chosen freely. Integrating once more we finally obtain
\begin{align}
\label{eq:Asolflat} A(u) = A_0 - \frac{u}{\ell_f} - \frac{\varrho^2}{24(d-1)} \left[ \frac{\Delta u}{\ell_f} + 2 \log \cosh \left( \frac{\Delta}{2} \frac{u-\bar{u}}{\ell_f} \right) - \frac{1}{2 \cosh^2 \left( \frac{\Delta}{2} \frac{u-\bar{u}}{\ell_f} \right)} \right] \, ,
\end{align}
with $A_0$ an integration constant. We can evaluate the Kretschmann invariant which for the ansatz \eqref{metricansatz} gives
\be
R_{\m\n;\r\s}R^{\m\n;\r\s}=4d(\ddot A+\dot A^2)^2+2d(d-1)\dot A^4={\frac{d}{(d-1)^2}}\left(2(d-1)\dot A^2-\dot\f^2\right)^2+2d(d-1)\dot A^4 \, ,
\label{b36}\ee
where in the second step we have used \eqref{c3}.  One can easily check that for the solutions considered here, this is regular everywhere. Therefore, we have succeeded in finding analytic solutions for $\f(u)$ and $A(u)$ for flat AdS domain-walls, that admit a thin-wall limit in the sense described above, which is reached for $\Delta \rightarrow + \infty$.

Finally, using \eqref{c4} (with $R^{(\zeta)}$ set to zero) and inserting our solutions for $\dot{\f}$ and $\dot{A}$ from \eqref{eq:phidotansatzflat}, \eqref{eq:Adotsolflat} we can arrive at an expression for a potential $V$ that admits this flat domain-wall solution. To write the potential as a function of $\f$ we invert \eqref{eq:phisolflat} to eliminate $u$ in favour of $\f$. To remove clutter, here we set $\f_f=0$ such that $\bar{\f}=\f_t /2$ and $\varrho=\f_t$. After some algebra one finds:
\begin{align}
\centering
\nonumber V(\f)= &-\frac{d(d-1)}{\ell_{f}^2}-\frac{\Delta(d-\Delta)}{2\ell_{f}^2}\f^2+\frac{\Delta(d-3\Delta)}{3\ell_{f}^2 \f_t}\f^3-\frac{\Delta^2\left(8-8d+d\f_t^2 \right)}{16(d-1)\ell_{f}^2 \f_t^2}\f^4 \\
&+\frac{d\Delta^2}{12(d-1)\ell_{f}^2\f_t}\f^5-\frac{d\Delta^2}{36(d-1)\ell_{f}^2\f_t^2}\f^6 \, .  \label{com3}
\end{align}
We also compute the corresponding solution $W(\f) = - 2(d-1) \dot{A}$. This can obtained from our expression for $\dot{A}$ in \eqref{eq:Adotsolflat} and eliminating $u$ in favour of $\f$ by inverting \eqref{eq:phisolflat}. Setting again $\f_f=0$ we find:
\begin{align}
\label{eq:Wflatanalytic} W_{\textrm{flat}}(\f) = \frac{1}{\ell_f} \left[2(d-1) + \frac{\Delta}{2} \f^2 -\frac{\Delta}{3 \f_t} \f^3 \right] \, ,
\end{align}
where we added the subscript `flat' to indicate that this describes a flat AdS domain-wall.

The potential \eqref{com3} has the following features:
\begin{itemize}
\item It has two AdS extrema at $\f_f=0$ and $\f_t$ with $V(\f_f) > V(\f_t)$. The extremum at $\f_t$ is always a minimum. The extremum at $\f_f=0$ is a maximum for $\Delta < d$ and a minimum for $\Delta > d$. In the latter case there is an intermediate maximum at some $\f \in [\f_f,\f_t]$.
\item The potential has three parameters that we are free to control: $\ell_f$, $\Delta$ and $\f_t$. The parameter $\ell_f$ adjusts the value of the potential at $\f_f=0$. By dialing $\Delta$ we can adjust the height of the barrier between $\f_f$ and $\f_t$. By choosing $\f_t$ we can set the separation in $\f$ between the two extrema at $\f_f=0$ and $\f_t$. Alternatively, an adjustment in $\f_t$ can also be used to set the value of the potential at that minimum through the relation \eqref{eq:Vparameterconstraint}.
\end{itemize}

The above solution also has an interpretation in terms of a holographic RG flow for a field theory on flat space(-time), as we briefly describe in the following.
\begin{itemize}
\item The above solutions correspond to RG flows from a UV fixed point at the extremum of $V$ at $\f_f=0$ to an IR fixed point at the extremum $\f_t$. The identification of $\f_f$ with the UV fixed point follows from the fact that when $\f_f$ is reached as $u \rightarrow -\infty$, simultaneously one finds $e^A \rightarrow \infty$, as expected for a UV fixed point. Similarly, when $\f_t$ is reached as $u \rightarrow +\infty$, one finds $e^A \rightarrow 0$, as required for an IR fixed point. One can also check that for $u \rightarrow -\infty$ the solution exhibits the expected asymptotic behaviour near a UV fixed point, as detailed in e.g.~\cite{Kiritsis:2016kog}.
\item For $\Delta > d/2$ one finds that the parameter $\Delta$ is related to $V''(\f_f)$ via \eqref{eq:Deltageneraldef}, as can be checked explicitly. This implies that in this case, $\Delta$ corresponds to the scaling dimension of the operator perturbing the UV CFT associated with the AdS extremum at $\f_f$, consistent with our use of $\Delta$ throughout this work. Note that this includes the case where $\f_f$ is a minimum of $V$, i.e.~$\Delta >d$, which is the situation of primary interest in this work. Comparing \eqref{eq:Wflatanalytic} with (\ref{eq:Wmreg}, \ref{eq:Wpreg}) we also conclude that for $\Delta > d/2$ the solution constructed here is a $(+)$-branch solution according to the classification introduced in sec.~\ref{sec:CDLasHoloRG}.
\item  For $0 < \Delta < d/2$ the relation between $V$ and $\Delta$ is not \eqref{eq:Deltageneraldef}, but instead (note the minus sign)
\begin{align}
\Delta \equiv \frac{d}{2} \left(1 - \sqrt{1 +\frac{4 (d-1)}{d} \frac{V''(\f_{\textrm{ext}})}{|V(\f_{\textrm{ext}})|}} \right) \, .
\end{align}
Therefore in this case the solution corresponds to the $(-)$-branch,
i.e.~the UV CFT is perturbed by the presence of a source for a
relevant operator.
\end{itemize}

\noindent \textbf{The thin-wall limit of flat domain-wall solutions:} The solutions derived above have been designed to admit a thin-wall limit in the sense described at the beginning of this section, which is reached for $\Delta \rightarrow + \infty$. In the following, we discuss which potentials $V$ correspond to thin-walled flat domain walls. We can distinguish two cases:

\begin{enumerate}
\item Consider taking $\Delta \rightarrow + \infty$ with the parameters $\ell_f$ and $\varrho$ (or $\f_t$ for $\f_f=0$) fixed. The effect on $V$ in \eqref{com3} is that the height of the barrier separating the two minima diverges, which is best checked numerically. However, at the same time the potential difference between the two minima $V(\f_f) - V(\f_t)$ also diverges. This follows from \eqref{eq:Vparameterconstraint}, which implies that $\ell_t \rightarrow 0$ for $\Delta \rightarrow + \infty$ with $\ell_f$ and $\varrho$ fixed. Therefore, for finite separation $\f_t-\f_f$, the extreme thin-wall limit of a flat AdS domain-wall requires a potential with diverging barrier and a diverging potential difference between the minima.
\item Alternatively, we can also consider the extreme thin-wall limit with the potential difference between the two minima fixed. That is, while increasing $\Delta$ we keep both $\ell_f$ and $\ell_t$ constant. The potential barrier between the minima again diverges for $\Delta \rightarrow + \infty$. This can e.g.~be seen analytically by evaluating $V(\f_t/2)$, which for $\Delta \rightarrow +\infty$ can be expanded as
\begin{align}
V(\f_t/2) = \frac{3(d-1)}{8} \frac{\ell_f-\ell_t}{\ell_f^2 \ell_t} \Delta + \mathcal{O}(\Delta^0) \, .
\end{align}
From \eqref{eq:Vparameterconstraint} it further follows that for fixed $\ell_f$ and $\ell_i$, the separation $\varrho$ between the minima decreases as $\Delta$ is increased. That is, we can reach the extreme thin-wall limit with a finite potential difference between the minima at the expense of the minima getting infinitesimally close to one another in $\f$-space.
\end{enumerate}

Although we obtained the result using a specific interpolation, any other interpolation produces qualitatively similar results.

\noindent \textbf{The thin-wall limit from a holographic viewpoint --}
So far in this section $\Delta$ has been mainly treated as a free parameter. However, in the holographic interpretation of flat domain-wall solutions as holographic RG flows, $\Delta$ is the scaling dimension of the scalar operator in  the UV CFT. Therefore, there may be bounds on $\Delta$ coming from consistency conditions on the field theory side. The thin-wall limit looks particularly challenging from the dual field theory point of view. This corresponds to a diverging scaling dimension, $\Delta \rightarrow \infty$, implying that the UV CFT is deformed by an operator that is de-facto infinitely irrelevant, but even a finite but large value of $\Delta$ may be problematic from the field-theory point of view. For example, results for 3d CFTs from the application of conformal bootstrap ideas have shown that their lowest lying operator must have scaling dimensions $\Delta \lesssim 7$ (for parity even operators) or $\Delta \lesssim 12$ (for parity odd operators) \cite{Dymarsky:2017yzx}. Therefore, field theories with excessively large values for $\Delta$ may not be physical and thus, the thin-wall limit may never arise in practice. One way of avoiding this would be to construct solutions that permit a thin-wall limit that does not rely on $\Delta \rightarrow \infty$. We leave this as an interesting question for future work.

\subsection{Flat domain-walls approximating $O(D)$-instantons } \label{sec:O4asflat}
In this section we  argue that the analytic results for flat domain-walls can be used to derive approximate expressions for certain $O(D)$-instantons. We begin by comparing $O(D)$-symmetric solutions and flat domain-walls, which are both captured by the ansatz \eqref{metricansatz}. The difference between $O(D)$-symmetric solutions and flat domain-walls is in the geometry of the slices at fixed coordinate $u$: In the former case a fixed-$u$ slice describes a $d$-sphere while in the latter case it is $d$-dimensional Euclidean space. That is, for $O(D)$-instantons fixed-$u$ slices are curved while for domain-walls they are flat.

The physical radius of the $d$-sphere slices in the $O(D)$-symmetric case is given by $r=\alpha e^{A(u)}$. Near the centre of the $O(D)$ bubble, i.e.~for $r \rightarrow 0$, the corresponding $d$-sphere is highly curved and we hence expect the behaviour of a $O(D)$-instanton to depart significantly from that of a flat domain-wall solution. This is indeed the case: at some finite value $u_0$, one reaches the centre of the $O(D)$ bubble, while a flat domain-wall continues until $u \rightarrow + \infty$. In contrast, for large radii, $r \rightarrow \infty$, the curvature of the $d$-sphere slices becomes vanishingly small and consequently only affects the solution at subleading order. At large radius the behaviour of a $O(D)$-instanton is therefore well-approximated by a corresponding flat domain-wall. The picture that emerges is, that at sufficiently large radius, any $O(D)$-instanton behaves like a flat domain-wall, but departs from it more and more as the radius is reduced.

An interesting situation occurs if the wall of an $O(D)$-instanton, i.e.~the radial interval in which the interpolation effectively happens, is located entirely in the region where the $O(D)$-solution can still be well-approximated by a flat domain-wall. This is not just a theoretical exercise,  as in section \ref{sec:num} we shall find numerical examples that exhibit this. Consider for example an $O(D)$-instanton that is well-approximated deep into the interior of the bubble by a flat domain wall solution of the type derived in section \ref{sec:flat}. In this case we can use the analytic expressions from section \ref{sec:flat} to describe the $O(D)$-instanton including the wall. For such solutions we can then, for example, give an approximate expression for the `thin-ness' parameter $\eta$ defined in \eqref{eq:etadefOD}, i.e.
\begin{align}
\eta_{\textrm{flat}} = \frac{e^{A_{\textrm{flat}}(u_{\textrm{out}})} - e^{A_{\textrm{flat}}(u_{\textrm{in}})}}{e^{A_{\textrm{flat}}(\bar{u})}} \, ,
\end{align}
where the subscript `flat' indicates that this is an approximate expression calculated using the exact solution for a flat domain-wall.

We can give an explicit expression for $\eta_{\textrm{flat}}$ for the flat domain-wall solution derived in the previous section. For $u_{\textrm{out}}$ and $u_{\textrm{in}}$ we use the implicit definition that we introduced in \eqref{eq:phiinoutdefOD} for the $O(D)$-symmetric case, but with $\f_0$ replaced by $\f_t$. Then, by employing the analytic solution for $\f(u)$ in \eqref{eq:phisolflat} we can solve for $u_{\textrm{out}}$ and $u_{\textrm{in}}$ explicitly. Using this and inserting for $A_{\textrm{flat}}(u)$ with \eqref{eq:Asolflat}, after some algebra we arrive at
\begin{align}
\label{eq:etaflatexplicit} \eta_{\textrm{flat}} = 2 \,  e^{-\frac{\ell_f /\ell_t -1}{4\Delta}\left[ \gamma^2 -2 \log (1-\gamma^2)\right]} \sinh \bigg( \frac{1+ \ell_f / \ell_t}{\Delta} \tanh^{-1} \gamma \bigg) \, ,
\end{align}
where we also used \eqref{eq:Vparameterconstraint} to eliminate $\f_t$. Recall that  $\gamma$ is the parameter introduced in \eqref{eq:phiinoutdefOD} specifying the extent of the wall in $\f$-space. In all numerical examples we set $\gamma=0.76$. We shall return to the expression in \eqref{eq:etaflatexplicit} when studying thin-walled $O(D)$-instanton solutions in section \ref{sec:ODthinwall}. {The reason that \eqref{eq:etaflatexplicit} will be useful there is that the potentials used in this section, will correspond to deformations of the potential \eqref{com3}, for which the flat domain-wall solution was derived.}


\section{O(4)-instantons in a sextic potential} \label{sec:num}

In this section we present a strategy for designing potentials that admit $O(D)$-instanton solutions.

We shall then obtain numerical solutions for $O(D)$-instantons in a family of potentials constructed using this method. We aim at answering the following questions:
\begin{itemize}
\item How to construct potentials that admit $O(D)$-instantons?
\item What kind of potentials admit \emph{thin-walled} $O(D)$-instantons?
\item What is the instanton action and hence the decay rate for the various solutions?
\end{itemize}
A challenge for constructing $O(D)$-instanton solutions is that they do not exist for generic potentials as shown in section \ref{sec:CDLnongeneric}. Therefore, to find $O(D)$-instanton solutions one first needs to identify potentials that do admit them. Interestingly, this can be done by starting with a potential admitting a flat domain wall solution, and then deforming it, as we argue next.

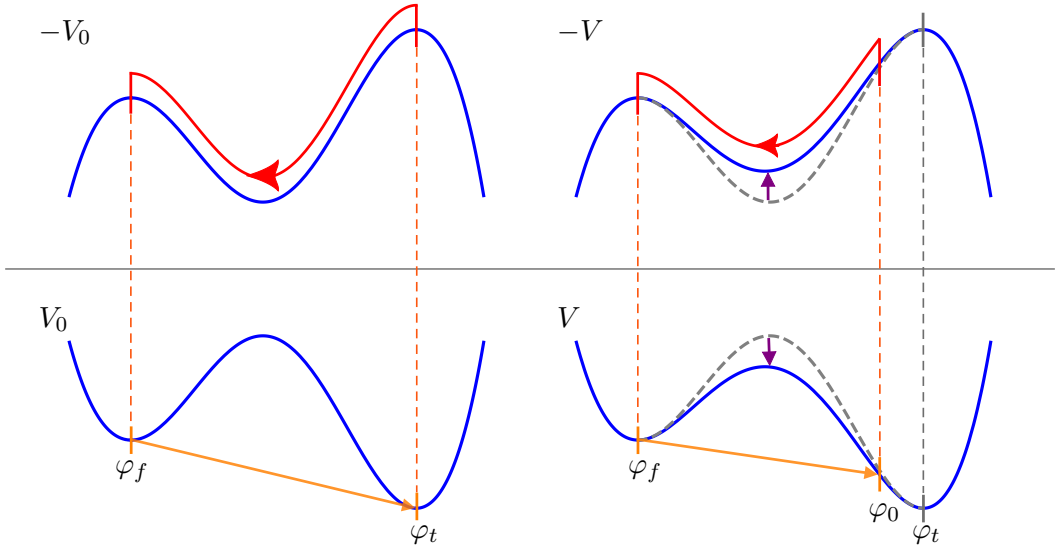
\begin{figure}[t]
\centering
\begin{overpic}
[width=0.95\textwidth]{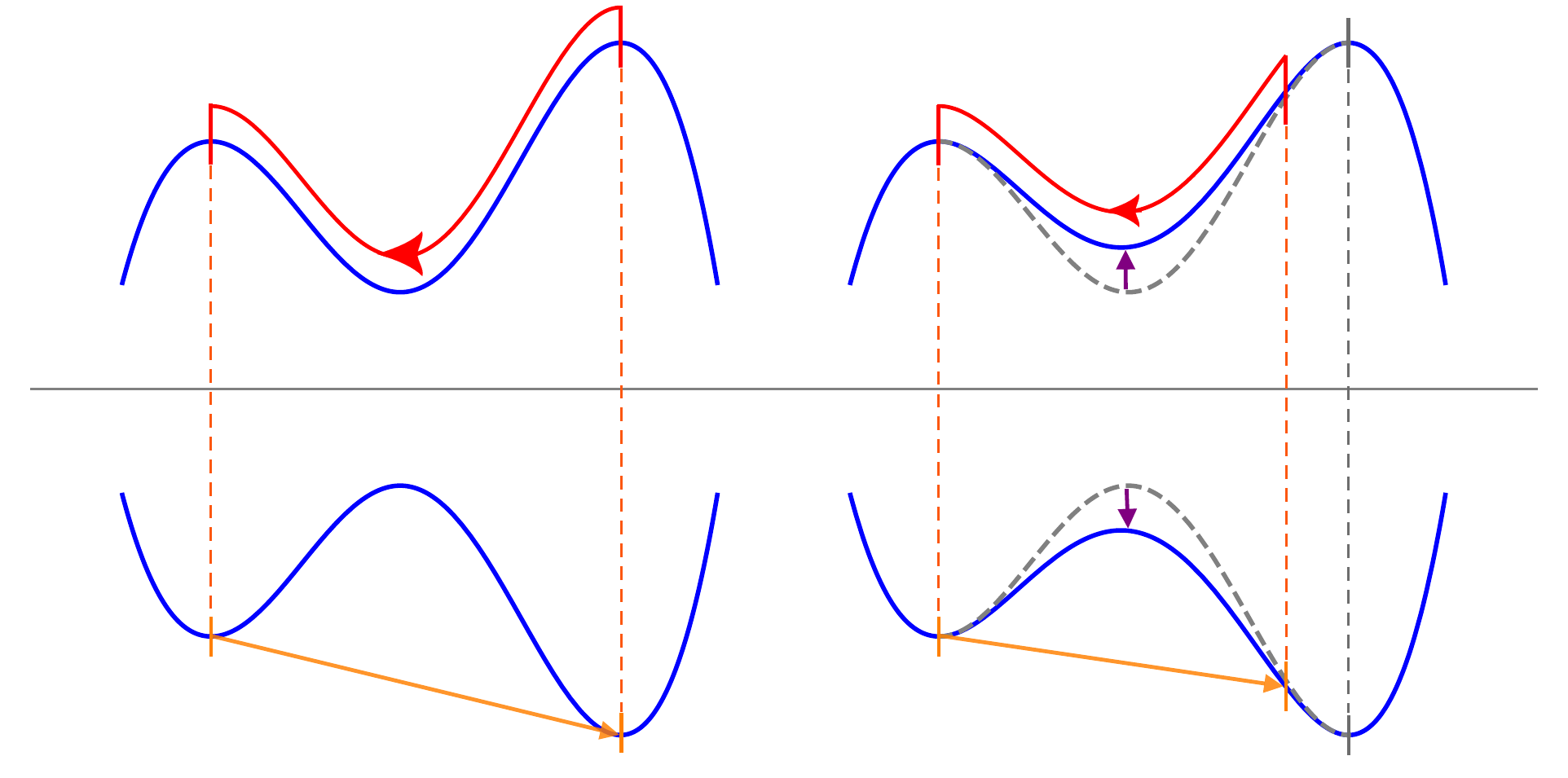}
\put(5,19.5){$\Vz$}
\put(5,46){$-\Vz$}
\put(52.5,19.5){$V$}
\put(52.5,46){$-V$}
\put(12,6){$\ff$}
\put(39,0){$\ft$}
\put(59,6){$\ff$}
\put(85,0){$\ft$}
\put(81,2){$\f_0$}
\end{overpic}
\caption{\textbf{Left:} Potential $V_0$ that admits a flat domain-wall solution interpolating between the two minima at $\f_f$ and $\f_t$, indicated by the orange arrow. In the mechanical picture in the inverted potential $-V_0$ this corresponds to a trajectory for a particle that is released from rest at $\f_t$ and coming to rest again at $\f_f$ (red trajectory). \textbf{Right:} Potential $V$ that differs from $V_0$ only by exhibiting a lower barrier separating the two minima at $\f_f$ and $\f_t$. In turn the inverted potential $-V$ exhibits a shallower valley separating $\f_f$ and $\f_t$ than $-V_0$. A particle released from $\f_t$ in $-V$ will overshoot $\f_f$ as the shallower valley leads to a lower velocity and hence less friction. By reducing the initial potential energy and releasing the particle lower down the slope at some $\f_0$, a trajectory can be found so that the particle neither over- nor undershoots, but comes to rest exactly at $\f_f$ (red trajectory). In the Euclidean picture, this corresponds to an $O(D)$-instanton describing tunnelling from $\f_f$ to $\f_0$ (orange arrow).}
\label{fig:algorithm}
\end{figure}

\subsection{A class of potentials permitting $O(D)$-instantons} \label{sec:potentialstrategy}
Here we present a strategy for constructing potentials that admit $O(D)$-instanton solutions for tunnelling from an AdS minimum. An important instrument in the construction of the argument will be the mechanical analogue to Euclidean tunnelling/domain-wall solutions already introduced in section \ref{sec:CDLnongeneric}. This is the observation that the equation of motion \eqref{c5} is identical to that for a particle of unit mass performing damped motion in the inverted potential. In section \ref{sec:CDLnongeneric} we also identified a sufficient condition for a potential with two non-degenerate minima $\f_f$ and $\f_t$ to admit $O(D)$-tunnelling solutions. In particular, an $O(D)$-instanton describing tunnelling from the false vacuum $\f_f$ will exist, if in the mechanical picture, the particle overshoots $\f_f$ if released from rest at $\f_t$ in the inverted potential. In the following,  we describe how a potential satisfying this condition can be constructed. An equivalent statement is that the potential admits
flat skipping flows.

We begin with a potential $\Vz(\f)$ with a false vacuum at $\ff$, a true vacuum at $\ft$, and \emph{admitting a flat domain-wall solution interpolating between the two}. This is a convenient starting point, as such a potential $V_0$ can be easily reverse-engineered from a given domain-wall solution. The idea is to propose some interpolating solution for $\f(u)$, solve for the associated geometry and then use the equations of motion to deduce the corresponding potential. This is for example what we have done in section \ref{sec:flatthinwall}, arriving at a sextic potential admitting flat domain-walls.

The inverted potential $-\Vz$ has two maxima at $\ff$ and $\ft$ separated by a trough. Both $V_0$ and $-V_0$ are depicted on the left of fig.~\ref{fig:algorithm}. The flat domain-wall solution has the following interpretation on the mechanical side. It implies that if the particle is released from rest at $\ft$, it will roll through the valley and come to rest again exactly at $\ff$. This is indicated by the red trajectory on the left in fig.~\ref{fig:algorithm}. The shape of the potential $\Vz$ is just right so that there is just enough friction to make the particle come to rest at $\ff$, avoiding an overshoot or an undershoot.

Now consider a potential $V$, that possesses the same minima as $\Vz$, but exhibits a smaller barrier separating $\ff$ and $\ft$. That is, to obtain $V$ we deform $\Vz$ by just lowering the barrier, as indicated by the purple arrow on the bottom right in fig.~\ref{fig:algorithm}. In the inverted potential $-V$, shown on the top right in fig.~\ref{fig:algorithm}, the lowered barrier of $V$ manifests itself as a shallower trough between the two maxima. Consider once more releasing the particle from rest at $\ft$, but now in the inverted potential $-V$. As the trough separating the two maxima is now shallower, the particle will reach a lower terminal velocity at the bottom compared to its trajectory in $-\Vz$. As the friction term in the equation of motion for the particle, \eqref{eq:EOMparticle}, is proportional to the velocity $\dot{\f}$, we expect that there will not be sufficient friction to stop the particle before arriving at $\ff$. As a result, we expect that the particle will overshoot $\ff$, therefore realising the condition for the existence of an $O(D)$-instanton solution.

Then, following the arguments laid out in section \ref{sec:CDLnongeneric}, the particle can be made to come to rest at $\ff$, if it is released from some value $\f= \f_0$ further down the the inverted potential, as shown by the red trajectory on the right of fig.~\ref{fig:algorithm}. On the Euclidean side this will correspond to an $O(D)$-instanton solution describing tunnelling from $\ff$ to $\f_0$.

What we observe is that for all example potentials $V$ constructed according to the instruction above, we always succeeded in finding an $O(D)$-instanton solution. We take this as evidence that the strategy presented above is indeed a practical method for constructing potentials $V$ that will admit $O(D)$-instantons describing tunnelling from $\ff$.

Interestingly, the above arguments can also be used to construct potentials with a false minimum at $\ff$, that we expect to be stable against tunnelling via $O(D)$-instantons. Consider again a potential $\Vz$ admitting a flat domain-wall, but now we deform it by increasing the barrier between $\ff$ and $\ft$, leaving the minima untouched. The inverted potential now has a deeper trough between the maxima, leading to a higher terminal velocity at the bottom and hence increased friction. This will cause all possible trajectories permitted by \eqref{c5} in the inverted potential to undershoot, and no starting point can be found so that the particle comes to rest at $\ff$. In the Euclidean setting, this implies that there should not exist any $O(D)$-symmetric interpolating solutions starting at $\ff$.

However, we cannot exclude that $\ff$ may still be unstable with respect to instantons with less symmetry, which exhibit more complicated equations of motions, and where our simple mechanical analogue need not apply.

\subsection{Numerical results: Tunnelling from AdS minima} \label{sec:numresults}
In this section we set $D=d+1=4$ throughout and present numerical $O(4)$-instanton solutions for tunnelling from an AdS minimum. We begin by specifying the types of potentials $V$ that will be employed here and which are constructed following the strategy in sec.~\ref{sec:potentialstrategy}. In the following, we again set $\ff=0$.

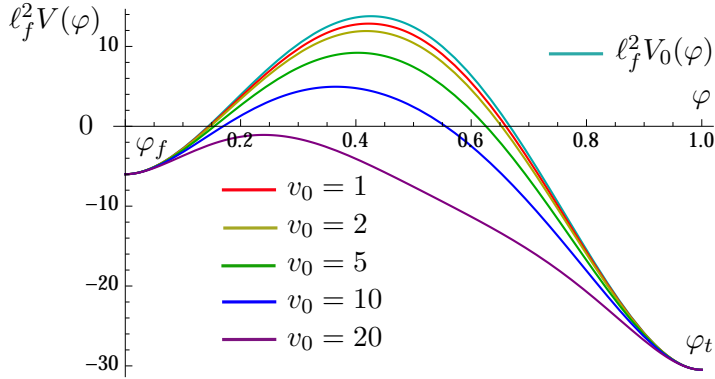
\begin{figure}[t]
\centering
\begin{overpic}
[width=0.55\textwidth]{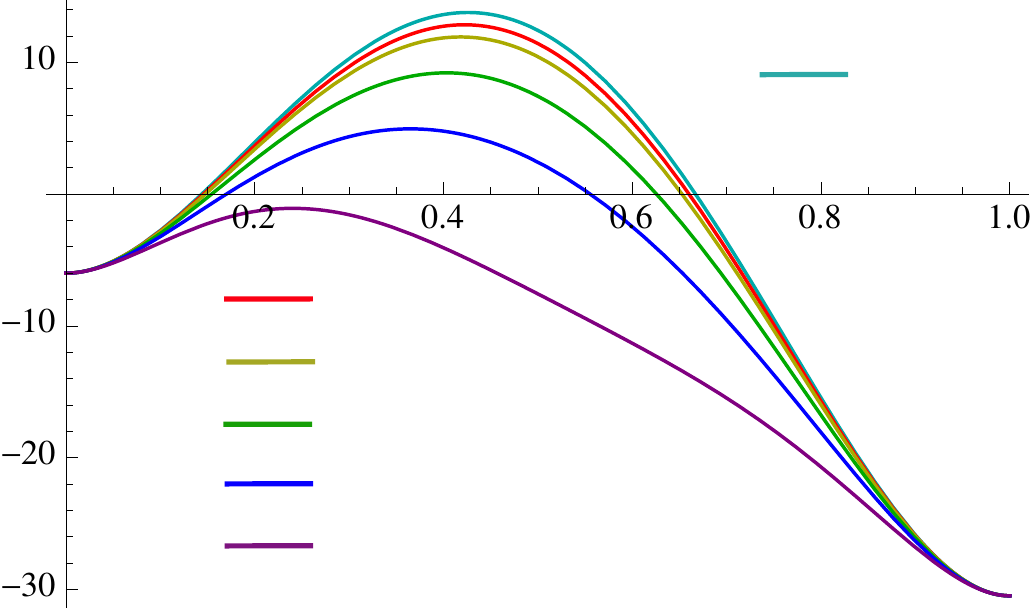}
\put(96,44){$\f$}
\put(8,36.5){$\f_f$}
\put(95,5){$\f_t$}
\put(-12,56){$\ell_f^2 V(\f)$}
\put(-1,39){$0$}
\put(84,51){$\ell_f^2 V_0(\f)$}
\put(32,29.3){$v_0=1$}
\put(32,23.3){$v_0=2$}
\put(32,17.3){$v_0=5$}
\put(32,11.3){$v_0=10$}
\put(32,5.3){$v_0=20$}
\end{overpic}
\caption{Potential $V(\f)$ in \protect\eqref{eq:Vnum} in units of $\ell_f^{-2}$ with $\Delta=30.1$ and $\f_t=1$ for different values of $v_0$. Note that the barrier separating the two AdS minima exhibits a region with $V >0$ for all example potentials shown except for $v_0=20$.}
\label{fig:PotEx30p1}
\end{figure}

The potentials $V$, are obtained as deformations of a potential $\Vz$ exhibiting a flat domain-wall solution. To be explicit, for $\Vz$ we use the sextic potential derived in sec.~\ref{sec:flatthinwall}, admitting a flat domain-wall with $\tanh$-profile for $\f(u)$, and which we reproduce here for $d=3$ for convenience:
\begin{align}
\Vz(\f) = &-\frac{6}{\ell_{f}^2}-\frac{\Delta(3-\Delta)}{2\ell_{f}^2}\f^2+\frac{\Delta(1-\Delta)}{\ell_{f}^2 \f_t}\f^3+\frac{\Delta^2\left(16-3\f_t^2 \right)}{32\ell_{f}^2 \f_t^2}\f^4 \nonumber \\
& +\frac{\Delta^2}{8\ell_{f}^2\f_t}\f^5-\frac{\Delta^2}{24\ell_{f}^2\f_t^2}\f^6 \  \label{eq:V0num}
\end{align}
This has two AdS minima at $\ff=0$ and $\ft$ separated by a barrier. According to the strategy in sec.~\ref{sec:potentialstrategy}, we can arrive at a potential $V$ admitting $O(4)$-instanton solutions by defining
\begin{align}
\label{eq:Vnum} V(\f) = \Vz(\f) + v(\f) \, ,
\end{align}
with $v(\f)$ a contribution that will lower the barrier while leaving the minima unaffected. To be specific, here we choose
\begin{align}
\label{eq:vdef} v(x) = - \frac{64 \, v_0}{\ell_f^2} \, \frac{\f^3 (\ft -\f)^3}{\ft^6} \, ,
\end{align}
with $v_0$ constant.\footnote{The factor $64 \ft^{-6}$ is chosen for convenience such that $v(\ft/2) = -v_0 / \ell_f^2$, which is the approximate value by which the barrier is lowered.} Having a holographic interpretation of our solutions in mind, an important quantity is the curvature of the potential at the minima, setting the dimensions of the operators deforming the associated CFTs. By having chosen $v(x)$ to only give cubic and higher contributions when expanded about $\ff=0$ and $\ft$, the operator dimensions are determined by $\Vz$ alone. Using \eqref{eq:Deltageneraldef} it is easy to verify that the dimension of the operator perturbing the CFT at $\ff=0$, i.e.~the UV CFT, is given by $\Delta$.

For illustration, in fig.~\ref{fig:PotEx30p1} we plot a family of potentials given by \eqref{eq:Vnum} obtained from a potential $\Vz$ with $\Delta=30.1$ and $\f_t=1$ for different choices of $v_0$. Note, that the barrier separating the AdS minima, typically exhibits a region with $V >0$.\footnote{{In this work we also discuss the interpretation of $O(D)$-instantons as holographic RG flows. It may therefore seem worrying at first, that the potential exhibits a region with $V>0$, raising questions about the applicability of holography \'a la AdS/CFT in this context. This worry is unfounded as the holographic RG flow solutions obtained here will always interpolate between regions of the potential with $V<0$ where a holographic interpretation in the spirit of AdS/CFT is possible.}}

\begin{table}
\begin{center}
\begin{tabular}{| c | c | c | c | c |}
\hline
\multicolumn{5}{ | c | }{$\Delta=30.1 \, , \qquad \ft=1 \, , \qquad \eta_{\textrm{flat}}=0.211$ } \\ \hline \hline
$v_0$ & $\f_0$ & $\mathcal{R}$ & $\eta$ & $B / (M \ell_f)^2$   \\ \hline
1 & \multicolumn{1}{ l |}{0.9999999999999996} & 0.36 & 0.216 & 123 \\ \hline
2 & \multicolumn{1}{ l |}{0.99999999999995} & 0.73 & 0.222 & 73 \\ \hline
5 & \multicolumn{1}{ l |}{0.99999999996} & 1.90 & 0.243 & 31 \\ \hline
10 & \multicolumn{1}{ l |}{0.999999985} & 3.9 & 0.287 & 13 \\ \hline
20 & \multicolumn{1}{ l |}{0.999965} & 8.3 & 0.466 & 3.5 \\ \hline
\end{tabular}
\end{center}
\caption{Numerical data for $O(4)$-instanton solutions in a family of potentials \protect\eqref{eq:Vnum} with $\Delta=30.1$, $\f_f=0$, $\f_t=1$. Data for five potentials is shown that differ in their value of $v_0$. For the $O(4)$-instantons the tunnelling end point $\f_0$, the dimensionless boundary curvature $\mathcal{R}$ from \protect\eqref{eq:curlyRdef}, the thin-ness parameter $\eta$ and the instanton action $B$ are recorded.}
\label{tab:30p1phit1}
\end{table}

In the following, we derive numerically $O(4)$-instanton solutions for various potentials $V(\f)$ given by \eqref{eq:Vnum}. In practice this is done by solving the relevant equations from a tentative tunnelling end point $\f_0$, shooting towards $\ff$. If the trial solution over- or undershoots $\ff$, the point $\f_0$ is adjusted accordingly. This is to be iterated until a solution is found that approximates the desired solution coming to rest at $\ff$ up to the desired accuracy. For a given example solution, we then record the value of $\f_0$, extract the dimensionless curvature $\mathcal{R}$ given in \eqref{eq:curlyRdef}, compute the thin-ness parameter $\eta$ defined in \eqref{eq:etadefOD} and calculate the instanton action $B$ given in \eqref{eq:acdiff}.

\begin{figure}[t]
\centering
\begin{subfigure}{.5\textwidth}
 \centering
   \begin{overpic}
[width=1.0\textwidth]{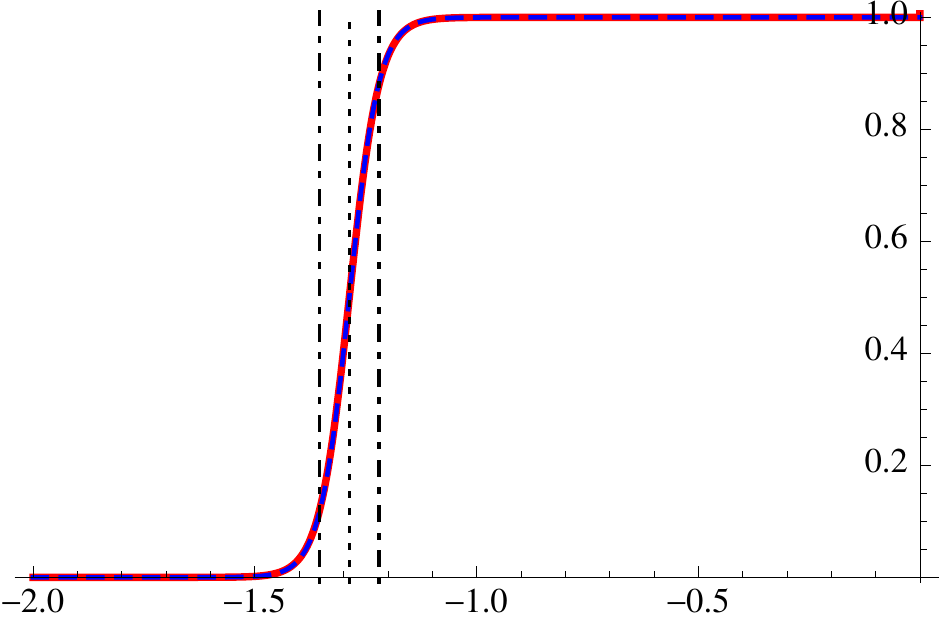}
\put(3,62){$\f(u)$}
\put(93,-2){$u / \ell_f$}
\end{overpic}
\caption{\hphantom{A}}
\label{fig:phiDeltap30p1Phit1p0DeltaV1}
\end{subfigure}%
\begin{subfigure}{.5\textwidth}
 \centering
   \begin{overpic}
[width=1.0\textwidth]{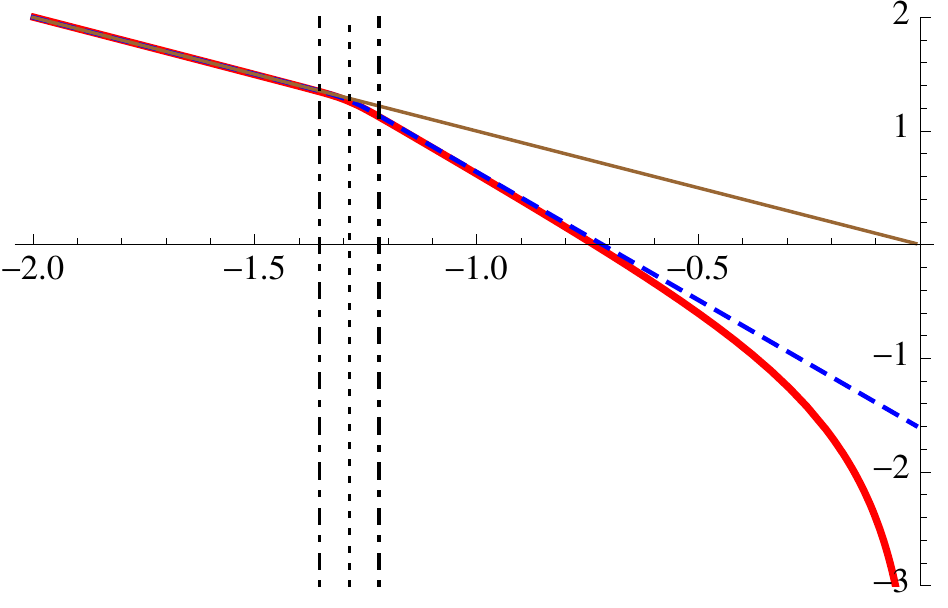}
\put(3,52){$A(u)$}
\put(100,38){$u / \ell_f$}
\end{overpic}
\caption{\hphantom{A}}
\label{fig:ADeltap30p1Phit1p0DeltaV1}
\end{subfigure}%
\caption{$\f(u)$ \textbf{(a)} and $A(u)$ \textbf{(b)} for an $O(4)$-instanton solution (\textbf{red}) and a flat domain wall solution (\textbf{blue, dashed}). The $O(4)$-instanton is obtained for the potential $V(\f)$ given in \protect\eqref{eq:Vnum} with $\Delta=30.1$, $\f_t=1$ and $v_0=1$ and was found to have dimensionless curvature $\mathcal{R}=0.36$. The flat domain wall solution arises from the potential $V_0(\f)$ given in \protect\eqref{eq:V0num} with the same values of $\Delta$ and $\f_t$. The integration constants in $A(u)$ were adjusted such that for $u \rightarrow -\infty$, the functions $A(u)$ asymptote to $- u / \ell_{f}$, which is denoted by the brown line. The vertical dot-dashed lines demarcate the wall, defined as the interval $[u_{\textrm{out}}, u_{\textrm{in}}]$ with $\f(u_{\textrm{out}}) = 0.12 \f_0$ and $\f(u_{\textrm{in}}) = 0.88 \f_0$, consistent with $\gamma=0.76$ in \protect\eqref{eq:phiinoutdefOD}. The dotted vertical line indicates $\bar{u}$, defined as the locus $\f(\bar{u}) = 0.5 \f_0$. Note that the $O(4)$-instanton solution for $A(u)$ is effectively indistinguishable from the solution $A_{\textrm{flat}}(u)$ at the locus of the wall.}
\label{fig:AphiDeltap30p1Phit1p0DeltaV1}
\end{figure}

\begin{figure}[t]
\centering
\begin{subfigure}{.5\textwidth}
 \centering
   \begin{overpic}
[width=1.0\textwidth]{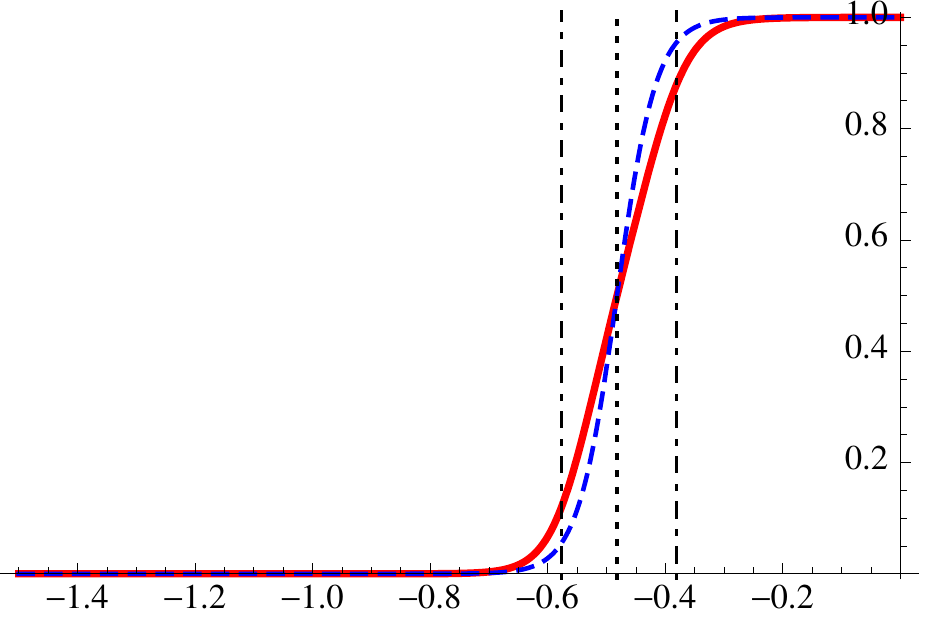}
\put(3,62){$\f(u)$}
\put(93,-2){$u / \ell_f$}
\end{overpic}
\caption{\hphantom{A}}
\label{fig:phiDeltap30p1Phit1p0DeltaV20}
\end{subfigure}%
\begin{subfigure}{.5\textwidth}
 \centering
   \begin{overpic}
[width=1.0\textwidth]{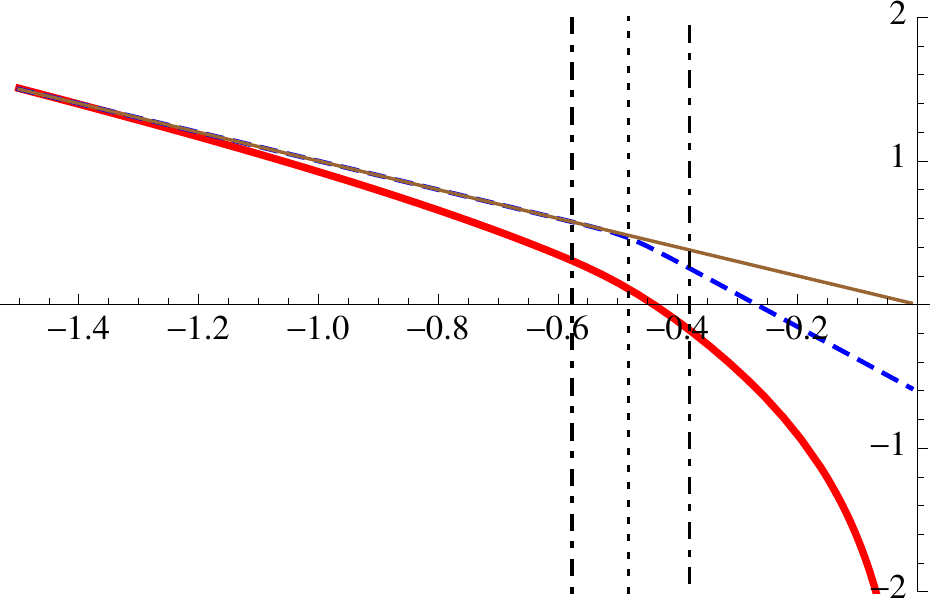}
\put(3,62){$A(u)$}
\put(100,32){$u / \ell_f$}
\end{overpic}
\caption{\hphantom{A}}
\label{fig:ADeltap30p1Phit1p0DeltaV20}
\end{subfigure}%
\caption{$\f(u)$ \textbf{(a)} and $A(u)$ \textbf{(b)} for an $O(4)$-instanton solution (\textbf{red}) and a flat domain wall solution (\textbf{blue, dashed}). The $O(4)$-instanton is obtained for the potential $V(\f)$ given in \protect\eqref{eq:Vnum} with $\Delta=30.1$, $\f_t=1$ and $v_0=20$ and was found to have dimensionless curvature $\mathcal{R}=8.3$. The flat domain wall solution arises from the potential $V_0(\f)$ given in \protect\eqref{eq:V0num} with the same values of $\Delta$ and $\f_t$. The integration constants in $A(u)$ were adjusted such that for $u \rightarrow -\infty$ the functions $A(u)$ asymptote to $- u / \ell_f$, which is denoted by the brown line. The vertical dot-dashed lines demarcate the wall, defined as the interval $[u_{\textrm{out}}, u_{\textrm{in}}]$ with $\f(u_{\textrm{out}}) = 0.12 \f_0$ and $\f(u_{\textrm{in}}) = 0.88 \f_0$, consistent with $\gamma=0.76$ in \protect\eqref{eq:phiinoutdefOD}. The dotted vertical line indicates $\bar{u}$, defined as the locus $\f(\bar{u}) = 0.5 \f_0$. Note that the $O(4)$-instanton solution for $A(u)$ differs significantly from the solution $A_{\textrm{flat}}(u)$ at the locus of the wall.}
\label{fig:AphiDeltap30p1Phit1p0DeltaV20}
\end{figure}

For the one-parameter family of potentials with $\Delta=30.1$ and $\f_t=1$ plotted in fig.~\ref{fig:PotEx30p1} the numerical results are shown in table \ref{tab:30p1phit1}.\footnote{The quantities $\eta$ and $\eta_{\textrm{flat}}$ depend on the parameter $\gamma$ trough the definition of $u_{\textrm{in}}$ and $u_{\textrm{out}}$ in \eqref{eq:phiinoutdefOD}. Here and throughout this work we choose $\gamma=0.76$.} We make the following observations:
\begin{itemize}
\item The lower the barrier between the minima (i.e.~the larger $v_0$), the further the point $\f_0$ is removed from the true vacuum at $\ft=1$.\footnote{This is a relative observation. Note that in absolute terms $\f_0$ is very close to the true vacuum $\ft=1$ in all cases displayed in table  \ref{tab:30p1phit1}.} This can be understood using the mechanical analogue introduced in section \ref{sec:CDLnongeneric}: the lower the barrier, the weaker friction becomes and hence $\f_0$ needs to be further away from $\ft$ to avoid overshooting beyond $\ff$.
\item We further observe that the closer $\f_0$ to $\ft$, the smaller $\mathcal{R}$ and $\eta$ and the larger $B/ (M \ell_f)^2$, i.e.~schematically:
\begin{align}
\nonumber \f_0 \rightarrow \ft : \quad \mathcal{R} \downarrow \, , \quad B/ (M \ell_f)^2 \uparrow \, .
\end{align}
The trends for $\mathcal{R}$ and $B$ can be understood intuitively. The flat domain-wall solution can be interpreted as the $\mathcal{R} \rightarrow 0$ limit of an $O(4)$-instanton with $B \rightarrow \infty$ as consequence of its stability. The observations above are then consistent with the fact that $O(4)$-instantons asymptote towards a flat domain-wall for $\f_0 \rightarrow \ft $.
\item Another observation is that $O(4)$-instantons with thicker walls have a lower instanton action. Therefore,  a false vacuum permitting an $O(4)$-instanton with a thicker wall is more prone to decay than a false vacuum permitting an $O(4)$-instanton with a thinner wall.
\end{itemize}

We find it instructive to also display the solutions for $\f(u)$ and $A(u)$ for a few example cases. Hence, in fig.~\ref{fig:AphiDeltap30p1Phit1p0DeltaV1} we display $\f(u)$ and $A(u)$ for the $O(4)$-instanton solution from table \ref{tab:30p1phit1} with $v_0=1$. These are given by the red curves in fig.~\ref{fig:AphiDeltap30p1Phit1p0DeltaV1} while in blue we overlay $\f(u)$ and $A(u)$ for the flat domain-wall solution solution in the potential $\Vz$ with the same values of $\Delta$ and $\f_t$. The vertical dot-dashed lines define the inner and outer limits of the wall, defined via \eqref{eq:phiinoutdefOD} with $\gamma=0.76$. The vertical dotted line indicates the centre of the wall where $\f$ has interpolated half way between $\f_f$ and $\f_0$. The brown line in fig.~\ref{fig:ADeltap30p1Phit1p0DeltaV1} corresponds to $-u/ \ell_f$, to which $A(u)$ asymptotes to for $u \rightarrow -\infty$.

The main observation here  is that the flat domain-wall solution effectively coincides with the $O(4)$-symmetric solution well into the interior of the $O(4)$-bubble: Firstly, the two solutions for $\f(u)$ are near-indistinguishable throughout; secondly, the solutions for $A(u)$ only start departing from one another significantly for $u \gtrsim -0.6$. This example therefore explicitly realises the scenario conjectured in section \ref{sec:O4asflat}, i.e.~the case of a $O(4)$-instanton that can be well-approximated by a corresponding flat domain-wall solution.

\begin{table}
\begin{center}
\begin{tabular}{| c | l | c | c | c |}
\hline
\multicolumn{5}{ | c | }{$\Delta=50.1 \, , \qquad \ft=1 \, , \qquad \eta_{\textrm{flat}}=0.159$ } \\ \hline \hline
$v_0$ & \multicolumn{1}{ c|}{$\f_0$} & $\mathcal{R}$ & $\eta$ & $B / (M \ell_f)^2$   \\ \hline
0.5 & 0.999999999999999999999999990 & 0.089 & 0.160 & 373 \\ \hline
 1 & 0.999999999999999999999998 & 0.18 & 0.160 & 246 \\ \hline
 2 & 0.9999999999999999999993 & 0.36 & 0.162 & 157 \\ \hline
 5 & 0.999999999999999998 & 0.90 & 0.167 & 80 \\ \hline
10 & 0.9999999999999993 & 1.9 & 0.176 & 44 \\ \hline
\end{tabular}
\end{center}
\begin{center}
\begin{tabular}{| c | l | c | c | c |}
\hline
\multicolumn{5}{ | c | }{$\Delta=30.1 \, , \qquad \ft=0.25 \, , \qquad \eta_{\textrm{flat}}=0.137$ } \\ \hline \hline
$v_0$ & \multicolumn{1}{ c|}{$\f_0$} & $\mathcal{R}$ & $\eta$ & $B / (M \ell_f)^2$   \\ \hline
0.05 & 0.2499999999999999999999999995 & 0.16 & 0.142 & 27 \\ \hline
0.1 & 0.249999999999999999999999991 & 0.33 & 0.144 & 16 \\ \hline
0.2 & 0.24999999999999999999985 & 0.67 & 0.151 & 9.3 \\ \hline
0.3 & 0.24999999999999999993 & 1.03 & 0.159 & 6.4 \\ \hline
0.4 & 0.249999999999999994 & 1.41 & 0.167 & 4.8 \\ \hline
0.5 & 0.2499999999999998 & 1.80 & 0.177 & 3.8 \\ \hline
\end{tabular}
\end{center}
\caption{Numerical data for $O(4)$-instanton solutions in two families of potentials \protect\eqref{eq:Vnum} with $\Delta=50.1$, $\f_f=0$, $\f_t=1$ and $\Delta=30.1$, $\f_f=0$, $\f_t=0.25$, respectively. Data for potentials differing in their value for $v_0$ is shown. For the $O(4)$-instantons the tunnelling end point $\f_0$, the dimensionless boundary curvature $\mathcal{R}$ from \protect\eqref{eq:curlyRdef}, the thin-ness parameter $\eta$ and the instanton action $B$ are recorded.}
\label{tab:thinnerwalls}
\end{table}

In fig.~\ref{fig:AphiDeltap30p1Phit1p0DeltaV20}, the red curves show $\f(u)$ and $A(u)$ for the $O(4)$-instanton solution from table \ref{tab:30p1phit1} with $v_0=20$. The blue curves again denote $\f(u)$ and $A(u)$ for the flat domain-wall solution solution in the potential $\Vz$. Here, the flat domain-wall solution already departs from the $O(4)$-instanton solution well outside the wall and is hence not a good approximation for the $O(4)$-instanton. This is not surprising, as for $v_0=20$ the potential $V$ is significantly altered compared to $\Vz$ as can be seen in fig.~\ref{fig:PotEx30p1}.

We now return to the results in table \ref{tab:30p1phit1}, focusing on how the thin-ness parameter $\eta$ varies across this family of solutions. Note that $\eta$ decreases as $v_0$ is lowered. More precisely, as $v_0$ is decreased the thin-ness parameter $\eta$ asymptotes towards the value $\eta_{\textrm{flat}}$, which appears to set the lower bound for $\eta$ in this family of potentials. This can be understood with the help of the observations made from the plots in figs.~\ref{fig:AphiDeltap30p1Phit1p0DeltaV1} and \ref{fig:AphiDeltap30p1Phit1p0DeltaV20}. There we observed that if $v_0$ is decreased, the $O(D)$-instanton solutions is increasingly well-approximated by the flat domain-wall solution that exists in the potential $\Vz$. As a result, the thin-ness parameter $\eta$ approaches the value $\eta_{\textrm{flat}}$ obtained for the flat domain-wall solution.

In table \ref{tab:thinnerwalls} we collect data for two additional one-parameter families of potentials corresponding to the parameter choices $(\Delta=50.1, \f_t =1)$ and $(\Delta=30.1, \f_t =0.25)$, respectively. The trends observed here in table \ref{tab:30p1phit1} as $v_0$ is increased are also reproduced by these additional examples.

\subsection{Thin-walled $O(4)$-instantons} \label{sec:ODthinwall}

 Thin-walled $O(4)$-instanton solutions have played an important role in the study of tunnelling processes. The reason is the existence of the `thin-wall approximation' of Coleman and de Luccia \cite{CdL}, which allows for an analytic computation of relevant quantities such as the decay rate. In this section we  explore thin-walled $O(4)$-instanton solutions without relying on the approximation scheme of Coleman and de Luccia. Instead, here we  use results from the previous section to develop an analytic understanding of the thin-wall limit. In particular, we shall be interested in the requirements on potentials to admit thin-walled  $O(4)$-instantons. We then compare our results with the findings of Coleman and de Luccia.

The results here-obtained will be derived for potentials $V(\f)$ that fall into the category defined in \eqref{eq:Vnum}, i.e.~potentials based on the sextic potential $V_0(\f)$ in \eqref{eq:V0num} with the barrier separating the two minima lowered by an amount $-v_0/\ell_f^2$. At the end of this section we then argue how we expect our findings to generalise beyond this class of potentials. In fact, we explain why we expect our main results to be robust and reveal universal properties of thin-walled instanton solutions.

\begin{figure}[t]
\centering
\begin{overpic}[width=0.5\textwidth,tics=10]{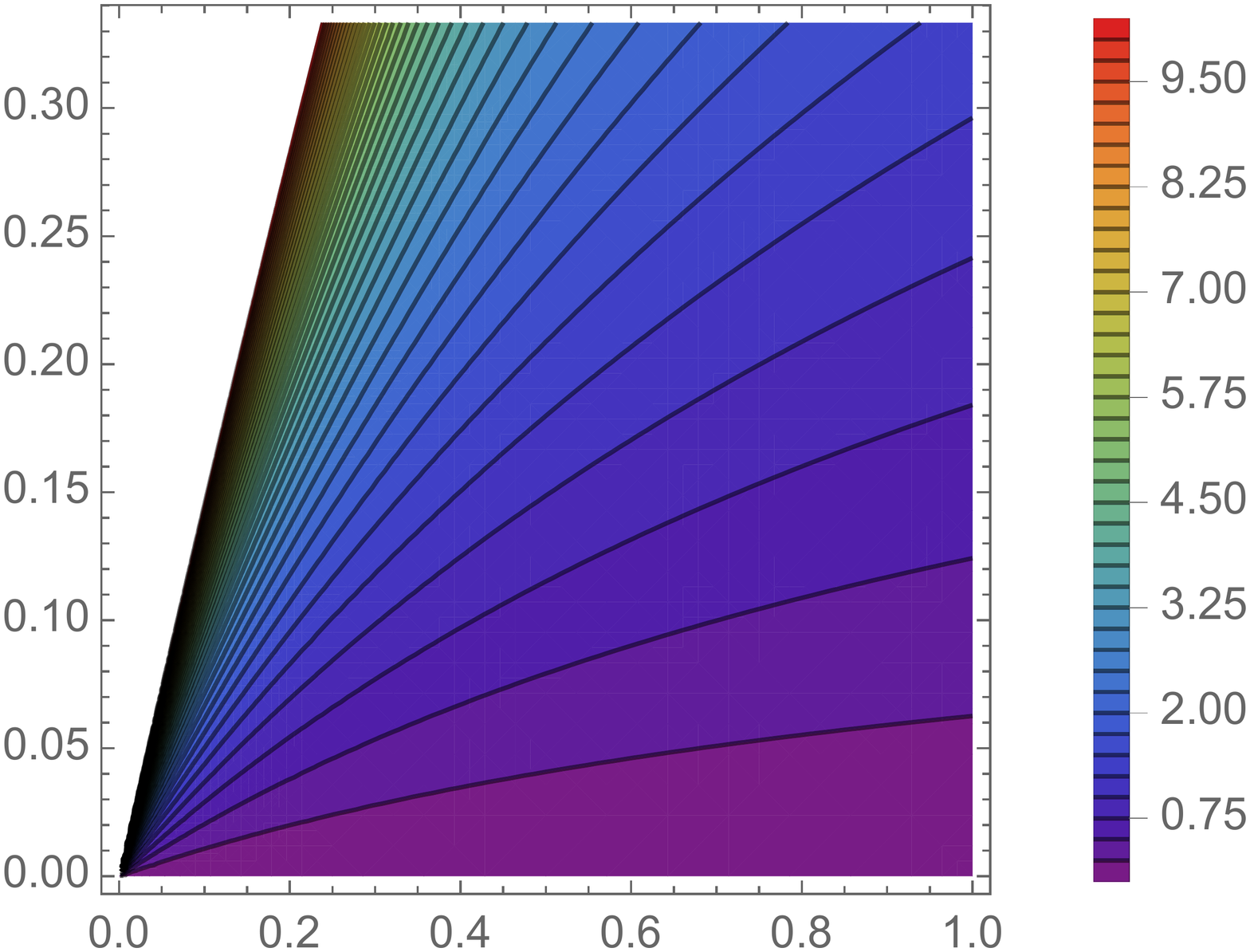}
\put(86,5){$\eta_{\textrm{flat}}$}
\put(6,35){$\frac{1}{\Delta}$}
\put(45,-1){$\ell_t / \ell_f$}
\end{overpic}
\caption{Contours of constant $\eta_{\textrm{flat}}$ as given in \protect\eqref{eq:etaflatexplicit} on the $(\ell_t/\ell_f)$-$1/\Delta$-plane for $\gamma=0.76$. The locus $\eta_{\textrm{flat}} \rightarrow 0$ is reached for $\Delta \rightarrow \infty$ with $\Delta \ell_t /\ell_f \rightarrow \infty$.}
\label{fig:etacontours}
\end{figure}

How can we then construct a thin-walled $O(4)$-instanton in the family of potentials \eqref{eq:Vnum}? First, consider a subfamily of potentials of type \eqref{eq:Vnum} with a \emph{fixed choice} for $V_0(\f)$ but different values of $v_0$. As argued before, every such potential will admit an $O(D)$-instanton and we can calculate the corresponding `thin-ness' parameter $\eta$ defined in \eqref{eq:etadefOD}. From the results in section \ref{sec:numresults}, recall that for all $O(D)$-instantons in this subfamily of potentials, the parameter $\eta_\textrm{flat}$ given in \eqref{eq:etaflatexplicit} represents the lower bound for $\eta$, i.e.~$\eta > \eta_{\textrm{flat}}$.\footnote{Recall that the value of $\eta_{\textrm{flat}}$ is unique for a given choice of $V_0(\f)$.} More precisely, by letting $v_0 \rightarrow 0^+$ we found that $\eta \rightarrow \eta_{\textrm{flat}}^+$. That is, we can construct an $O(D)$-instanton with a value of $\eta$ arbitrarily close to $\eta_{\textrm{flat}}$ by choosing $v_0$ sufficiently small.

It hence follows that to construct an $O(D)$-instanton solution with a small value of $\eta$ we must consider subfamilies with sufficiently small $\eta_{\textrm{flat}}$ in the first place. That is, $\eta$ can be made small to the extent that we can make $\eta_{\textrm{flat}}$ small.

Here we shall  be interested in the extreme thin-wall limit $\eta \rightarrow 0$. From the above we conclude that such solutions can only occur in families of potentials that exhibit $\eta_{\textrm{flat}} \rightarrow 0$. In \eqref{eq:etaflatexplicit} we have an expression for $\eta_{\textrm{flat}}$ as a function of the two dimensionless parameters $\Delta$ and $\ell_f / \ell_t$.\footnote{The expression \eqref{eq:etaflatexplicit} also contains the parameter $\gamma$, whose value determines which part of the interpolating solution we define as the wall. Its choice, while to some extent a matter of personal taste, is made once and for all and hence $\gamma$ is not an adjustable parameter.} Note that $\ell_f/\ell_t >1$ as a consequence of our choice $V(\f_f) > V(\f_t)$. The main observation then is that $\eta_{\textrm{flat}} \rightarrow 0$ can \emph{only} be obtained for $\Delta \rightarrow \infty$, that is $\eta_{\textrm{flat}} \rightarrow 0$ cannot be reached at finite $\Delta$. One further finds that for $\eta_{\textrm{flat}} \rightarrow 0$ to be successfully attained $\Delta$ has to grow faster than $\ell_f/ \ell_t$, i.e.~$\Delta \tfrac{\ell_t}{\ell_f} \rightarrow \infty$. These results are summarised in fig.~\ref{fig:etacontours} where we plot contours of constant $\eta_{\textrm{flat}}$ on the $(\ell_t/\ell_f)$-$1/\Delta$-plane.

 We are now in a position to describe the corresponding potentials $V$ that permit $O(D)$-instantons in the extreme thin-wall limit $\eta \rightarrow 0$. We can proceed without specifying $v_0$ in \eqref{eq:vdef} as we find that $v_0$ will only enter the subsequent expressions at subleading order for $\Delta \rightarrow \infty$. Here we can work for general $D=d+1$ so instead of using the expression \eqref{eq:V0num} for $V_0(\f)$ valid for $d=3$ we revert to the general expression in \eqref{com3}. We make the following observations:
\begin{itemize}
\item In the limit $\eta_{\textrm{flat}} \rightarrow 0$ the separation between minima vanishes, i.e.~$\varrho \rightarrow 0$ or equivalently $\f_t \rightarrow \f_f$. This follows from \eqref{eq:Vparameterconstraint} and the requirement that for $\eta_{\textrm{flat}} \rightarrow 0$ we need both $\Delta \rightarrow \infty$ and $\Delta \tfrac{\ell_t}{\ell_f} \rightarrow \infty$. Contrast this with flat domain walls, where there were two ways of obtaining a thin flat domain wall, summarised in the enumerated list close to the end of section \ref{sec:flatthinwall}. In case (1) $\varrho$ remains finite while in case (2) $\varrho \rightarrow 0$. For $O(D)$-instantons there is only one way, but it coincides with case (2) of the flat domain wall analysis.
\item The barrier separating the two minima becomes infinitely tall compared to the potential separation $V_0(\f_f)-V_0(\f_t)$ between false and true minimum. We can e.g.~compute (again using \eqref{eq:Vparameterconstraint} to substitute for $\f_t)$:

\begin{align}
\label{eq:barrierdifferenceratio} \frac{V \big(\tfrac{\f_f+\f_t }{2} \big) - V(\f_f)}{V(\f_f) -V(\f_t)} \underset{\substack{\Delta \rightarrow \infty \, , \\ \Delta \frac{\ell_t}{\ell_f} \rightarrow \infty}}{=} \frac{3 \ell_t \, \Delta}{8d(\ell_f+\ell_t)} + \mathcal{O}(\Delta^0) \, ,
\end{align}
which diverges in the relevant limits $\Delta \rightarrow \infty$ and $\Delta \tfrac{\ell_t}{\ell_f} \rightarrow \infty$ as postulated.
\item Another observation is that in the thin-wall limit the dimensions $\Delta(\f_f)$  and $\Delta(\f_t)$ of the operators perturbing the false vacuum and true vacuum CFTs diverge. By definition, what we refer to as $\Delta$ is the operator dimension for the false vacuum CFT. For the true vacuum CFT the operator dimension can be calculated from \eqref{eq:Deltageneraldef} so that we find.
\begin{align}
\Delta(\f_f)=\Delta \, , \qquad \Delta(\f_t)= \frac{\Delta \, \ell_t}{\ell_f \Big(1 - \tfrac{\ell_t}{\ell_f} \Big)} \, .
\end{align}
These indeed diverge in the thin-wall limit $\Delta \rightarrow \infty$ and $\Delta \tfrac{\ell_t}{\ell_f} \rightarrow \infty$.
\end{itemize}

\noindent\textbf{Comparison with Coleman and de Luccia:} Our results from the family of sextic potentials \eqref{eq:Vnum} imply that thin-walled $O(D)$-instantons exist in potentials that exhibit a large barrier separating false and true minima compared to the potential difference between the minima. How does this compare to criteria for applicability of the thin-wall approximation as laid out by Coleman and de Luccia? In \cite{CdL} Coleman and de Luccia write that the thin-wall approximation applies ``in the limit of a small energy difference between the two vacuums.'' As the energy difference is dimensionful, it can only be small compared to another quantity with the same dimensions. The only other appropriate quantity in \cite{CdL} is the barrier height\footnote{Other possible dimensionful quantities for comparison are the absolute values of the potential at the minima. However in \cite{CdL} either $V(\f_f)=0$ or $V(\f_t)=0$ so that these values are either zero or given by the potential difference itself.} and hence our conditions agree.
We also provide a more quantitative comparison in appendix \ref{app:CDLcompare}. However, note that our analysis implies that the criterion of \cite{CdL}  is not sufficient for an instanton solution to exist. As we have argued in section \ref{sec:CDLnongeneric}, $O(D)$-instantons do not exist in generic potentials and this also holds for thin-walled ones. What we find is that if a potential admits a thin-walled $O(D)$-instanton, the potentials will be of the type required by \cite{CdL}, but the converse is not generically true.
\vspace{0.1cm}

\noindent\textbf{Thin-walled $O(D)$-instantons in more general potentials:} In the class of potentials in \eqref{eq:Vnum}, instanton solutions are thin-walled to the extent that $\Delta$ is chosen large. Here we describe to what extent we expect this finding to be valid beyond this class of potentials. An instanton solution is thin-walled if the interpolation in $\f$ effectively happens over a small radial interval $r_\textrm{out}-r_\textrm{in}$ when compared to the radius of the bubble $\bar{r}$. We observe that in this case the interpolation is also `rapid' in terms of the coordinate $u$, i.e.~the interpolation between $\f_f$ and $\f_0$ occurs in a smaller $u$-interval the thinner-walled the solution, compare e.g.~the solutions in figs.~\ref{fig:phiDeltap30p1Phit1p0DeltaV1} (thinner wall) and \ref{fig:phiDeltap30p1Phit1p0DeltaV20} (thicker wall).

In the following, it will be useful to invoke the mechanical picture of tunnelling in terms of a particle crossing a well, introduced at the beginning in section \ref{sec:CDLnongeneric}, with $t=-u$ the time variable. In this formulation, a thin-walled solution corresponds to the particle staying at rest at $\f_0$ for an extended period of time, before experiencing a short period of strong acceleration followed by short period of strong deceleration, before coming to rest again at $\f_f$. An important observation is that the initial acceleration from rest and final deceleration to come to rest again are effectively controlled by $V'$, which follows from the equation of motion \eqref{c5} or, equivalently, \eqref{eq:EOMparticle}. For the particle to stay effectively at rest at $\f_0$ for an extended period of time, one thus requires $V'(\f_0) \approx 0$, which implies that $\f_0$ needs to be close to the extremum at $\f_t$. That is, thinner-walled instantons deposit $\f$ closer to the true minimum, which is indeed what we observe for the examples in tables \ref{tab:30p1phit1} and \ref{tab:thinnerwalls}. Now, to achieve strong initial acceleration and final deceleration, we require that $|V'|$ becomes large \emph{immediately} once the particle departs from $\f_0$ or approaches $\f_f$. By definition, a change in $V'$ due to a step in $\f$ is quantified by the curvature $V''$ of the potential. Thus, a strong initial acceleration and strong final deceleration is achieved if $|V''(\f_0)| \approx |V''(\f_t)|$ and $|V''(\f_f)|$ are `large' (in appropriate units). Using \eqref{eq:Deltageneraldef}, this can be translated into the requirement $\Delta(\f_f) \gg 1$ and $\Delta(\f_t) \gg 1$, i.e.~the dimensions of the operators perturbing both the false vacuum and true vacuum CFTs need to be large. This is indeed what we observed when analysing the thin-wall limit of instantons in the family of sextic potentials \eqref{eq:Vnum}.

Identifying thin-walled instantons as solutions with a period with strong acceleration follows by a period strong deceleration, we can define a general condition on potentials to admit such solutions. As stated above, we require $|V'|$ to grow quickly once the particle departs from $\f_0$ or approaches $\f_f$. The most generic way of achieving this is to have $|V''(\f_f)|$ and $|V'(\f_t)'|$ large or, equivalently, $\Delta(\f_f) \gg 1$ and $\Delta(\f_t) \gg 1$. One could in principle also imagine the case where $|V''(\f_f)|$ and $|V'(\f_t)'|$ are small but some higher derivative is extremely large at and close to $\f_f$ and $\f_t$. This would eventually also lead to a strong acceleration/deceleration, but this is an even more artificial and tuned choice than just having $\Delta(\f_f) \gg 1$ and $\Delta(\f_t) \gg 1$ large.

\vspace{0.1cm}

\noindent \textbf{The thin-wall limit from a holographic point of view --} As already discussed at the end of section \ref{sec:flatthinwall} one may ask to what extent $\Delta$ can be taken large. The reason is that for the AdS-to-AdS-tunnelling considered here $\Delta$ also has an interpretation as the scaling dimension deforming the UV CFT associated with the false vacuum. Results from the conformal bootstrap program applied to 3d CFTs show that the scaling dimension of the lowest-lying scalar operator is bounded in consistent theories \cite{Dymarsky:2017yzx}. This in turn would imply that consistency of the dual field theory sets a limit on how thin-walled $O(D)$-instantons can be. One way of avoiding this would be to construct solutions that permit a thin-wall limit that does not rely on $\Delta \rightarrow \infty$. We leave this as an interesting question for future work.

\begin{figure}[t]
\centering
\begin{overpic}[width=0.7\textwidth,tics=10]{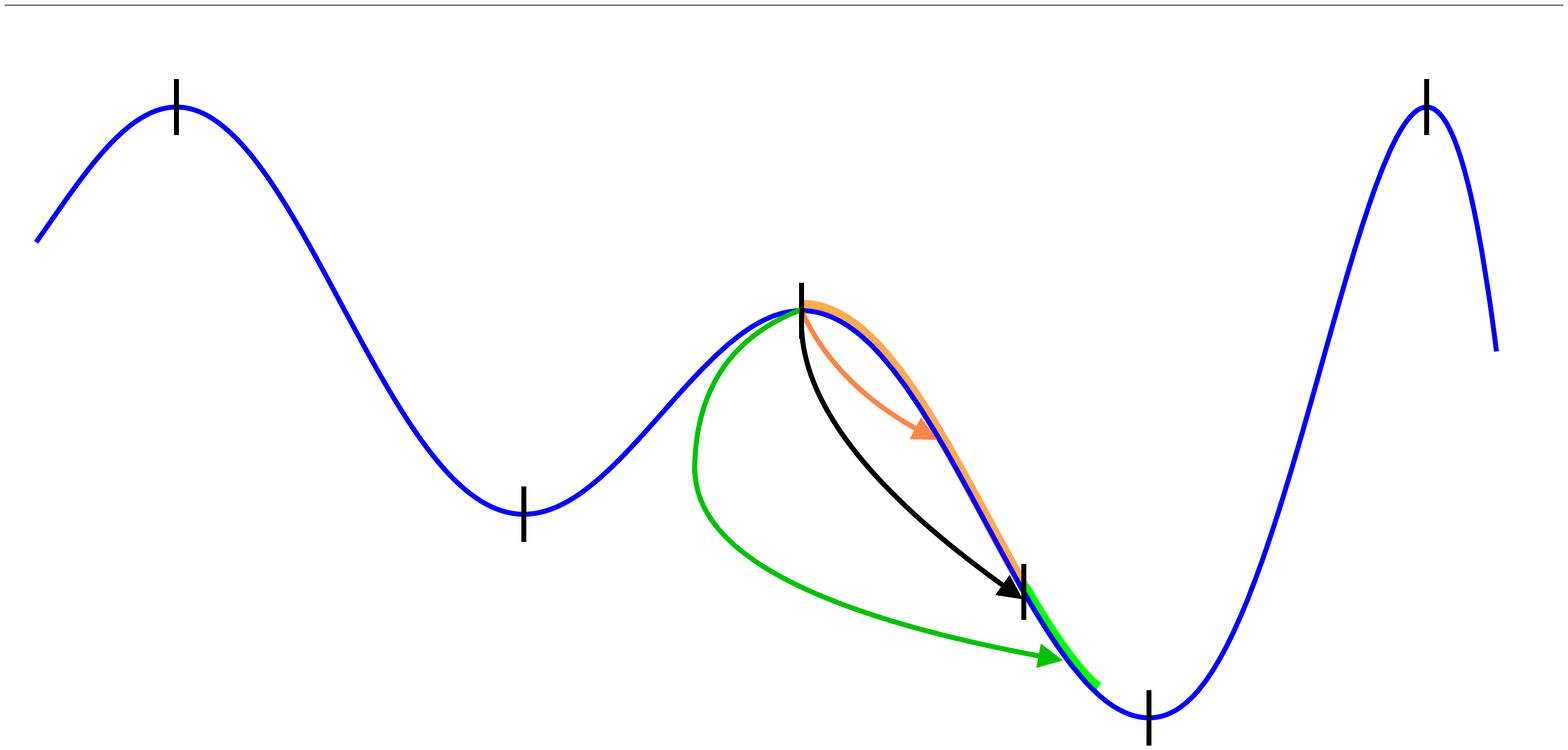}
\put(-1,56){$0$}
\put(71.5,9){$\f_3$}
\put(88,46){$\f_4$}
\put(11.5,46){$0$}
\put(32,21){$\f_1$}
\put(-1,37){$V(\f)$}
\put(49,42){$\f_2$}
\put(63.5,25){$\f_0$}
\end{overpic}
\caption{Cartoon of a degree-12 polynomial potential $V(\f)$ used in sec.~\protect\ref{sec:maxresults}. The explicit expression for $V(\f)$ is given in \cite{Kiritsis:2016kog}. The potential admits an $O(D)$-instanton solutions describing tunnelling from the AdS maximum at $\f_2$ (black arrow), corresponding to a holographic RG flow of $(+)$-type. At the same time it admits non-bouncing (orange arrow) and bouncing (green arrow) RG flow solutions of $(-)$ type from $\f_2$. The end point $\f_0$ sits at the interface of possible end points for non-bouncing (orange region) and bouncing solutions (green region).}
\label{fig:poly12cartoon}
\end{figure}

\subsection{Numerical results: Tunnelling from AdS maxima} \label{sec:maxresults}
Here we  consider $O(D)$-instanton solutions describing tunnelling from $\f_f$, where $\f_f$ is now an AdS maximum of $V$. There is one important difference when $\f_f$ is a maximum compared to a minimum: Following the discussion in sec.~\ref{sec:CDLasHoloRG}, for $\f_f$ describing a maximum there exist two types of $O(D)$-symmetric solutions with $\f(u \rightarrow - \infty) =\f_f$, which are the $(-)$-branch solutions and the $(+)$-branch solutions, while for $\f_f$ describing a minimum the only possible solutions are $(+)$-branch solutions. However, as discussed before in sec.~\ref{sec:instaction} only the $(+)$-branch solutions permit an interpretation as instantons.

In sec.~\ref{sec:CDLnongeneric} we employed the fact that $(+)$-branch solutions only have one free parameter $\mathcal{R}$ near $\f_f$ to argue that $(+)$-branch solutions do not exist in generic potentials. While there it was assumed that $\f_f$ is a minimum of $V$, nothing in the argument based on the number of free parameters hinged on this fact, and hence the argument also holds for $\f_f$ describing a maximum of $V$. Hence, for a generic potential $V$ with an AdS maximum at $\f_f$, one does not expect $O(D)$-instantons describing tunnelling from $\f_f$ to generically exist.

\begin{table}
\begin{center}
\begin{tabular}{| c | c | c | c | c || c | c | c | r |}
\hline
 $\f_1$ & $\f_2$ & $\f_3$ & $\f_4$ & $\Delta(\f_2)$ & $\f_0$ & $\mathcal{R}$ & $\eta$ & $B / (M \ell_f)^2$  \\ \hline \hline
0.74 & 0.81 & 1.75 & 2.80 & 1.83 & 0.9214 & 66.46 & 2.51 & 0.03 \\ \hline
0.74 & 0.81 & 1.75 & 2.80 & 1.86 & 0.9296 & 62.80 & 2.48 & 0.03 \\ \hline
0.79 & 0.85 & 1.75 & 2.66 & 2.10 & 1.0464 & 46.38 & 2.27 &  0.06 \\ \hline
0.79 & 0.85 & 1.75 & 2.66 & 2.11 & 1.0521 & 45.89 & 2.26 & 0.06 \\ \hline
\end{tabular}
\end{center}
\caption{Choices of model parameters $(\f_1, \f_2, \f_3, \f_4 , \Delta(\f_2) )$ used to construct a degree-12 polynomial (explicit expression in \cite{Kiritsis:2016kog}) admitting $O(4)$-instanton solutions describing tunnelling from a maximum at $\f_2$. The values of the tunnelling end point $\f_0$, the dimensionless boundary curvature $\mathcal{R}$ from \protect\eqref{eq:curlyRdef}, the thin-ness parameter $\eta$ from \protect\eqref{eq:etadefOD} and the instanton action $B$ for the respective $O(4)$-instanton solutions are also recorded.}
\label{tab:ODplusfrommax}
\end{table}

Therefore, as for the case of tunnelling from an AdS minimum, tunnelling from an AdS maximum only occurs in potentials that are sufficiently `tuned'. Here we could once more consider the sextic potential in \eqref{eq:Vnum}, which allows for $O(D)$-instantons describing tunnelling from the UV fixed point at the maximum between $\f_f$ and $\f_t$. However, to also show examples that do not rely on the potential \eqref{eq:Vnum}, in the following we will consider a family of potentials hat can be written as a degree-12 polynomial, and that has been previously analysed in \cite{Kiritsis:2016kog} in the context of exotic RG flows. $O(D)$-symmetric solutions of $(+)$-type were observed to exist in this family in \cite{Ghosh:2017big}, including such flows emanating from a maximum. \footnote{These are the flows ending at $\f_*$ in Figure 30 of \cite{Ghosh:2017big}.} For the explicit expression for $V$ we refer readers to \cite{Kiritsis:2016kog} and instead show a cartoon in fig.~\ref{fig:poly12cartoon}. The potential can be defined by specifying values for $(\f_1, \f_2, \f_3, \f_4)$, which describe the loci of four extrema of $V$. Without loss of generality one further extremum is fixed at $\f=0$. In addition, there is the freedom to pick the value of $\Delta$ at one extremum of choice. We once more set $D=d+1=4$.

In table \ref{tab:ODplusfrommax} we record four choices of values for $(\f_1, \f_2, \f_3, \f_4 , \Delta(\f_2) )$ for which the potential permits $O(4)$-solutions of $(+)$-type from a maximum at $\f_2$. That is, in this example $\f_2$ plays the role of $\f_f$. Here $\Delta(\f_2)$ corresponds to the value of $\Delta$ computed from \eqref{eq:Deltageneraldef} at the maximum $\f_2$. In table \ref{tab:ODplusfrommax} we further record the value of $\mathcal{R}$ for the $O(D)$-solution, the thin-ness parameter $\eta$ as well as the value of the instanton action $B$.

 Note that for all examples in table \ref{tab:ODplusfrommax} we have $\eta >1$. These are thick-walled solutions with the interpolation between $\f_f$ and $\f_0$ happening gradually. This is consistent with our findings of sec.~\ref{sec:ODthinwall}: There, we observed that in this class of potentials, thin-walled solutions require $\Delta \rightarrow \infty$. However, for $\f_f$ to be a maximum of $V$, the value of $\Delta$ is bounded as $\Delta < d$. Therefore, we expect that tunnelling solutions from an AdS maximum can never be of the thin-walled type in this class of potentials. This is supported by the examples in table \ref{tab:ODplusfrommax}.

Interestingly, we observe that potentials that permit tunnelling solutions from an AdS maximum also permit so called `bouncing' flows from the same AdS maximum. For example, this is the case for the degree-12 polynomial potentials considered here and studied in \cite{Ghosh:2017big}. This can be understood as follows. The tunnelling solutions corresponds to a holographic RG flow on the $(+)$-branch, i.e.~a flow with vanishing value $j=0$ for the UV source. Since here the UV fixed point is an AdS maximum, in addition to the $(+)$-branch solution it will also permit a family $(-)$-branch solutions for any value $j \neq 0$ of the UV source. Now consider such a $(-)$-branch solution with $j=\epsilon$ with $\epsilon$ arbitrarily small but non-zero. By continuity, the corresponding $(-)$-branch solution will only depart slightly from the $(+)$-branch solution with $j=0$. In particular, by choosing $\epsilon$ sufficiently small the end point of the $(-)$-branch flow can be made arbitrarily close to $\f_0$, the end point of the $(+)$-branch solution. Alternatively, we can consider a solution with $j=- \epsilon$. Again, by choosing $\epsilon$ sufficiently small we can ensure that the end point of this  $(-)$-branch solution is sufficiently close to $\f_0$. Now note that for the $(-)$-branch solution with $j=\epsilon$ the flow departs the UV fixed point towards the right (larger values of $\f$), while for $j=\epsilon$ the flow departs the UV fixed point towards the left (smaller values of $\f$). For these two flows to both end in the vicinity of $\f_0$ at least one of the flows needs to reverse direction (and also pass the UV fixed point one more time), therefore constituting an exotic flow of bouncing type. Note that in general, this will be a bouncing flow for a QFT defined on $S^d$. Thus we conclude that solutions describing tunnelling generically correspond to the interface in solution space between bouncing and non-bouncing holographic RG flows, see again fig.~\ref{fig:poly12cartoon} for illustration. Another way of phrasing our finding is that the existence of bouncing flows for theories on $S^d$ is in one-to-one correspondence with the existence of an $O(D)$-instanton describing tunnelling from a maximum.


\section{The Lorentzian continuation} \label{sec:analyticcont}

In this section we will consider the Lorentzian geometry,  as a solution to the bulk equations of motion, subject to the initial conditions specified by the instanton solution.

\subsection{Lorentzian solutions inside and outside of the bubble} \label{sec:insideoutside}
To describe the Lorentzian space-time after a tunnelling event mediated via an $O(D)$-instanton, we have to analytically continue the Euclidean $O(D)$-symmetric tunnelling solution to Lorentzian signature. How to do this has already been explained in the foundational work \cite{CdL}. To briefly review this, here we find the conventions in \cite{Dong:2011gx} useful.

For an $O(D)$-instanton the metric can be written as in \eqref{eq:metricxirho} with $\zeta_{\mu \nu}$ a metric on the sphere $S^d$. In the following it will be useful to specify coordinates on $S^d$ to write \eqref{eq:metricxirho} as:
\begin{align}
\label{eq:metricbefore} ds^2= d \xi^2 + \rho^2(\xi) \alpha^2 \left(d \theta^2 + \sin^2 (\theta) d \Omega_{d-1}^2 \right) = d \xi^2 + r^2(\xi) \left(d \theta^2 + \sin^2 (\theta) d \Omega_{d-1}^2 \right) \, ,
\end{align}
where in the second step we introduced $r(\xi) = \alpha \rho(\xi)$ corresponding to the physical radius of a shell at coordinate locus $\xi$.

To describe the complete Lorentzian space-time after the tunnelling
event two analytic continuations will be required. The reason is the
following. Consider the center of the CdL bubble at $r(0)=0$ as the
origin of the Lorentzian coordinate system. As we later show, the
first analytic continuation will allow us to only access points that
are space-like separated from the origin. This is also referred to as
the region `outside the bubble'.  To access the time-like separated region we shall require a different analytic continuation. To find the space-time in this region, denoted as `inside the bubble', we need to solve the corresponding equations of motion. The Euclidean solution only provides the initial conditions in this case.

To describe the region `outside the bubble' the angle $\theta$ on $S^d$ is continued as \cite{Dong:2011gx}:
\begin{equation} \label{eq:theta}
 \theta=\frac{\pi}{2}+i\chi \ .
 \end{equation}
Substituting this into \eqref{eq:metricbefore} the metric becomes
\begin{equation}
ds^2_{\textrm{out}}=d\xi^2+ r^2(\xi) \left[ -d\chi^2+\cosh^2(\chi) d\Omega^2_{d-1} \right].  \ \label{extmetric}
\end{equation}
The expression multiplying $r^2(\xi)$ can be identified as a metric of $d$-dimensional de Sitter space with unit radius. One can check that for this continuation, the equations of motion for the metric and scalar field are unchanged compared to the Euclidean case, i.e.~they are still given by \eqref{c3}--\eqref{c5}. As the equations of motion are unchanged, this implies that the Euclidean solution itself can be continued into this region. That is, the expression $r(\xi)$ in \eqref{extmetric} and also $\f(\xi)$ are just given by the solutions obtained for the Euclidean instanton. As one can easily confirm, continuing the Euclidean solution results in the entire bubble including wall and bubble exterior being mapped into the region described by \eqref{extmetric}.

To access the region `inside the bubble' we instead continue as \cite{Dong:2011gx}:
\begin{equation}
 \xi=i \tau , \quad \theta=i\eta  \, ,
 \end{equation}
 and define
 \begin{align}
 a(\tau) = -i r(i \tau)
 \end{align}
Inserting this into \eqref{eq:metricbefore} we find:
\begin{align}
ds^2_{\textrm{in}} =-d\tau^2+a^2(\tau) \left( d\eta^2+\sinh^2\eta  \, d\Omega^2_{d-1}\right) \  \label{intmetric} \, ,
\end{align}
which describes a FRW universe with hyperbolic slicing with scale factor $a(\tau)$. In Euclidean signature, the dilaton $\f=\f(\xi)$ was just a function of $\xi$. In this Lorentzian continuation we instead have $\f=\f(\tau)$.  The equations of motion are now given by
\begin{align}
2(d-1) \ddot{\hat{A}}+\dot{\f}^2+\frac{2}{d} R^{(\zeta)} e^{-2 \hat{A}}=0 \label{9.6} \ ,  \\
-d(d-1)\dot{\hat{A}}^2+\frac{1}{2} \dot{\f}^2 +V(\f)+R^{(\zeta)} e^{-2 \hat{A}}=0 \label{9.7} \ ,  \\
\ddot{\f}+d\dot{\hat{A}}\dot{\f}+V'(\f)=0 \ . \label{9.8}
\end{align}
where we introduced the scale factor $\hat{A}(\tau)$ via $a(\tau)=\alpha e^{\hat{A}(\tau)}$ and a dot $\dot{}$ now denotes a derivative with respect to $\tau$. The difference with the equations of motion \eqref{c3}-\eqref{c5} in the Euclidean case is that the sign in front of every appearance of $V$ and $V'$ is flipped compared to \eqref{c3}-\eqref{c5}.

As we show shortly, the metric \eqref{extmetric} describes the space-time for the space-like-separated region from the center of the bubble at $r(0)=0$ whereas the metric \eqref{intmetric} is valid in the time-like separated region. For a continuous space-time and solution for the scalar, the two metrics and the scalar field have to be matched at the center of the bubble. The relevant matching conditions are
\begin{align}
\nonumber & \f(\xi=0)=\f(\tau=0), \quad \frac{d}{d\xi}\f(\xi)|_{\xi=0}=\frac{d}{d\tau}\f(\tau)|_{\tau=0},  \\
& r(\xi=0)=a(\tau=0), \quad \ \frac{d}{d\xi} r(\xi)|_{\xi=0}=\frac{d}{d\tau}a(\tau)|_{\tau=0}\ .
\label{match}\end{align}
One can show that the curvature invariants (both `outside' and `inside the bubble') can be written as functions of $\dot{\f}$ \cite{Ghosh:2017big}.  Therefore, the conditions \eqref{match} imply that all the geometric quantities are continuous across $r(0)=0$.

In the region `inside the bubble' the equations of motion differ from those in the Euclidean setting and hence the Euclidean solution itself cannot be continued into this region. Instead, we have to solve afresh for $a(\tau)$ and $\f(\tau)$. However, the Euclidean instanton will provide the initial conditions for this analysis via the matching conditions \eqref{match}.

\noindent \textbf{Causality structure of the Lorentzian continuation --} To close this section, we construct the Penrose diagram for the continued Lorentzian space-time. To this end,  it is helpful to bring the Lorentzian space-time into the form
\begin{equation}
ds^2= f^2(\xi) \left(-dt^2+dx^2+dy^2+dz^2 \right) \label{min1} \, ,
\end{equation}
describing a manifestly conformally flat space-time. We begin by showing how the metric \eqref{extmetric} `outside the bubble' can be brought into the form \eqref{min1}. To this end we define
 \begin{align}
t&=\sigma(\xi) \sinh\chi \, , \label{trafo1} \\
x&=\sigma(\xi) \cosh\chi \cos\psi \,  \label{trafo2} \\
y&=\sigma(\xi) \cosh\chi \sin\psi\cos\phi \, , \label{trafo3} \\
z&=\sigma(\xi) \cosh\chi \sin\psi\sin\phi \, . \label{trafo4}
\end{align}
Inserting this into \eqref{min1} one finds:
\begin{equation}
ds^2=f^2(\xi) \left[\sigma'^2(\xi)d\xi^2+ \sigma^2(\xi) \left( -d\chi^2+\cosh^2(\chi) \, d\Omega^2_{2} \right) \right] \label{min2} \, ,
\end{equation}
which we can identify as \eqref{extmetric} as long as
\begin{equation}
f^2(\xi)\sigma'^2(\xi)=1, \quad  f^2(\xi)\sigma^2 (\xi)=r^2(\xi) \ .
\end{equation}
These conditions are satisfied for
\begin{equation}
\sigma(\xi)=\sigma_0 e^{\int_{\xi_0}^\xi \frac{d\xi'}{r(\xi')}} \, , \qquad f(\xi) =\frac{r(\xi)}{\sigma(\xi)} \, ,\label{sigmadef}
\end{equation}
where $\xi_0$ is a fiducial initial  point and  $\sigma_0$ can be
chosen at will.

We can also confirm that the point $r(\xi=0)=0$ in the coordinates
\eqref{extmetric} coincides with the origin $t=x=y=z=0$ in
\eqref{min1}. To this end, note that for $\xi \rightarrow 0$ one has
$r(\xi) \rightarrow \xi$, which follows from
e.g.~\eqref{eq:AIR}. Using \eqref{sigmadef} this implies that
\be \label{xitozero}
\sigma(\xi) \simeq C \xi \qquad \xi\to 0,
\ee
where $C$ is a constant which can be set to unity by an appropriate
choice of $\sigma_0$ in  equation (\ref{sigmadef}). From
(\ref{trafo1}-\ref{trafo4}) we see that $t=x=y=z \rightarrow 0$ as $\xi\to
0$.  To conclude, the metric  \eqref{extmetric} `outside the bubble' can be brought successfully into the form \eqref{min1} via the coordinate transformations \eqref{trafo1}--\eqref{trafo4}.

What can we learn from this? Note that from the definitions \eqref{trafo1}--\eqref{trafo4} it follows that
\begin{equation}
 x^2+y^2+z^2-t^2=\sigma^2(\xi)>0 \  .
\end{equation}
i.e.~the region that can be accessed via the coordinates $t,x,y,z$ as defined in \eqref{trafo1}--\eqref{trafo4} are the points that are space-like separated from the origin / center of the bubble. Note that  $t,x,y,z$ cover the space-time given in \eqref{extmetric} termed the region `outside the bubble'. Here, we find that this region fills the part of space-time that is space-like separated from the origin / center of the bubble. Recall that the Euclidean solution itself is continued into this region.

We now turn to the metric \eqref{intmetric} `inside the bubble' and show how it can be brought into conformally flat form. However, as the coordinate $\xi$ has been eliminated from \eqref{intmetric} in favour of $\tau$, instead of \eqref{min1} for the conformally flat metric we write
\begin{equation}
ds^2= g^2(\tau) \left(-dt^2+dx^2+dy^2+dz^2 \right) \label{min3} \, ,
\end{equation}
To bring this into the form \eqref{intmetric} we now define
\begin{align}
& t=\omega(\tau) \cosh\eta \, ,\label{trafo5} \\
& x=\omega(\tau)\sinh\eta\cos\psi \, , \label{trafo6}\\
& y=\omega(\tau)\sinh\eta \sin\psi\cos\phi \, , \label{trafo7}\\
& z=\omega(\tau) \sinh\eta \sin\psi\sin\phi \, , \label{trafo8}
\end{align}
which inserted into \eqref{min2} gives
  \begin{equation}
 ds^2=g^2(\tau)\left[-\omega'^2(\tau)d\tau^2+\omega^2(\tau) \left( d\eta^2+\sinh^2(\eta) \, d\Omega^2_{2}\right) \right] \ . \label{min4}
 \end{equation}
 This reduces to \eqref{intmetric} for
 \begin{equation}
 g^2(\tau) \omega'^2(\tau)=1, \quad g^2(\tau)\omega^2(\tau)=a^2(\tau) \ .
 \end{equation}
 which can be solved as
 \begin{equation}
 \omega(\tau)=\omega_0 e^{\int_{\tau_0}^\tau \frac{d\tau'}{a(\tau')}} \, , \qquad g(\tau) = \frac{a(\tau)}{\omega(\tau)} \, ,
 \end{equation}
with $\tau_0$ is a fiducial point where $\omega$ takes the value
$\omega_0$.

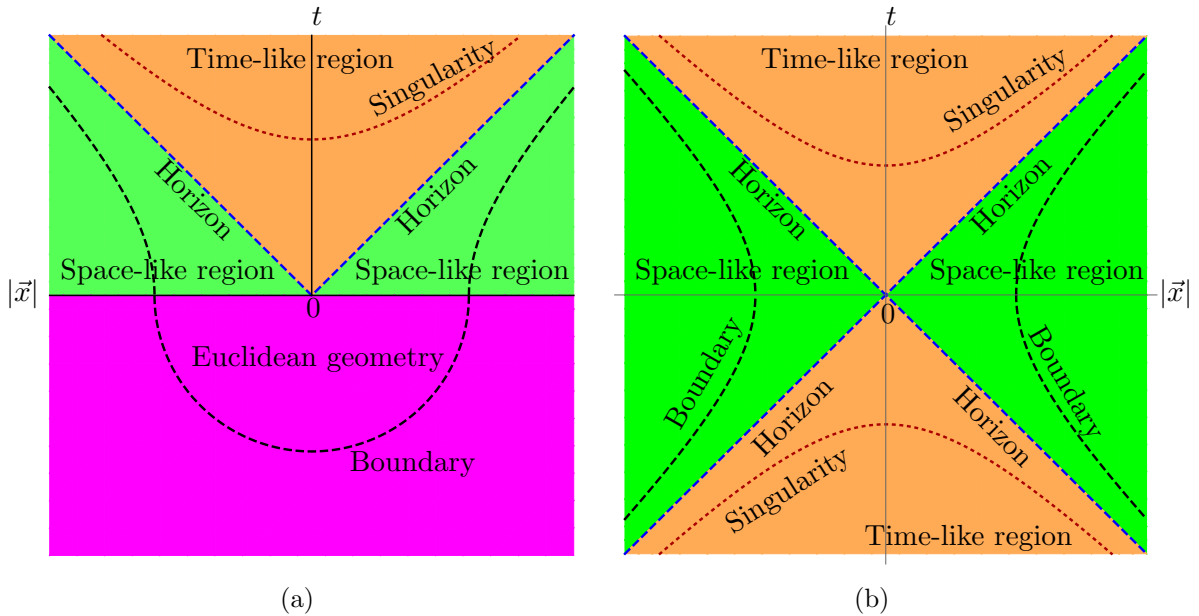
\begin{figure}[t]
\centering
\begin{subfigure}{.5\textwidth}
 \centering
   \begin{overpic}
[width=0.95\textwidth]{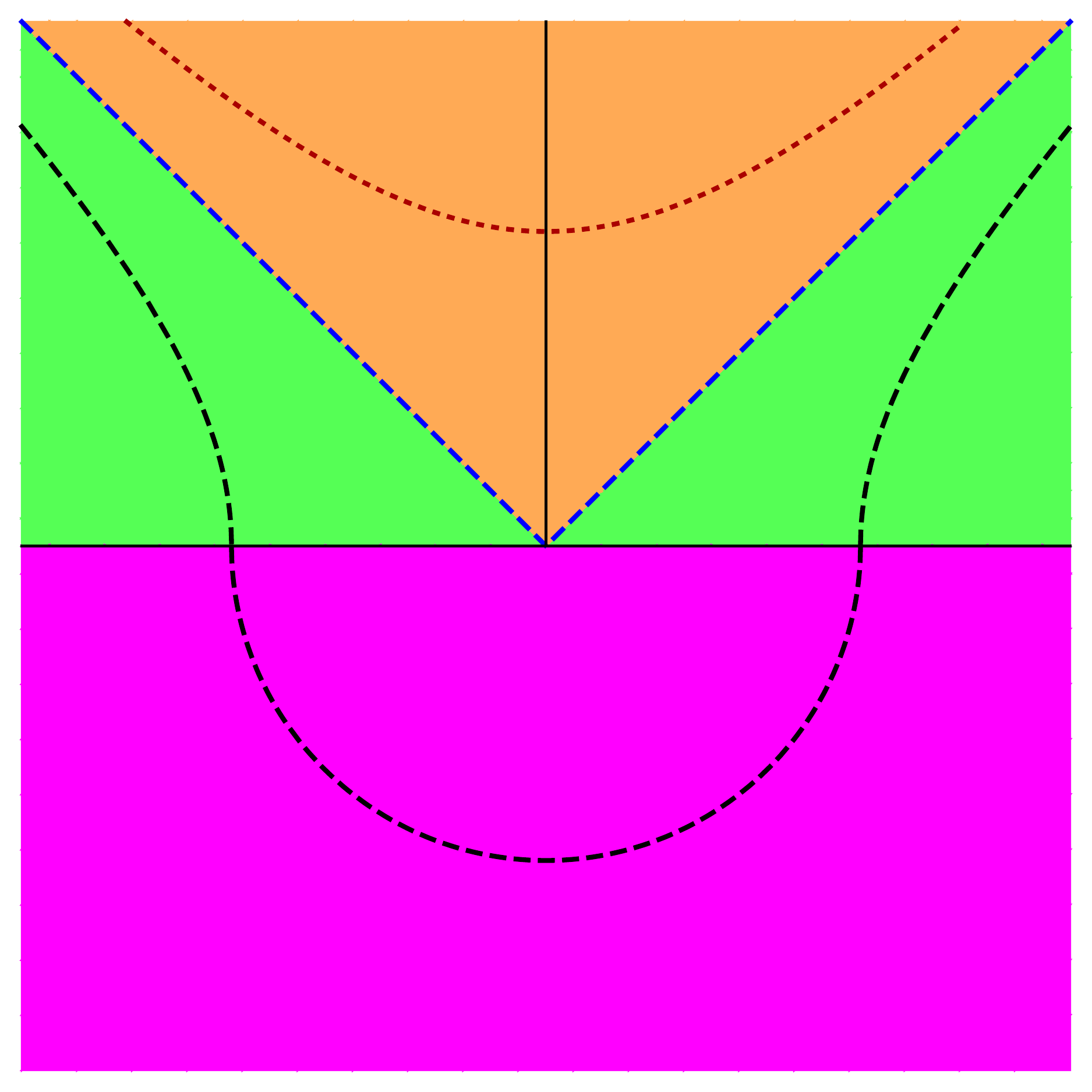}
\put(28,37){Euclidean geometry}
 \put(65,61){\rotatebox{45}{Horizon}}
  \put(21,73){\rotatebox{-45}{Horizon}}
  \put(49,46){$0$}
  \put(-6,49){$|\vec{x}|$}
  \put(50,100){$t$}
  \put(27,92){Time-like region}
   \put(4,53){Space-like region}
    \put(58,53){Space-like region}
   \put(60,83){\rotatebox{28}{Singularity}}
   \put(57,17.5){Boundary}
\end{overpic}
\caption{\hphantom{A}}
\label{Penrose}
\end{subfigure}%
\begin{subfigure}{.5\textwidth}
 \centering
   \begin{overpic}
[width=0.95\textwidth]{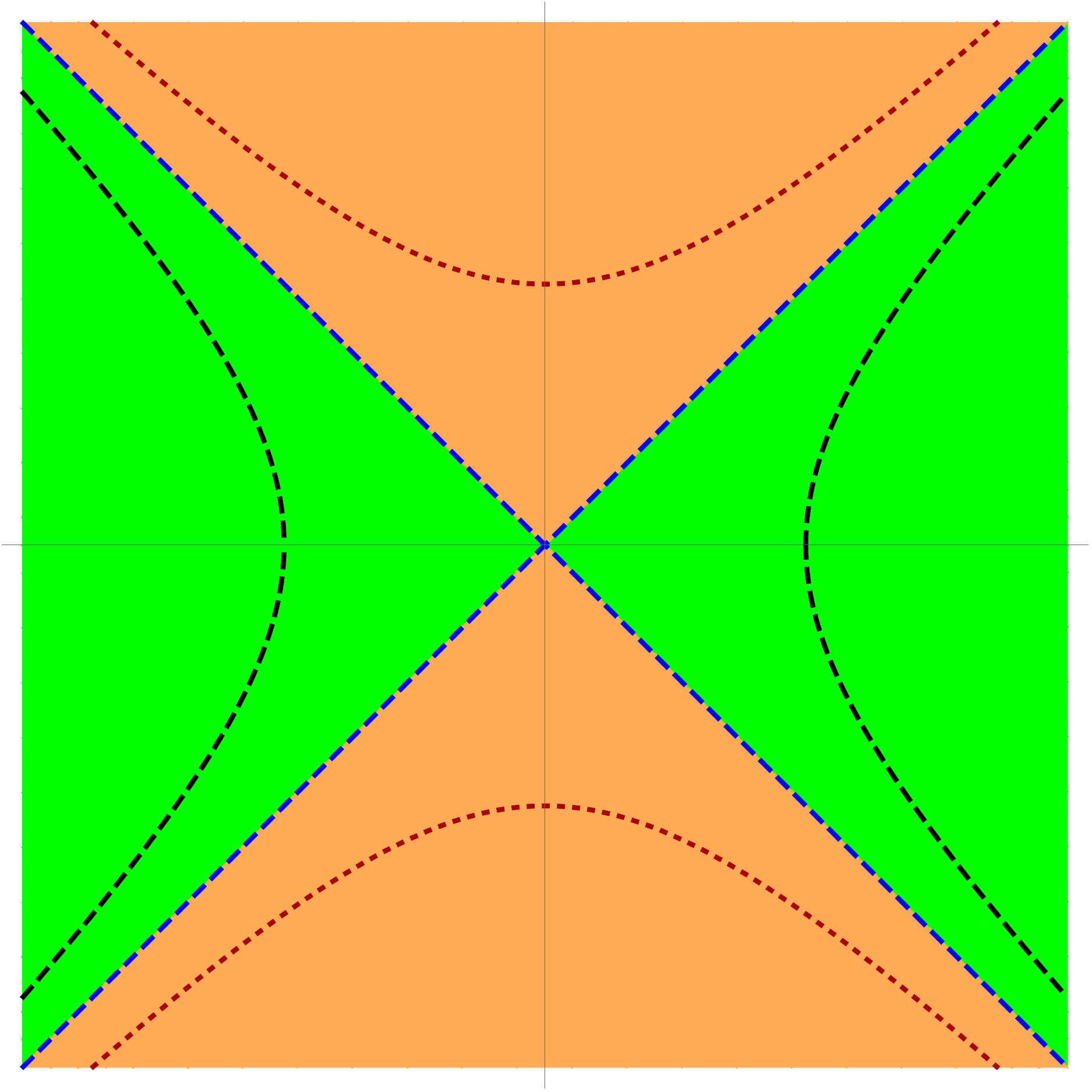}
 \put(65,61){\rotatebox{45}{Horizon}}
  \put(21,73){\rotatebox{-45}{Horizon}}
  \put(49,45){$0$}
  \put(100,49){$|\vec{x}|$}
  \put(50,100){$t$}
  \put(27,92){Time-like region}
   \put(4,53){Space-like region}
    \put(58,53){Space-like region}
   \put(60,80){\rotatebox{32}{Singularity}}
\put(20,6){\rotatebox{32}{Singularity}}
   \put(9,25){\rotatebox{60}{Boundary}}
   \put(75,43){\rotatebox{-60}{Boundary}}
\put(25,20){\rotatebox{45}{Horizon}}
\put(62,31){\rotatebox{-45}{Horizon}}
\put(46,4){Time-like region}
\end{overpic}
\caption{\hphantom{A}}
\label{fig13}
\end{subfigure}%
\caption{\textbf{(a):} Space-time diagram of the CdL geometry including the Lorentzian continuation. The purple region is the Euclidean geometry. Green and orange are space-like and time-like regions from the origin respectively. The boundary is denoted as the dashed black line whereas the blue dashed line denotes the horizon. Inside the time-like region, there is a singularity which is denoted as the red dashed line. This singularity is  hidden behind the horizon. \textbf{(b)}: Space-time diagram for a holographic RG flow for a QFT on dS$_d$, shown for comparison with the CdL case. The causal structure is identical to two copies of the Lorentzian continuation of the CdL geometry, connected at the surface $t=0$. The solution in the green regions has been previously studied in \cite{Ghosh:2017big}. The solution in the orange regions can be computed using the same method as in the CdL case. By analogy, we once more expect singularities in the orange regions.}
\label{fig:2Penrose}
\end{figure}

For $t,x,y,z$ as defined in \eqref{trafo5}--\eqref{trafo8} we now have
 \begin{equation}
 x^2+y^2+z^2-t^2=-\omega^2(\tau) <0 \ .
 \end{equation}
i.e.~the coordinates $t,x,y,z$ cover only the region that is time-like separated from the origin. This implies that the space-time described by \eqref{intmetric} describes the region that is time-like separated from the origin / center of the bubble.

We now have all ingredients to draw the Penrose diagram for the CdL geometry including the Lorentzian continuation. This is given by the plot of the space-time as the $|\vec{x}|$-$t$-plane and is shown in fig.~\ref{Penrose}. On the lower half of the plot, we show the Euclidean solution, while the upper half shows the Lorentzian continuation with $t>0$. The center of the CdL bubble is located at $t =0 = |\vec{x}|$. In the region that is space-like separated from the origin, (shown in green), the space-time is described by the metric \eqref{extmetric}. As remarked before, the entire CdL bubble solution is mapped into this region. The Euclidean bubble can be described as a series of concentric shells labelled by the coordinate $\xi$, with $\f(\xi)$ and $r(\xi)$ varying across shells. In the Lorentzian continuation, surfaces of constant $\xi_\star$ correspond to the hyperbolic curves
\begin{align}
|\vec{x}|^2 - t^2 = \sigma^2(\xi_\star) \, ,
\end{align}
as follows from \eqref{trafo1}--\eqref{trafo4}. In fig.~\ref{Penrose} we plot one such surface corresponding to the boundary of the space-time. {As the boundary is only reached for $\xi \rightarrow \infty$, for illustration purposes we plot the boundary contour for some $\xi$ that is finite but very large.} The corresponding surface is shown as a black dashed line in fig.~\ref{Penrose}. The region that is time-like separated from the origin (shown in orange) is described by the metric \eqref{intmetric}. Note that any event that occurs in this region, cannot reach the green space-like separated region and the surface $t=|\vec{x}|$ acts as a horizon for the interior region. In the time-like separated region, we also display one further feature that we have not discussed yet, but will be the focus of the next section. In this region, at some time $\tau_\star$, the space-time will generically exhibit a big crunch, as is known since the foundation paper \cite{CdL}.\footnote{The crunching is generically observed whenever the $O(D)$-instanton deposits the field in a region of the potential with $V<0$ as is the case here.} This big crunch singularity at $\tau_\star$ corresponds to
\be
t^2-|\vec{x}|^2=\omega^2(\tau_{\star}) > 0 \, ,
\ee
which is a hyperbola in the $|\vec{x}|$-$t$-plane and shown as the dotted red line in fig.~\ref{Penrose}.

\subsection{Big crunch singularities after tunnelling to AdS} \label{sec:bigcrunch}
In the foundational work of Coleman and de Luccia \cite{CdL} it was observed that when tunnelling to an AdS region of the potential the subsequent evolution leads to a big crunch of the space-time in the region `inside of the bubble', i.e.~the space-time described by \eqref{intmetric}. While in \cite{CdL} this was examined in the context of the thin-wall approximation, the existence of the big crunch does not rely on the applicability of this approximation, but is a generic outcome of tunnelling to a region with $V<0$, as is also emphasised by Tom Banks in e.g.~\cite{Banks:2002nm}.

Here we briefly present arguments for the generic existence of a big crunch singularity, after tunnelling to an AdS region, and comment on the significance of the singularity for the holographic interpretation of the tunnelling solution as an RG flow.

To understand the appearance of the big crunch, it will be useful to consider an explicit example. To this end,  we  revisit one example of an $O(D)$-instanton from section \ref{sec:numresults} and compute the continuation in the region described by \eqref{intmetric}.

\begin{figure}[t]
\centering
\begin{subfigure}{.5\textwidth}
 \centering
   \begin{overpic}
[width=0.95\textwidth]{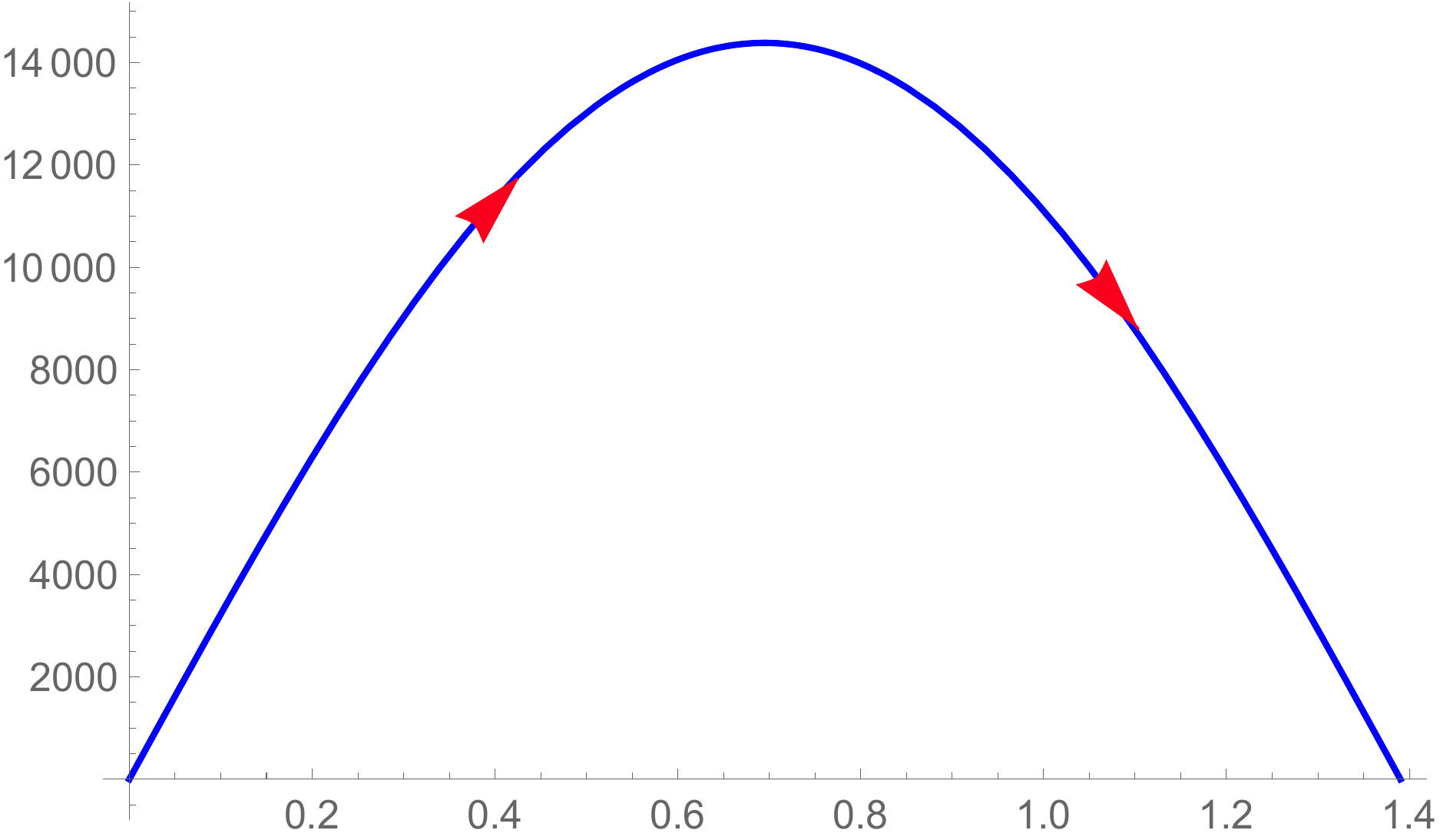}
\put(100,2){$\tau$}
\put(9,56){$\frac{a(\tau)}{\alpha} = e^{\hat{A} (\tau)}$}
\end{overpic}
\caption{\hphantom{A}}
\label{fig:Ahat1}
\end{subfigure}%
\begin{subfigure}{.5\textwidth}
 \centering
   \begin{overpic}
[width=0.95\textwidth]{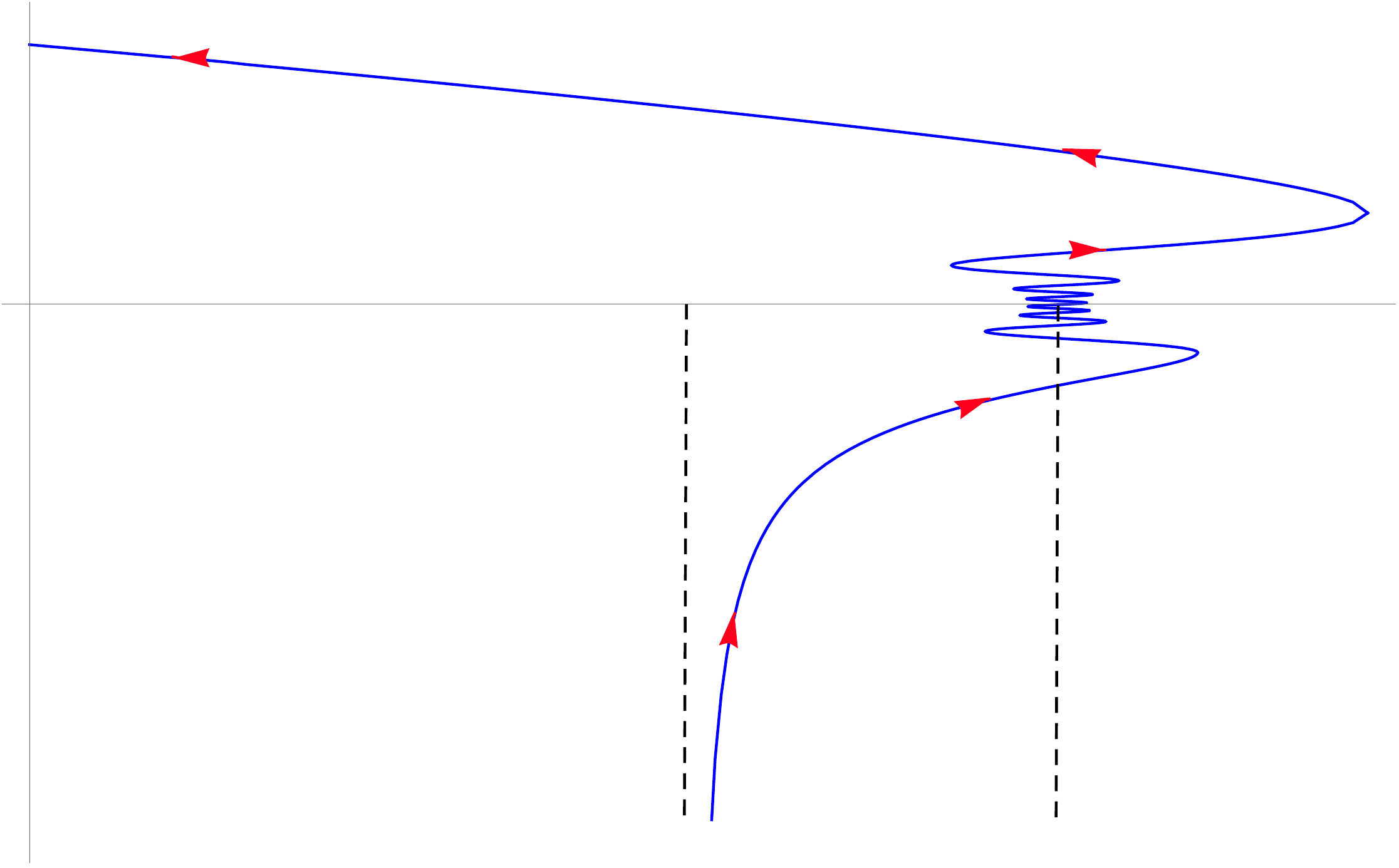}
\put(96,36){$\f$}
\put(46,43){$\f_0$}
\put(70,0){$\f=\f_t$}
\put(-2,63){$\hat{W}(\f)$}
\put(15,60){$\hat{W}$ goes to infinity}
\end{overpic}
\caption{\hphantom{A}}
\label{fig:What1}
\end{subfigure}%
\caption{Plots of $a(\tau)/\alpha$ (\textbf{(a)}) and $\hat{W}(\f)$ (\textbf{(b)}) for the Lorentzian continuation of an $O(D)$-instanton in the space-time region described by \protect\eqref{intmetric}. Here the $O(D)$-instanton interpolates between $\f_f=0$ to $\f_0=0.999999985$ in the potential \protect\eqref{eq:Vnum} with $\Delta=30.1$, $\f_t$ and $v_0=10$. Red arrows show the direction of time flow. The scale factor $a(\tau)$ increases until it reaches a maximum before decreasing and reaching zero again. The plot for $\hat{W}(\f)$ shows that the dilaton $\f$ oscillates about the true minimum of the potential at $\f_t=1$, first with a decreasing amplitude, then with increasing amplitude, until it diverges as $|\f| \rightarrow \infty$ when $a(\tau)$ drops to zero, where $W(\f) \rightarrow \infty$ also diverges.}
\label{fig:AWhat1}
\end{figure}

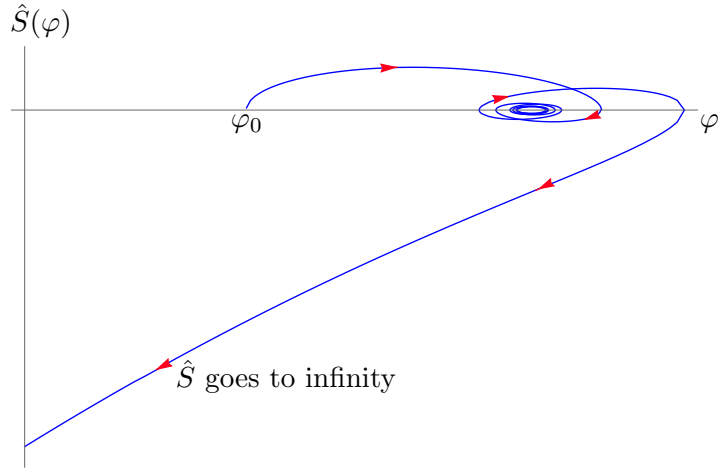
\begin{figure}[t]
\centering
\begin{overpic}
[width=.6\textwidth]{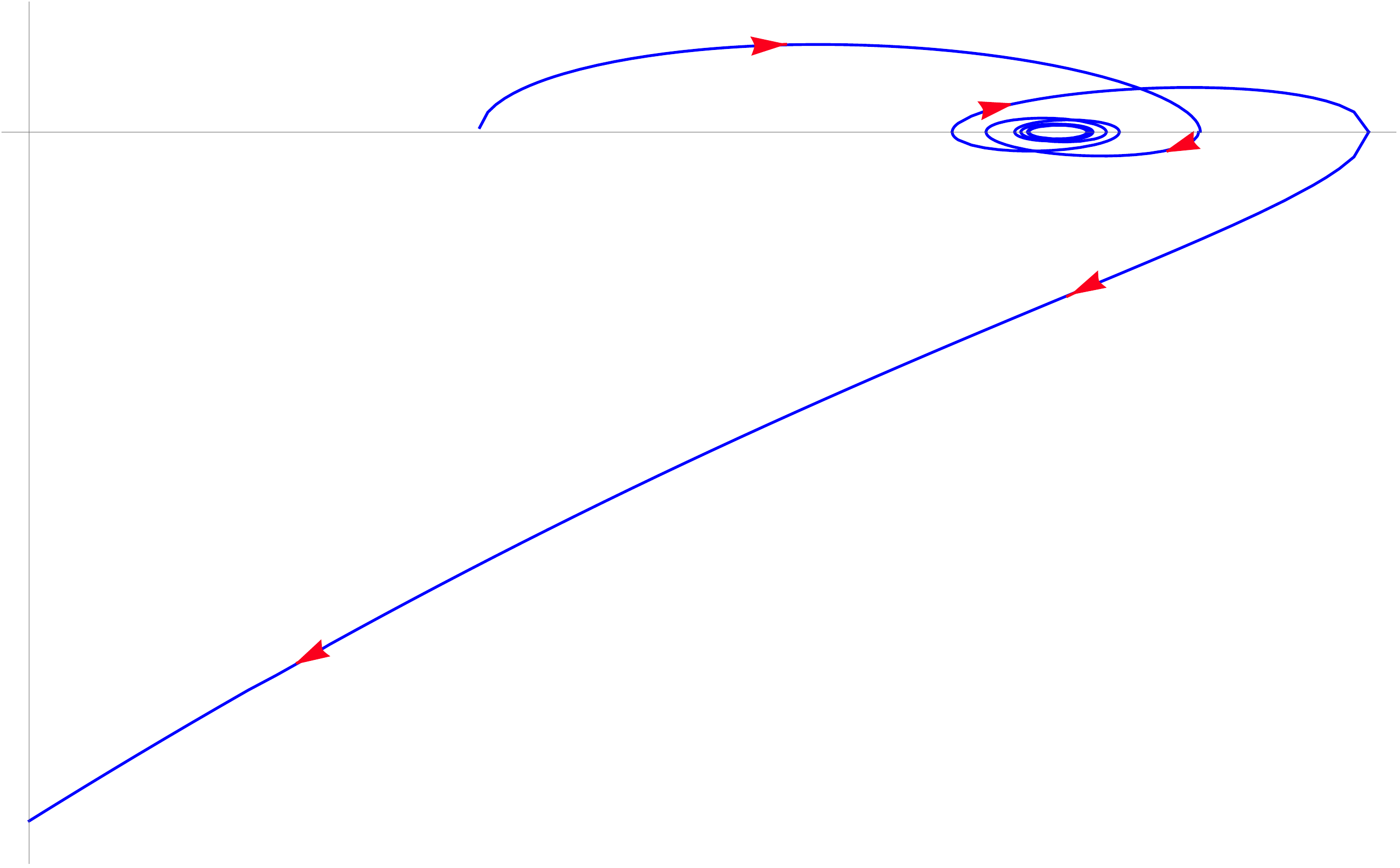}
\put(100,50){$\f$}
\put(0,64){$\hat{S} (\f)$}
\put(32,50){$\f_0$}
\put(24,12){$\hat{S}$ goes to infinity}
\end{overpic}
\caption{Plot of $\hat{S}(\f)$ for the Lorentzian continuation of an $O(D)$-instanton in the space-time region described by \protect\eqref{intmetric}. Initially, after starting from rest at $\f_0$ the scalar field $\f$ oscillates about the minimum at $\f_t$ leading to the cyclic trajectory in $\hat{S}$. Eventually, the amplitude of this oscillation increases and $\f$ is ejected from the potential well with $|\hat{S}|$ diverging. }\label{Shat1new}
\end{figure}

To find the solution in \eqref{intmetric} we need to solve \eqref{9.6}--\eqref{9.8} with the initial conditions given by \eqref{match}. That is, the initial conditions for $a(\tau)$ and $\f(\tau)$ in the Lorentzian continuation are set by the behaviour of $r(\xi)$ and $\f(\xi)$ or equivalently $A(u)$ and $\f(u)$ at the center of the Euclidean bubble geometry. The behaviour of $A(u)$ and $\f(u)$ near the center has been recorded in \eqref{eq:AIR} and \eqref{eq:phiIR} which via \eqref{match} imply that:
\begin{align}
\label{eq:aphitauinitial} a(\tau) \underset{\tau \rightarrow 0}{=} \tau + \mathcal{O}(\tau) \, , \qquad \f(\tau) \underset{\tau \rightarrow 0}{=}  \f_0 - \frac{V'(\f_0)}{2(d+1)} \tau^2 + \mathcal{O}(\tau^3) \, .
\end{align}
Here, in addition to solving for $a(\tau)$ and $\f(\tau)$, we introduce a new set of dynamical quantities that will satisfy first order differential equations. To this end, parallel to what we did in the Euclidean case in \eqref{eq:defWc}, we introduce the quantities
\begin{align}
\label{9.9} \hat{W}(\f)=-2(d-1) \dot{\hat{A}} \ , \quad \hat{S}(\f)=\dot{\f} \, , \quad \hat{T}(\f)=R^{(\zeta)} e^{-2 \hat{A}} \, .
\end{align}
In terms of these, the equations of motion \eqref{9.6}--\eqref{9.8} just become \eqref{eq:EOM4}--\eqref{eq:EOM6}, but with the sign in front of every instance of $V$ and $V'$ flipped. Using \eqref{eq:EOM4} and substituting into (\ref{eq:EOM5},\ref{eq:EOM6}) we can eliminate $\hat{T}$ to arrive at the following two equations:
\begin{align}
\frac{d}{2(d-1)} \hat{W}^2 +(d-1)\hat{S}^2 -d \hat{S}\hat{W}'-2 V=0 \ ,  \label{9.12} \\
\hat{S}\hat{S}'-\frac{d}{2(d-1)}\hat{W}\hat{S}+V'=0 \label{9.13}  \ .
\end{align}
One can further show that the initial conditions \eqref{eq:aphitauinitial} imply the following boundary conditions for $\hat{W}$ and $\hat{S}$:
\begin{align}
\hat{W}(\f) \underset{\f \rightarrow \f_0}{=}  \frac{\hat{W}_0}{\sqrt{|\f-\f_0|}} + \mathcal{O} \big( |\f-\f_0|^0\big) \, , \quad \hat{S}(\f) \underset{\f \rightarrow \f_0}{=} \hat{S}_0 \sqrt{|\f-\f_0|} + \mathcal{O} \big( |\f-\f_0|\big) \, ,
\end{align}
with
\begin{equation}
\hat{S}_0=\sqrt{\frac{2 |V'(\f_0)|}{d+1}}, \qquad \hat{W}_0=-(d-1)\hat{S}_0 \ .  \label{9.17}
\end{equation}

\noindent \textbf{A numerical example --} We now solve for $a(\tau)$, $\hat{S}(\f)$ and $\hat{W}(\f)$ for the Lorentzian continuation in the region \eqref{intmetric}, in the aftermath of the $O(D)$-instanton admitted by the potential given in \eqref{eq:Vnum} with $\Delta=30.1$, $\f_t$ and $v_0=10$. This was shown in section \ref{sec:numresults} to describe tunnelling from $\f_f=0$ to $\f_0=0.999999985$ (see e.g.~table \ref{tab:30p1phit1}). In fig.~\ref{fig:AWhat1} we display the results for $a(\tau)$ (fig.~\ref{fig:Ahat1}) and $\hat{W}(\f)$ (fig.~\ref{fig:What1}).  In fig.~\ref{Shat1new} the result for $\hat{S}$ is shown. We observe that $a(\tau)$ first grows, before reaching a maximum and then decreasing again until it vanishes again at $\tau_\star$ which we identify as a big crunch singularity. The fact that $a(\tau)$ initially increases follows from the initial conditions \eqref{eq:aphitauinitial} which initially set $a=0$, $\dot{a} >0$. The fact that the curve for $a(\tau)$ eventually turns around is a direct consequence of the equation of motion \eqref{9.6}. Rearranging, the equation implies that $\ddot{a} \leq 0$, so that $\dot{a}$ can only decrease. This eventually turns $\dot{a}$ negative, $\dot{a} <0$, at which point the crunch becomes unavoidable. The solution for $\hat{W}(\f)$ is a multivalued function. This results from the fact that the solution for $\f(\tau)$ oscillates about the minimum at $\f_t=1$. Intuition into the behaviour of $\f(\tau)$ can be obtained from \eqref{9.8}, which can be identified as the Newtonian problem of a particle performing motion in the potential $V$ (not the inverted potential in this case), with $\dot{a}>0$ providing positive friction, and $\dot{a}>0$ negative friction (anti-friction). From fig.~\ref{fig:What1} we see that the amplitude of oscillation initially decreases as a result of positive friction ($\dot{a}>0$). When $a(\tau)$ reaches the maximum, i.e.$\dot{a}=0$ we have $\hat{W}=0$ following from \eqref{9.9}. After that, the amplitude of the oscillation increases after $\dot{a}$ changes sign to act as anti-friction. Eventually the anti-friction is so strong that the field $\f$ is ejected towards infinity for $\tau \rightarrow \tau_\star$, with $\hat{W} \rightarrow + \infty$ also diverging. This interpretation is also supported by the behavior of $\hat{S}$ as shown in fig.~\ref{Shat1new}. By definition, $\hat{S} \sim \dot{\f}$ vanishes whenever $\f$ comes to rest. The oscillation of $\f$ about $\f_t=1$ therefore manifests itself as the cyclic trajectory in fig.~\ref{Shat1new}. Ultimately, $\hat{S}$ diverges when anti-friction leads to $\f$ being ejected out of the potential well.

To demonstrate that the locus $\tau_\star$ where $a(\tau)$ crunches is indeed a singularity,  we compute the curvature scalar $R_{AB}R^{AB}$. In particular, in the region `inside the bubble' this is given by \cite{Ghosh:2017big}:
\begin{equation}
R_{AB}R^{AB}= d^2 \left( -\frac{\hat{S}^2}{2d}+\frac{V}{d(d-1)}\right)^2 +\frac{d V^2}{(d-1)^2} \, ,  \label{Riccisquare}
\end{equation}
which we have written as a function of $\hat{S}$ and $V$. This diverges as $R_{AB}R^{AB} \rightarrow \infty$ at the locus $a(\tau \rightarrow \tau_\star) = 0$ in virtue of $|\hat{S}| \rightarrow \infty$. Therefore, the fate of the space-time `inside the bubble' is indeed a big crunch singularity.

Are there examples where the crunch singularity can be avoided in principle? One can imagine that at $\tau =\tau_\star$ with $a(\tau_\star) = 0$ the field $\f$ is not ejected to infinity, but comes to rest at some locus $\f_{\star}=\f(\tau_\star)$. The geometry could be regular there as it is at $\tau=0$, where the scale factor also vanishes, $a(\tau=0)=0$, and one has $\f(\tau=0) = \f_0$. For a generic point $\f_\star$, i.e~for $V'(\f_\star) \neq 0$, this would correspond to
\begin{align}
\label{eq:insideIRsol} a(\tau) \underset{\tau \rightarrow \tau_\star}{=} - (\tau - \tau_\star) + \mathcal{O}\big( (\tau-\tau_\star)^3\big) \, , \quad \f(\tau) \underset{\tau \rightarrow \tau_\star}{=} \f_\star - \frac{V'(\f_\star)}{2(d+1)} (\tau-\tau_\star)^2 + \mathcal{O}\big( (\tau-\tau_\star)^3\big) \, ,
\end{align}
which one can confirm to be the regular solution to \eqref{9.6}--\eqref{9.8} that satisfies the boundary conditions $\f(\tau_\star)=\f_\star$, $\dot{\f}_\star=0$ and $a(\tau_\star)=0$. One observation is that for a given value $\f_\star$ the solution \eqref{eq:insideIRsol}, just like the solution for $\tau \rightarrow 0$ in \eqref{eq:aphitauinitial}, is completely determined by the boundary conditions. That is, for a given value $\f_\star$, a tentative solution $\big(a(\tau), \, \f(\tau) \big)$ has to approach the value $\f_\star$ in a unique way determined by \eqref{eq:insideIRsol} to end there in a regular way. However, in the interior region of the bubble the solution is determined by the initial conditions at $\tau=0$. Thus there is no further freedom to ensure that the solution can approach any value $\f_\star$ in `the right way' to end there in a regular fashion. This would require a tuned potential and thus does not happen generically.

What is not possible either is that the field settles at the minimum, i.e.~$\f(\tau \rightarrow \tau_\star) \rightarrow \f_t$ while at the same time $a(\tau \rightarrow \tau_\star) \rightarrow 0$. The argument is the same as that presented in section \ref{sec:CDLnogo}, where we show that the Euclidean tunnelling solution cannot deposit the field $\f$ exactly at the AdS minimum at $\f_t$. Note that this argument, both for the Lorentzian and the Euclidean case, only holds for $R^{(\zeta)} \neq 0$, which is the case for $O(D)$-instantons and their Lorentzian continuation.

Alternatively, one could try to avoid the crunch singularity by ensuring that the space-time expands forever. For this to happen, it must be avoided that $\dot{a}$ turns negative, as then the crunch becomes unavoidable due to $\ddot{A} \leq0$. For this to be realised, one requires that $\ddot{\hat{A}} \rightarrow 0$ while $\dot{a}$ is still positive or approaches zero. Note that from \eqref{9.6} it follows that for $\ddot{\hat{A}} \rightarrow 0$ one requires that both $\dot{\f} \rightarrow 0$ and $\hat{A} \rightarrow \infty$. Again, while we cannot exclude that such a solution may exist, it does not seem generic that all these conditions can be satisfied simultaneously in a generic potential. Indeed, in the example considered above, this did not occur. Hence, we conclude that while big crunch singularities may be avoided in principle, in practice they are unavoidable once a CdL process deposits the field $\f$ in an AdS region. On this topic, also see the discussion in \cite{Banks:2002nm}, where it is also argued that the big crunch singularity cannot be avoided in general.

\subsection{The interpretation of tunnelling and the big crunch singularity from the perspective of the boundary field theory\label{5.3}}

In this section we would like to ask again the question: what is the significance of the big crunch singularity, and more
generally of the CdL $O(D)$-symmetric instanton, from the point of
view of the (dual) boundary field theory\footnote{We remind the reader
  that $d$ denotes the dimension of the boundary and $D=d+1$ is the
  dimension of the bulk.}?

One possibility is that the singularity signals that the process is unphysical, i.e.~that decays into spaces with negative cosmological constant do not occur in quantum gravity, a viewpoint advocated in e.g.~\cite{Banks:2002nm, Aguirre:2006ap}.
Alternatively, if the $O(D)$-symmetric solution interpolates between a
dS and an AdS region, it was argued in
\cite{Aguirre:2006ap,Espinosa:2015zoa} that this solution should not
be interpreted as a tunnelling process but instead as a recurrence to a
low-entropy state represented by the crunching region (see also
\cite{Banks:2012hx} for a brief summary of this idea).

The holographic duality however casts a different light on the big
crunch singularity and on the tunneling process as a whole.
If one takes the point of view of the boundary field theory, the AdS-to-AdS tunnelling by the
$O(d+1)$-instanton  has  a consistent interpretation in terms of a
holographic RG flow for a Euclidean  field theory  defined on the sphere S$^d$ or (in the
Lorentzian continuation)  on $d$-dimensional de Sitter space dS$_d$. This
RG flow originates from a CFT, dual to the false-vacuum  AdS
extremum of the bulk potential at $\f=\f_f$. When  defined on dS$_d$,
this CFT admits a non-trivial RG flow with subleading scalar field
asymptotics of $\f_+$ type, as in equation (\ref{phisolution1}).  The
flow is driven by the vev $\< \mathcal{O}\>$   of the
operator dual to $\f$, given by equation (\ref{vvev}),  with the source $j$ set to zero. This operator is relevant if the false vacuum is a maximum and irrelevant if the false vacuum is a minimum.

By  equation
(\ref{eq:curlyRdef}), the value of the vev $\<\mathcal{O}\>$ is   fixed by the
value of the curvature $R^{(\zeta)}$ of the de Sitter space-time on which the
boundary theory is defined, because for a $W_+$-type flow the
parameter ${\cal R}$ is uniquely fixed to a specific value as explained at the end of section \ref{sec:CDLasHoloRG}.

If the big crunch introduced a pathology into the system, one would
expect this to manifest itself on the boundary. Hence, here we
examine to what extent the boundary is affected by the existence of
the crunch.  To this end, we return to the space-time diagram in
fig.~\ref{Penrose}. The $O(D)$-symmetric solution  is identified with
the holographic RG flow of the Euclidean theory. This is shown in
fig.~\ref{Penrose} as the Euclidean solution in the lower half. It  is
also mapped by analytical continuation into the green region that is
space-like separated from the origin. This region is dual to the RG
flow on dS$_d$ of the boundary theory. The dashed black line indicates
the UV boundary of the space-time, which is formally located at $\xi
\rightarrow \infty$, but is shown here for illustration purposes at
some large, but finite $\xi$.

From the Penrose diagram in figure \ref{Penrose} we can make two
observations:
\begin{enumerate}
\item The singularity and the horizon both reach the boundary only in
  the asymptotic de Sitter future $\chi \to +\infty$.
\item No information from the orange region in the future lightcone of
  the origin can reach the green region, nor the boundary. This then also applies to the big crunch singularity (shown as the red dotted line) as this is confined to the future lightcone of the origin.
\end{enumerate}
Therefore, from the boundary point of view, the big crunch singularity is cloaked by a horizon at $t=|\vec{x}|$ and hence the boundary is unaffected by the crunch.\footnote{AdS-to-AdS CdL processes have also been considered
  in \cite{Barbon:2010gn} from a holographic point of view. There,
  the CdL process  in the bulk was argued to lead to  a dual decay
  instability of the  boundary theory, dual  to global AdS. We will come back
  to this point in Section \ref{sec:spheretocylinder}.}

The considerations above may be taken as a hint that,  despite the existence of the big
crunch singularity,  a holographic interpretation of the
$O(D)$-instanton geometry is still perfectly possible, i.e.~it
successfully describes a holographic RG flow of a consistent
theory on de Sitter space-time. This point of view is also the one
advocated in \cite{Maldacena:2010un}. If the existence of the big
crunch does  introduce a pathology on the field theory side, it must do so in a more subtle way, which leaves an interesting direction for future work.

We therefore continue under the assumption that the existence of the
CdL solution does not constitute a pathology of the holographic theory, and we turn to discussing what its
existence implies for the boundary theory.

For one thing, the CdL instanton certainly does not describe a vacuum
 transition or decay process   on the field theory side. As we
 discussed above, the boundary theory sees only the exterior geometry,
 whose  boundary is de Sitter  space-time with a constant
 curvature and a non-zero but constant vev for an irrelevant operator
 (in the case the false vacuum is a minimum) which induces an RG
 flow. The boundary field theory is the false-vacuum QFT at all
 times, and no vacuum decay occurs.

What makes the situation interesting is that we have found two
distinct regular Lorentzian  geometries with
the same boundary asymptotics:
\begin{enumerate}
\item The solution with constant scalar field set to the false vacuum
  value,  $\f = \f_s$ and metric given by
  AdS$_{d+1}$ foliated by dS$_{d}$;
\item The vev-driven RG-flow  solution with non-trivial $\f =
  \f(\xi)$, and metric which deviates from  AdS$_{d+1}$ as we move
  towards the interior.
\end{enumerate}
In both solutions the source term in the near-boundary expansion of
the scalar field is vanishing, and the metric source is the same. Therefore  (1) and (2)
represent two different {\em semiclassical} states of the same dual QFT on de
Sitter, which we will denote by $|C_1\>$ and $|C_2\>$ .
They have the
property:
\be \label{vevs}
\< C_1 | \mathcal{O} | C_1 \> = 0, \qquad  \< C_2 | \mathcal{O} | C_2 \> = (M
\ell_f)^{(d-1)} (2\Delta - d) \f_+ \, ,
\ee
where $\mathcal{O}$ is the operator dual to the bulk scalar field.

The two semiclassical states have different symmetry properties:
$|C_1\>$ enjoys the  full AdS$_{d+1}$ invariance $SO(2,d)$,  whereas
$|C_2\>$ is only invariant under the  $SO(1,d)$ subgroup, i.e. the
group of  dS$_d$ isometry of the constant-$\xi$ slices.

An interesting point is that  both solutions have the same holographic
stress tensor. Indeed, in pure vev solutions the backreaction of the
running scalar on the metric appear at orders which are subleading
with respect to the terms giving the boundary stress tensor
\cite{Ghosh:2017big}. However the difference may be seen at the level of
higher-point correlation functions which will bear the signs of the spontaneously broken conformal invariance.\footnote{This phenomenon for a CFT defined on flat space was discussed in \cite{vev1,vev2}.}

So far we have described the field theory  interpretation of the two
different  Lorentzian solutions.   What is the interpretation of the
CdL instanton, and what is the meaning on the field theory side of the
tunneling rate $e^{-B}$, with $B$ given  in (\ref{eq:acdiff})?

The gravitational path integral with Euclidean AdS$_{d+1}$ boundary
conditions corresponds to the Hartle-Hawking state\footnote{The
 original no-boundary prescription of Hartle and Hawking is adapted here to
  AdS space-times, and becomes the condition that in the Euclidean path integral
  the geometry asymptotes to Euclidean AdS with no sources turned on \cite{eternal}.}
\cite{Hartle:1983ai}.  Specifying these boundary conditions in  Euclidean time
on the gravity side corresponds to putting the field theory in a
specific state,  which we call $|HH \>$. Although the boundary field theory is
not coupled to a dynamical  metric,  one can say that this state obeys
the ``no-boundary'' condition of \cite{Hartle:1983ai}, since the
geometry  on which the field theory lives is $S^d$ in the Euclidean past.\footnote{Hartle and Hawking argued that this
  state may represent the  vacuum  state of the gravity theory. If one
accepts this interpretation, then one may suppose that one can take this state as the
vacuum of the QFT on de Sitter. This is the case for example in a
(near) free QFT on de Sitter, where the Bunch-Davies vacuum does correspond to
the Hartle-Hawking state (see e.g.~\cite{Maldacena:2002vr})}

\begin{figure}[t]
\centering
\begin{overpic}
[width=\textwidth]{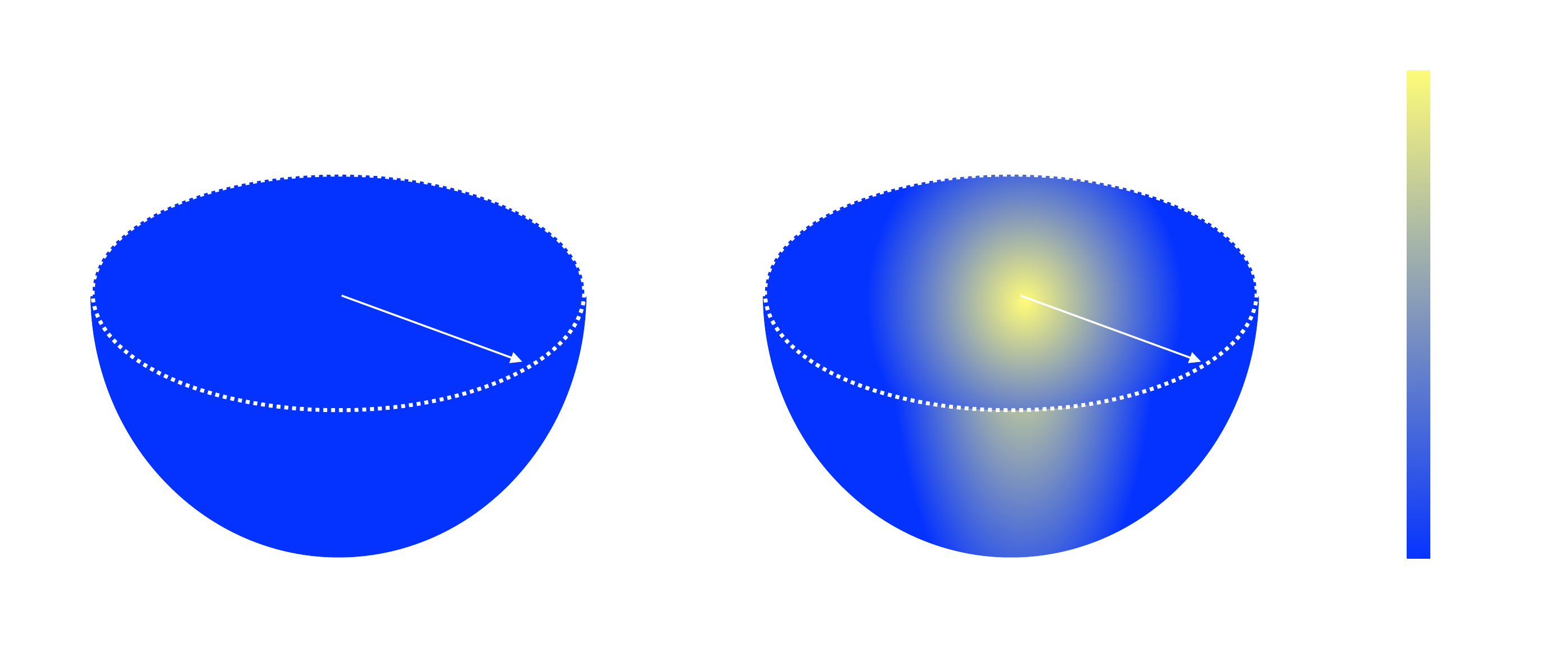}
\put(26,19){$\color{white}{\xi}$}
\put(70,19){$\color{white}{\xi}$}
\put(40,24){$\chi=0$}
\put(15,3){AdS solution}
\put(55,3){RG flow solution}
\put(92,7){$\f_f$}
\put(92,37){$\f_0$}
\end{overpic}
\caption{No-boundary Euclidean solutions with two different  spatial metric and dilaton at
  the fixed time-slice $\theta=\pi/2$ (Lorentzian time $\chi=0$). The
  left figure represents the  false-vacuum solution, in which the
  scalar field is constant,  $\f = \f_f$, and the metric is AdS$_{D}$ sliced
  by $S^d$. On  the right, the solution corresponding to the CdL
  instanton, a.k.a.~holographic RG flow on $S^d$, in which the scalar field flows from $\f_f$ (for $\xi\to
  +\infty$) to $\f_0$ at the center, and
  the metric deviates from AdS in the interior. }\label{fig:WF}
\end{figure}
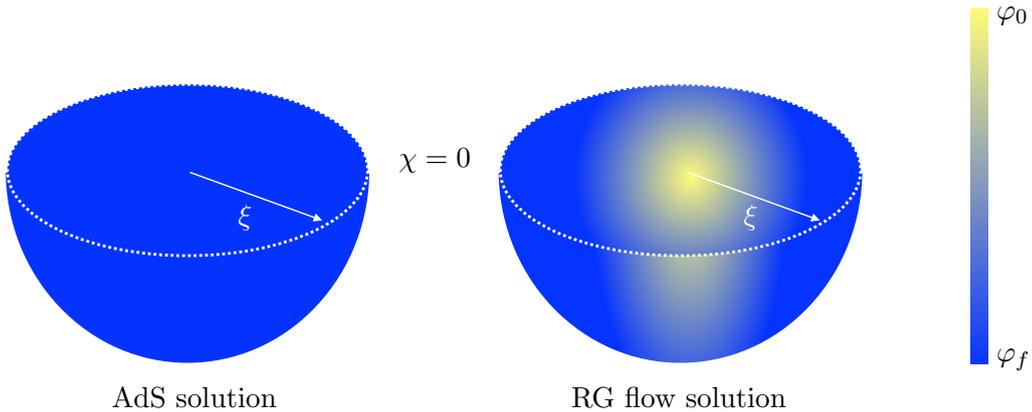

The  $|HH \>$ state  is not  a semiclassical
state, i.e.~it does not correspond holographically to a specific classical
geometry. In fact, we have just seen that there are at least two classical
solutions of the bulk theory which obey the Hartle-Hawking condition in
the Euclidean regime: these are the ones obtained by gluing solutions (1)
and (2) above  to the corresponding Euclidean solutions (the one with
trivial scalar and the one with running scalar) on  a space-like
hypersurface at $\chi=0$. In  each of
these two solutions,  we denote by $\gamma_{ab}^0(\xi)$ and $\f^0(\xi)$ the
metric and scalar field profile induced on the $\chi=0$ surface. The
two corresponding Euclidean geometries are represented in figure \ref{fig:WF}: on the left we have (the lower half of) the
Euclidean AdS false-vacuum solution with constant scalar field; on the
right (the lower half of) the CdL instanton with scalar field
interpolating from $\f_f$ to $\f_0$. Both solutions have the same
sources at the (conformal) boundary.

According to the standard interpretation of the Hartle-Hawking
prescription,  the quantity $e^{-S_E[\gamma^0,\f^0]}$ represents  (up to an
overall normalization) the  semiclassical approximation to the quantum  amplitude for obtaining the $(\gamma^0(\xi),\f^0(\xi))$
geometry in the HH state.  In other words, the
Hartle-Hawking condition provides a  state, i.e.~a wave-functional of
classical spatial geometries,   and  the
  instanton action is the probability of finding a given semiclassical
  configuration, in that state.\footnote{The same as in quantum
    mechanics
    $|\psi(x)|^2 = |\<\psi |x\> |^2$ is the probability (density) of finding the particle
    at $x$ in the state corresponding to the  wave-function $\psi$. In  our case, the classical position variable $x$ is replaced
  by  the classical  $d$-dimensional geometry and scalar field  on a space-like
  hypersurface.}

 Equivalently,
$e^{-S_E[\gamma^0,\f^0]}$ gives, in the semiclassical approximation,  the
  wave-function  $\Psi$  of the HH state (which is a functional of a
  fixed-time geometry), evaluated at the classical
  ``point''($\gamma_{ab}^0, \f^0$), which can be taken to be
  solution (1) or (2) respectively,
\be \label{HH1}
\Psi[\gamma^0_1,\f^0_1] =  \< HH | C_1 \> \propto e^{-S_E[\gamma^0_1,\f^0_1]} , \qquad \Psi[\gamma^0_2,\f^0_2]=\< HH | C_2 \> \propto e^{-S_E[\gamma^0_2,\f^0_2]}
\ee

According to this interpretation, the bounce action gives the relative
probability of ``measuring'' one of the two classical configurations
when the system is in the  HH state:
\be \label{HH2}
{\frac{|\< HH | C_2 \> |^2}{|\< HH | C_1 \> |^2}} = e^{- (S_2 - S_1)},
\ee
(up to subleading  $1/N^2$ corrections which correspond to quantum
effects in the bulk), where $S_2$ is the Euclidean bounce action and $S_1$ is
the  Euclidean false vacuum AdS action. The exponent on the right hand side of
equation (\ref{HH2}) is exactly the negative of the quantity  $B$ in (\ref{eq:acdiff}).

\begin{figure}[t]
\centering
\begin{overpic}
[width=0.5\textwidth]{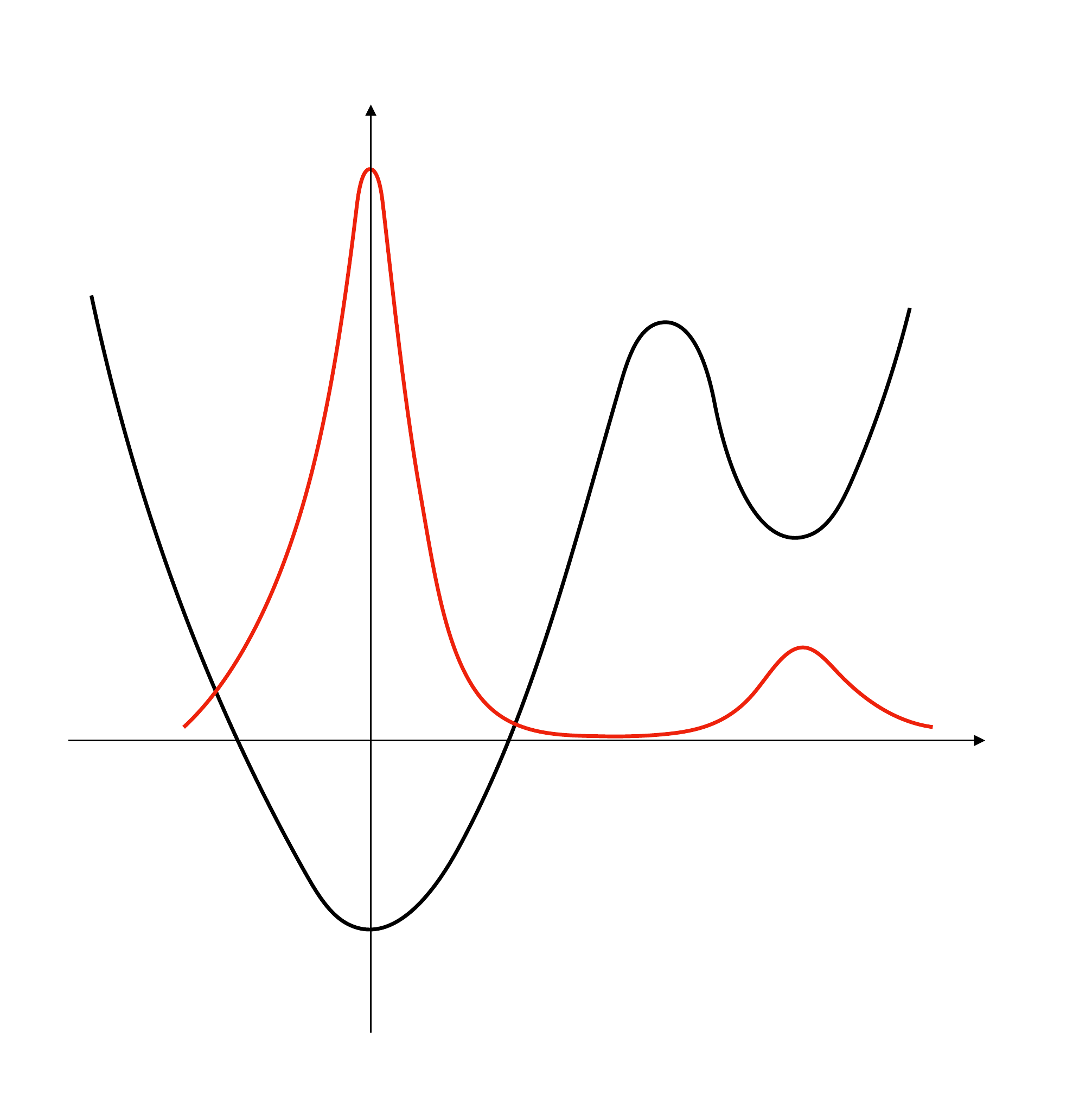}
\put(92,30){$\< \mathcal{O}\>$}
\put(10,70){${\cal V}$}
\put(40,50){$\color{red}{\Psi}$}
\end{overpic}
\caption{Quantum mechanical analogy of the space of classical
  no-boundary solutions  in terms of a double-well potential ${\cal V}$
  (black curve). The
  horizontal axis represents the space of solutions (characterized in
  this case by the vev of the operator $\mathcal{O}$ dual to $\f$). The two
  minima of the potential are the semiclassical solutions: The
  minimum on the left represents the AdS solution, the  one on the
  right the RG flow solution. The red curve is a sketch of the
  would-be ground-state wave-function of this quantum mechanical system. It should be remarked that the potential here is not the bulk scalar potential. }\label{fig:QM-WF}
\end{figure}

One way of visualizing the situation qualitatively (although with many
caveats) is by making the analogy with a  quantum mechanical system with
a potential having two (non-degenerate) minima, as in figure
\ref{fig:QM-WF}. Each minimum of the potential (black curve)
corresponds to  a classical state (the particle on the left or on the
right of the barrier), but  the ground state wave-function (red curve)
is
delocalized. If the energy difference between the two minima is
large, the wave-function will be mostly peaked on the lower-energy
minimum. A measurement of the position of the particle will have with
high probability the classical outcome ``the particle is localized on
the left''.

In our case, the peak in $\Psi$ on the right corresponds to the CdL instanton geometry,
which  has higher free energy than the pure AdS solution
 (i.e.~$B>0$,
as we have found in section \ref{sec:instaction}). Therefore the HH state has an (exponentially)
larger overlap with the AdS solution than with the RG
flow. Nevertheless, the full AdS symmetry is broken in the HH state, albeit by
exponentially small effects.

The quantum mechanical analogy cannot  be taken too seriously,
however. First of all, the free energy of the field theory on the
sphere is not the same as the eigenvalue of the
Hamiltonian. Furthermore, we are dealing with  a QFT in de Sitter,
which does not have a global time-like killing vector, so there is not really
a Hamiltonian to speak of, and the concept of a ground state is
rather vague. What we can say is that there is a state with higher
symmetry, which is the semiclassical state $|C_1 \>$. The HH state on the
other hand has  lower symmetry since it is a linear combination of (at
least) $|C_1 \>$ and $|C_2\>$. It would be very interesting if one could give an
independent characterization of the HH state from the field theory
side. This is  possible for example in perturbative field
theories on de Sitter (the Bunch-Davies vacuum).

To summarize this section, while the CdL instanton has a dynamical
interpretation in the bulk as computing a decay rate between the false
AdS vacuum and (something close to) the true AdS vacuum, the boundary
field theory interpretation is completely different, as it does not
describe a transition: it concerns only the ``false vacuum'' QFT,
and it computes  the relative
overlap of two states, which have a semiclassical counterpart each, with
a third state (the HH state) which does not. In this picture, there is
no obvious trace of the ``true vacuum'' AdS CFT in the boundary field
theory (as it is hidden behind the horizon).

\section{From the sphere to the cylinder}
\label{sec:spheretocylinder}

In the previous sections we discussed vacuum decay by bubble
nucleation described by a maximally symmetric instanton. As we have
emphasized, this only applies when the asymptotic boundary of the
Euclidean bulk space-time has the geometry of $S^d$. Consequently,
 the dual QFT is defined on a round sphere or,  in the
Lorentzian solution describing the expanding bubble, on de Sitter
space-time.

In this section we ask the question whether these solutions give any
information about vacuum stability for different asymptotic boundary
geometries. For example, can we say anything about the stability of a
dual QFT defined on $d$-dimensional Minkowski space-time or on a (generalized) cylinder?

The reason this question makes sense is that in AdS/CFT there exist bulk
diffeomorphisms which have various effects on the boundary QFT data.
In general, in  locally asymptotically AdS space-times, one can
identify  three types of bulk diffeomorphisms, depending on their
near-boundary asymptotics:
\begin{enumerate}
\item Diffeomorphisms  which change the leading near-boundary behavior of the
  metric and bulk fields. These transformations change the sources of
  the UV QFT (i.e.~the metric and coupling constants of relevant
  operators), therefore they are maps between different boundary
  theories.

\item  Diffeomorphisms  which preserve the leading boundary asymptotics (sources) but change the subleading  near-boundary
  behavior of the metric and bulk fields (i.e.~the vevs of the dual operators).
 These are maps which change
  the {\em state} of the boundary theory, but not the theory itself.

\item Diffeomorphisms which vanish fast enough close to the boundary, as to
  change neither the sources nor the vevs (therefore they do not
  affect the Fefferman-Graham expansion to any order).
  Such transformations are preserving the state of the dual QFT.
   As all data of the QFT are computed from vevs near the boundary, such transformations act trivially on all QFT data and in particular on all QFT correlation functions.  They do not
  have a corresponding description in the dual quantum field theory
  but are true gauge transformations (i.e.~redundancies) of the bulk gravitational description.

\end{enumerate}

In all of these cases, diffeomorphisms may be accompanied (and typically they do)  by the appearance
of horizons.

As we shall  discuss below, one can construct  type 1 diffeomorphisms which
connect the QFT on de Sitter space-time to the same field theory on
the (generalized) cylinder or flat space.

Consider  the case of an exact AdS geometry (with no bubble). There,
one can make a globally defined (as opposed to only asymptotically)
change of coordinates which transforms the $S^d$ (or $dS_d$) radial
slicing to a slicing by $R_t\times S^{d-1}$ or by flat space-time. Here, the CdL tunnelling geometries we considered are
asymptotically AdS in spherical slices, and hence one may define appropriate
coordinate transformations which, asymptotically, change the boundary
structure from $S^d$ to flat space-time or the generalized cylinder. The
corresponding geometry will still be a (regular) solution of Einstein equation,
and one may wonder, whether this solution still describes vacuum decay
of the same dual QFT theory, when defined on flat-space or on a
 cylinder.

In the case  we analysed so far,  the exterior solution, in which the
bubble expands, extends all the way to the infinite de Sitter
future. As $\chi \to \infty$, with $\chi$ the de Sitter time (as defined in equation
(\ref{extmetric})), both the null horizon separating the interior and
the exterior solution,  and the big-crunch  singularity in the
interior, approach the asymptotic boundary of the $(d+1)$-dimensional
space-time. The fact that this happens in the infinite future indicates
that the dual QFT on the boundary {\em can exist indefinitely}.

It may appear, however, that this situation changes drastically when we
go from the sphere to flat space or the cylinder. In fact, as we will
review below,  in the
exact AdS case,
  future infinity in the de Sitter slicing of AdS
corresponds to a {\em finite} time in both the cylinder (Global AdS)
and Minkowski slicing (Poincar\'e AdS). Therefore, it may seem that the
solutions we found would, upon an appropriate coordinate
transformation, yield solutions with different boundary conditions
which describe a QFT which ceases to exist after a finite time (when
the singularity reaches the boundary). This observation had been
already made by several authors, who  argued that the
existence of an $O(D)$-symmetric instanton in the  bulk implies a
runaway vacuum decay  at finite time for the corresponding dual field
theory defined on the cylinder \cite{Barbon:2010gn}.

Here,  we argue that this conclusion is too rash:
the $O(D)$-symmetric instanton, upon
coordinate transformation to a different slicing, does not in fact
describe a {\em spontaneous} decay  in a finite time of the dual field
theory. Rather, it describes a driven decay, not unlike what one would
obtain by turning on, in the UV CFT, a time-dependent source which
becomes singular at a finite time.

Below, we will carry out the analysis for the cylindrical boundary, but
similar conclusions can be reached for a flat-space boundary.

\subsection{AdS in different radial slicings}

Before we turn to the full vacuum decay problem, we review the
various ways to slice AdS$_{d+1}$ corresponding to different boundary
geometries.

Global   AdS$_{d+1}$ space-time with length $\ell$ is described by the metric:
\be \label{cyl-1}
\textrm{Cylinder slicing:} \quad  ds^2  = d\lambda^2 - \ell^2 \cosh^2 {\frac{\lambda}{\ell}} d\psi^2  + \ell^2 \sinh^2 {\frac{\lambda}{\ell}} d\Omega_{d-1}^2 \, ,
\ee
where $d\Omega_{d-1}^2$ is the metric on the unit $(d-1)$-sphere, and the
global time coordinate $\psi$ has the domain $(-\infty,+\infty)$. The boundary
region is reached as $\lambda \to +\infty$, where the metric asymptotes
to:
\be\label{cyl-2}
ds^2 \to  d\lambda^2 + e^{2\lambda/\ell}\ell^2 \left[-   d\psi^2  +
  d\Omega_{d-1}^2 \right] \, , \qquad \lambda \to +\infty \, .
\ee

The de Sitter slicing we have been using so far is given in equation
(\ref{extmetric}) (we will only need the exterior geometry, i.e.~the
one which is space-like separated from the center of the bubble),
which we record below:
\be\label{cyl-3}
\textrm{de Sitter slicing}: \quad ds^2  = d\xi^2 + \ell^2 \sinh^2 {\frac{\xi}{\ell}} \left[-d\chi^2 +
  \cosh^2{\chi} d\Omega_{d-1}^2 \right] \, .
\ee
The metric in the square brackets is the de Sitter metric in global coordinates, with time $\chi$
running from $-\infty$ to $+\infty$.   The radial coordinate  $\xi$  takes values in
$(0, +\infty)$. The asymptotic boundary is reached as $\xi \to
+\infty$, where the metric asymptotes to:
\be\label{cyl-4}
ds^2 \to  d\xi^2 + e^{2\xi/\ell} \ell^2 \left[-d\chi^2 +
  \cosh^2{\chi} d\Omega_{d-1}^2 \right] \, , \qquad \xi \to +\infty \, .
\ee
In writing (\ref{cyl-3}) we have chosen  coordinates so that the
asymptotic de Sitter scale is  $1/\ell$, but any other choice would
have been possible by a shift in $\xi$.

\begin{figure}[t]
\centering
\begin{overpic}
[width=.8\textwidth]{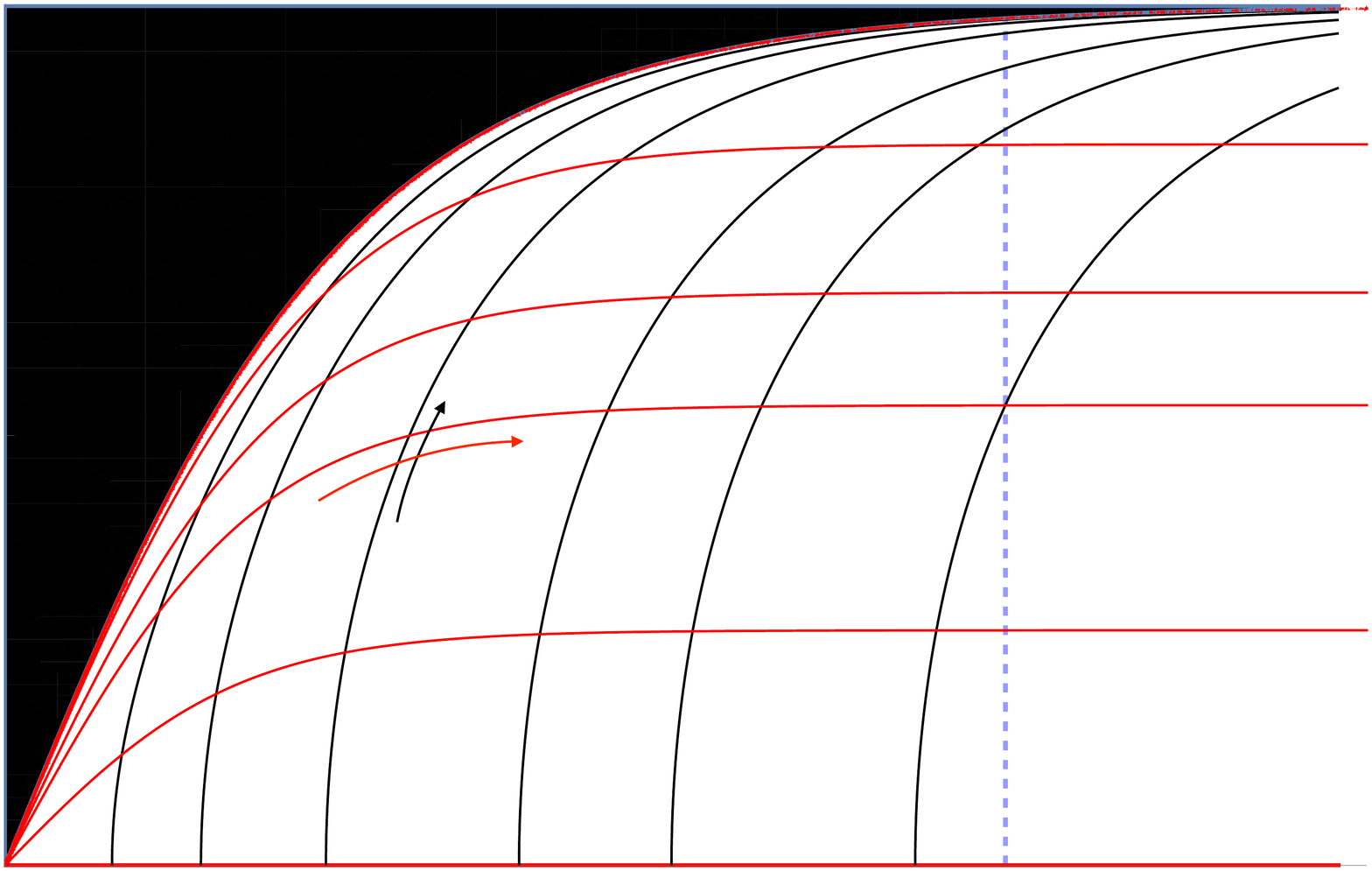}
\put(36,28){$\xi$}
\put(35,40){$\chi$}
\put(100,4){$\lambda$}
\put(1,67){$\psi$}
\put(-1,64){$\frac{\pi}{2}$}
\put(-1,34){$\frac{\pi}{4}$}
\put(69,0){$\lambda=\bar{\lambda}$}
\end{overpic}
\caption{Global vs.~de Sitter slicing of AdS. The Black region is the
  interior (not covered by the de Sitter slicing); Black curves are
  contours of constant $\xi$ (with $\xi = 0$ corresponding to the null surface
  which separates the interior from the exterior), red curves are
  contours of constant $\chi$ (with $\chi=0$ coinciding with the horizontal axis).
The dashed blue line is a fixed-$\lambda$ surface, which eventually
crosses all constant-$\xi$ curves as global time increases towards
$\pi/2$. }\label{fig:coordinates}
\end{figure}

The two metrics (\ref{cyl-1}) and (\ref{cyl-3})  are diffeomorphic to each other, and they are
connected by the coordinate transformation:
\be\label{cyl-5}
\left\{ \begin{array}{l} \sinh {\frac{\lambda}{\ell}} = \sinh {\frac{\xi}{\ell}}
    \cosh\chi \, , \\  \\
\tan \psi = \tanh {\frac{\xi}{\ell}} \sinh\chi \, , \end{array} \right.
\ee
and its inverse:
\be \label{cyl-6}
\left\{ \begin{array}{l} \sinh {\frac{\xi}{\ell}} = \cos
    \psi\left(\sinh^2 {\frac{\lambda}{\ell}} - \tan^2 \psi\right)^{1/2} \, , \\ \\
\displaystyle{\cosh\chi = \frac{\sinh{\frac{\lambda}{\ell}}}{\cos\psi \left(\sinh^2
    {\frac{\lambda}{\ell}} - \tan^2 \psi\right)^{1/2}}} \, . \end{array} \right.
\ee
The corresponding coordinate transformation in the Euclidean
signature is obtained  by using the identification (\ref{eq:theta}), which connects
the $R_{t_E}\times S^{d-1}$ and the  $S^d$ slicings of Euclidean AdS:
\be \label{cyl-7}
\left\{ \begin{array}{l} \sinh {\frac{\lambda}{\ell}} = \sinh {\frac{\xi}{\ell}}
    \sin \theta \, , \\ \\
\tanh t_E = \tanh {\frac{\xi}{\ell}} \cos \theta \, , \end{array}
\right.  \qquad \left\{ \begin{array}{l} \sinh {\frac{\xi}{\ell}} = \cosh
    t_E \left(\sinh^2 {\frac{\lambda}{\ell}} + \tanh^2 t_E\right)^{1/2} \, , \\ \\
\displaystyle{\sin\theta = \frac{\sinh{\frac{\lambda}{\ell}}}{\cosh t_E \left(\sinh^2
    {\frac{\lambda}{\ell}} + \tanh^2 t_E\right)^{1/2}}} \, . \end{array} \right.
\ee
Below we discuss several features of the two
slicings (we fix the coordinates on $S^{d-1}$, which are identified in the
two metrics) also summarized in figure \ref{fig:coordinates}.

\begin{enumerate}
\item The two radial coordinates $\xi$ and $\lambda$ (and in fact the full
  spatial geometries)  coincide at $\psi=\chi=0$. We can therefore
  identify the initial  spatial hypersurfaces at that point, and focus
  on the future  evolution, $\psi >0  $ and $\chi >0$.
\item The center of the de Sitter slicing $\xi=0$ coincides with the
  center of global AdS at $\psi=0$, and subsequently it describes the
  curve (for $\psi >0$):
\be \label{cyl-8}
\sinh \lambda/\ell = \tan \psi \, .
\ee
One can check that this is a null hypersurface, and it divides global AdS
in two regions, which we will call interior and exterior. The former
is the black region in figure \ref{fig:coordinates}, while the latter is the
clear region.
\item  The coordinate system $(\xi, \chi)$ in  (\ref{cyl-5})
covers only the exterior region $\sinh\lambda/\ell > \tan \psi$, and it
only extends in the future to times $\psi < \pi/2$ (see figure
\ref{fig:coordinates}). Notice that the boundary of AdS is completely
contained in the exterior region for all global times $\psi < \pi/2$.
The exterior region  ``shrinks'' as global time increases towards $\psi=\pi/2$.
The limiting  curve (\ref{cyl-8})  also corresponds to the infinite de Sitter future $\chi
 \to +\infty$ (except for the point $\lambda=\psi=0$). The ``end'' of
 global time, $\psi=\pi/2$,  is of course only a coordinate singularity, but to
 continue past this time one has to go to a different patch with  de
 Sitter slicing.
\item In the Euclidean case on the other hand, the coordinate
 system $(\xi, \theta)$ in (\ref{cyl-7}) covers the full Euclidean AdS. Here,  $\xi=0$
 is the single point $\lambda=0, t_E=0$ and there are no``hidden''
 regions. Notice that Euclidean time $t_E$ is not periodic, so this
 solution  describes a QFT at zero temperature.
\item A general $\xi = \bar{\xi}$ hypersurface is described by the curve:
\be\label{cyl-9}
\sinh {\frac{\lambda}{\ell}}  = \frac{1}{\cos \psi}\sqrt{\sin^2 \psi + {\sinh^2{\frac{\bar{\xi}}{\ell}}}} \, .
\ee
\item The  boundary of the de Sitter slicing $\xi\to +\infty$ is
  reached as $\lambda\to +\infty$ and $\psi$ fixed. In this limit, the
  coordinate transformation (\ref{cyl-5}-\ref{cyl-6}) simplifies to:
\be \label{cyl-10}
\left\{ \begin{array}{l} \lambda \simeq  \xi + \log \left(2\cosh\chi\right) \, ,
   \\  \\
\cos \psi \simeq \frac{1}{\cosh\chi} \, . \end{array}  \right.
\ee
\item The boundary of the cylinder slicing, $\lambda \to +\infty$, can
  instead be reached in two different ways: either sending $\xi \to
  +\infty$ with $\chi$ fixed, or by keeping $\xi = \bar{\xi}$ fixed and
  sending $\chi \to \infty$. The latter option corresponds to going
  towards the boundary by following one of the curves (\ref{cyl-9})
 as $\psi \to \pi/2$.
\item Any fixed-$\lambda$ hypersurface, no matter how far towards the
  boundary of global AdS,  will eventually cross the curve
  (\ref{cyl-8}) and cross over into the interior patch. In other words, the
  surface spanned by the  horizon $\xi=0$ will intersect
  any fixed large-$\lambda$ hypersurface  $\lambda=\bar{\lambda}$  at a time $\psi_{max}$
  approximately given by equation  (\ref{cyl-8}):
\be \label{cyl-11}
\psi_{\textrm{max}} \simeq \frac{\pi}{2} -  \frac{1}{2}e^{-\lambda/\ell} \, , \qquad \lambda/\ell \gg
1 \, .
\ee
\end{enumerate}
The last property is the crucial one for our purposes, when we go from
an exact AdS solution to the expanding bubble solution in the next
section.

\subsection{RG flow/vacuum bubble solution}

We shall now  discuss the case where the scalar field undergoes a
non-trivial radial flow. We start from the solution in dS slicing,
\be \label{cyl-12}
ds^2 = d\xi^2  + r^2(\xi) \left[-d\chi^2 +
  \cosh^2{\chi} d\Omega_{d-1}^2 \right], \qquad \f = \f(\xi),
\ee
where $r=e^A$ and  we chose the coordinate $\xi$ such that the boundary is at
$\xi\to +\infty$ and  the flow ends at $\xi=0$
in the interior.\footnote{In the notation of section \ref{sec:CDLanalytic},  we define $\xi = u_0-u$.} As we have seen in section \ref{sec:CDLanalytic}, the scalar field behaves asymptotically as
\be \label{cyl-13}
\f(\xi \to +\infty) \simeq \f_f + \f_+ \ell_f^\Delta e^{-\Delta \xi/\ell_f} +\ldots \, ,  \quad \textrm{and} \quad
\f(0) = \f_0 \, ,
\ee
where $\f_f$ is the UV minimum of the potential (the false vacuum) and
$\f_0$ is the value reached by the scalar field at the center of the
bubble.

In the UV, the metric has the same asymptotic behavior as in equation
(\ref{cyl-4}):
\be \label{cyl-14}
r(\xi \to +\infty)\simeq \ell_f e^{\xi/\ell_f} \, ,
\ee
Close to the
endpoint $\xi=0$, one can show\footnote{This can be done by
  integrating the first order equations (\ref{eq:defWc})  near the flow
  endpoint,  with the
  expansion  of the scalar functions $W$, $S$  and $T$  which can be found
  in Appendix F of \cite{Ghosh:2017big}.} that the scale factor behaves as:
\be \label{cyl-15}
r(\xi) =  \ell_0 \left[ \frac{\xi}{\ell_0 }+ \frac{1}{6} \left(\frac{\xi}{\ell_0}\right)^3 + O(\xi^5) \right] ,  \qquad \ell_0^2 = - \frac{d(d-1)}{V(\f_0)} \, .
\ee
Notice that equation (\ref{cyl-15}) is the same as the expansion to cubic order
of the scale factor $\sinh \xi/\ell$ appearing in the exact AdS solution (\ref{cyl-3}) ,
except that $\ell$ is replaced by $\ell_0$. This can be traced to the
fact that close to the IR endpoint, the bulk Ricci tensor is
approximately  \cite{Ghosh:2017big}:
\be\label{cyl-15ii}
R_{AB} \simeq \frac{V(\f_0)}{d-1} g_{AB}.
\ee

Since the UV and IR asymptotics are AdS-like (albeit with two
different scales), there must exist a coordinate transformation of the
form
\be \label{cyl-16}
\xi = F(\lambda, \psi) \, , \quad  \chi = G(\lambda, \psi) \, ,
\ee
which, in both asymptotic regions, implements the change from de
Sitter to cylinder slicing, i.e.~it approaches the transformation
(\ref{cyl-6}) both near the boundary $\xi \to \infty$  and close to the
endpoint $\xi\to 0$. Even without having an explicit form for this
transformation, from the previous considerations we  can conclude that
it will have the  following properties:
\begin{enumerate}
\item
The Lorentzian solution in the $\lambda, \psi$ coordinates will have the
general form\footnote{We can choose the functions $F$ and $G$ in
  (\ref{cyl-16}) such that the
  off-diagonal components vanish and set  the $\lambda\lambda$-component to one.}
\be \label{cyl-17}
ds^2 = d\lambda^2 - f^2(\lambda, \psi) d\psi^2 + g^2(\lambda, \psi)
d\Omega_{d-1}^2,  \qquad \f = \f(\lambda, \psi),
\ee
such that:
\be \label{cyl-18}
f(\lambda,\psi)\simeq g(\lambda,\psi) \simeq \ell_f e^{\lambda/\ell_f}, \quad  \lambda \to
\infty, \psi \; \text{fixed}. 
\ee
We took the above limit with $\psi$ fixed, because this corresponds,
in the pure AdS case, to going to the boundary of the spherical
slicing by sending $\xi \to +\infty$ as in (\ref{cyl-14}). It is
in this limit, that the coordinate transformation (\ref{cyl-16}) has to
approach (\ref{cyl-6})

The  metric  (\ref{cyl-17})
describes a complicated time-dependent solution of the equations of
motion, evolving   out of  an initial state which corresponds to the
nucleation of the bubble at $\psi=0$. The time-evolution is with
respect to a time-coordinate which matches onto the global
time-coordinate $\psi$ when we approach the boundary.

\item Close to the center of the bubble, $\xi=0$, the coordinate
  transformation can be made  to match the one we found in pure AdS,
  i.e.~the $\xi\to 0$ limit  of (\ref{cyl-6}) with   $\ell$ replaced
  by $\ell_0$. With this choice, the $\xi=0$ hypersurface which
  separates the exterior from the interior geometry will again
  be described in $(\lambda,\psi)$ coordinates by the null curve $\sinh
  \lambda/\ell_0 = \tan \psi$, which will again approach the boundary as
  $\psi \to \pi/2$.

\item Since the coordinate transformation (\ref{cyl-16})  reduces to
  (\ref{cyl-10}) as $\xi \to +\infty$, global time $\psi$ will again
  come to an end at $\psi=\pi/2$. However this time, unlike in the
  previous subsection,  the metric will have a
  true singularity in the interior patch, which will also approach the
  curve $\xi=0$ from the interior. This means that this time,
  $\psi=\pi/2$ is really the end of global time, and the geometry
  cannot be continued beyond this point. Does this mean that the dual
  QFT vacuum spontaneously decays in a finite time? To answer this
  question  we have to understand the fate of the bulk scalar.

\item
The near-boundary scalar field asymptotics can also be obtained
easily. From the original dS slicing (\ref{cyl-12}),  the scalar field is
constant along curves of constant $\xi$. Its near-boundary
limit  as $\xi\to +\infty$ is given in equation (\ref{cyl-13}) and
we can use the asymptotic
form of the coordinate transformation for large $\xi$, equation
(\ref{cyl-10}), to find:
\be \label{cyl-19}
\f(\lambda,\psi) \simeq \f_f  +\frac{\f_+}{(\cos \psi)^{\Delta} }
e^{-\Delta \lambda/ \ell_f} +\ldots \, , \qquad \lambda \to +\infty.
\ee
From the point of view of the dual QFT which  lives on $R_\psi
\times S^{d-1}$, these asymptotics represent a theory with the
relevant coupling set to zero , but with a time-dependent  vev of the
dual operator,
\be \label{cyl-19ii}
\langle \mathcal{O} \rangle \propto \frac{\f_+}{(\cos{\psi})^{\Delta}} \, .
\ee
This agrees with what  was already noticed in \cite{Barbon:2010gn}.

\end{enumerate}

From equations (\ref{cyl-19}-\ref{cyl-19ii}) it may seem  that the $\langle \mathcal{O} \rangle$ becomes infinite at
$\psi=\pi/2$. This would seal the fate of the  QFT. However,  the
crucial point is that the approximation (\ref{cyl-19}) becomes invalid
{\em before} we reach the final time. The reason is that for $\psi$ arbitrarily
close to $\pi/2$, the scalar field (\ref{cyl-19}) is not  a small
perturbation around the fixed point value $\f_f$, and the metric is not close to
the UV AdS metric.  The usual expansion in leading (source)
term and subleading (vev) term breaks down.

To make the situation clearer,  it is convenient to introduce a
cut-off at a finite radius. The cut-off surface should have the same geometry
as the one seen by the dual QFT. Therefore,  if we want information about the
fate of the theory on
the cylinder, we have to introduce a cut-off at fixed $\lambda =
\bar{\lambda}$. For convenience, we also define an energy cut-off:
\be \label{cyl-20}
\Lambda  = \frac{1}{\ell_f} e^{\bar{\lambda}/\ell_f} \, .
\ee
The cut-off geometry is $R\times S^{d-1}$, where $R$ is parametrized
by global time $\psi$. On this hypersurface, the scalar field takes
values:
\be \label{cyl-21}
\f_{\textrm{cut-off}}(\psi) = \f(\bar{\lambda}, \psi).
\ee
As $\psi$ increases,  the cut-off surface at fixed $\lambda$   will
approach the null trajectory spanned by the center of the bubble,
which is approximately  described by equation (\ref{cyl-9}) (with $\ell_0$
replacing $\ell$).
Along the way, the scalar field value will smoothly interpolate
between $\f_f$ and $\f_0$, and  only for small  $\psi$ (i.e.~as long
as $\f$ stays close to $\f_f$) will expression (\ref{cyl-19}) hold. In  a
time approximately given by
$\pi/2 - {1 /(\Lambda\ell_0)}$, the scalar
field on the cut-off surface will have moved close  to the value at
the center of the bubble, $\f_0$.

\begin{figure}[t]
\centering
\begin{overpic}
[width=0.8\textwidth]{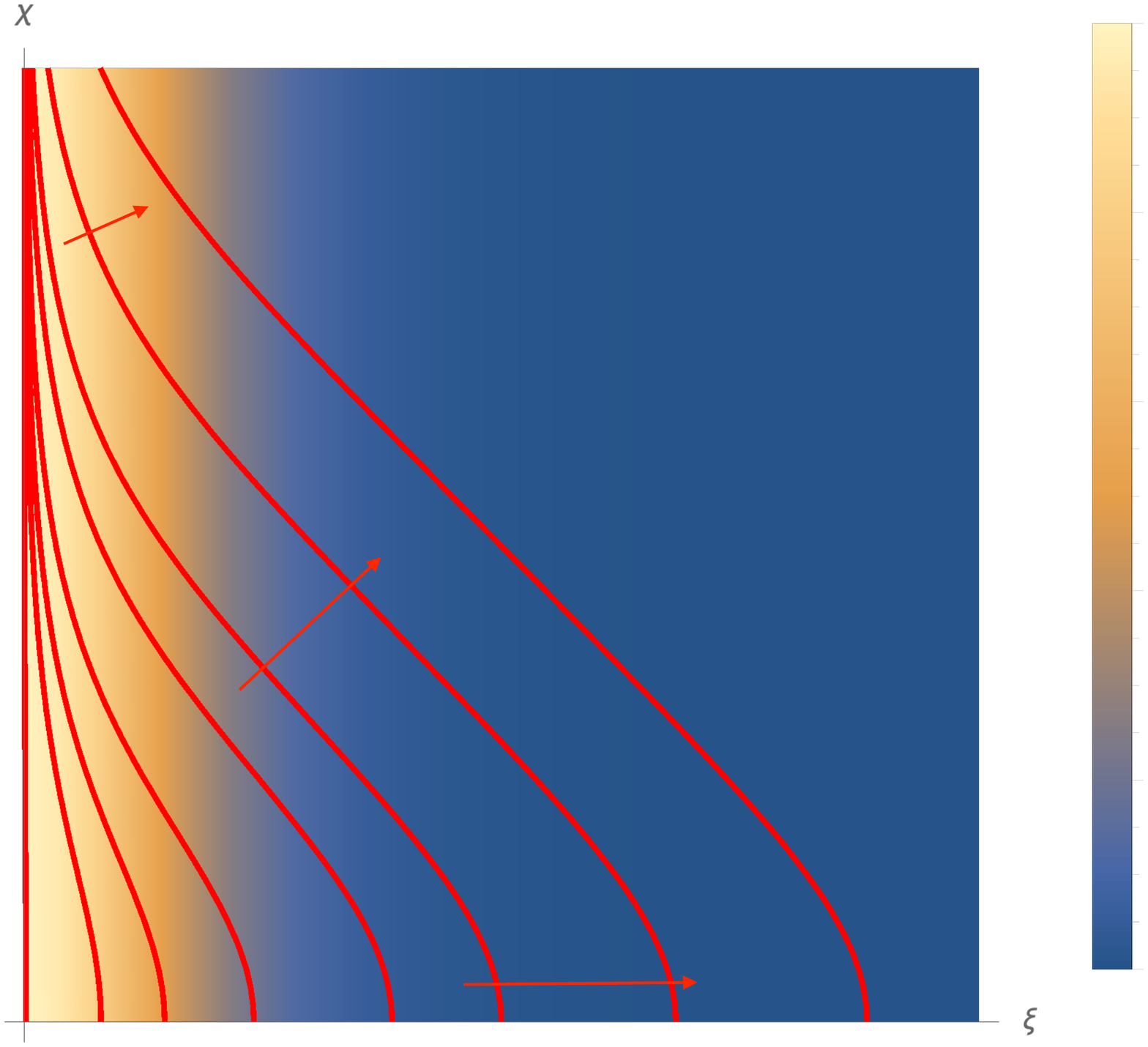}
\put(83,5){$\f_f$}
\put(83,68){$\f_0$}
\end{overpic}
\caption{This figure is a representation (not obtained from the actual
  solution) of the embedding of the constant-$\lambda$ hypersurfaces (thick red
  lines) in the de-Sitter sliced RG flow geometry. The red arrows
  indicate the direction of increasing $\lambda$. The color-shading
  indicates the value of the bulk scalar field,  which interpolates
  between $\f_f$ ($\xi \to +\infty$) and $\f_0$ ($\xi = 0$).}\label{fig:cutoff}
\end{figure}

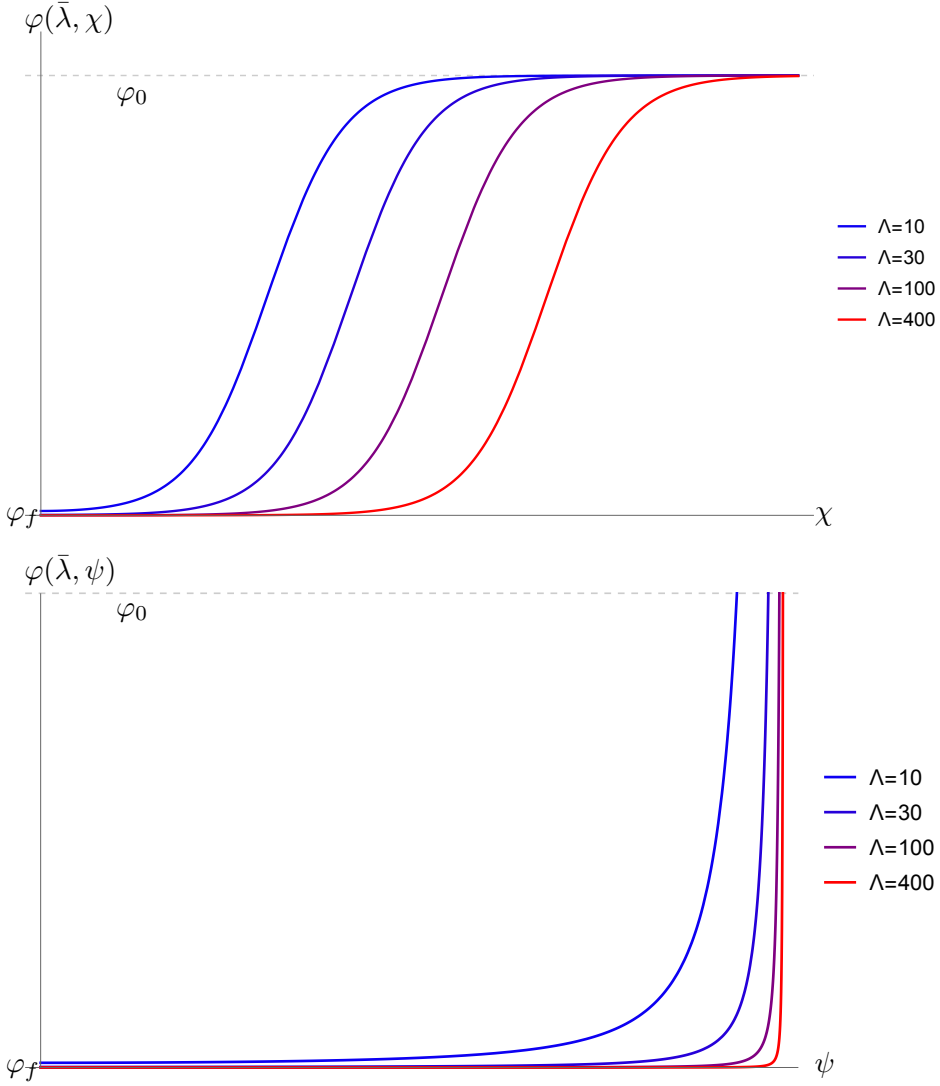
\begin{figure}[h]
\centering
\begin{overpic}
[width=0.8\textwidth]{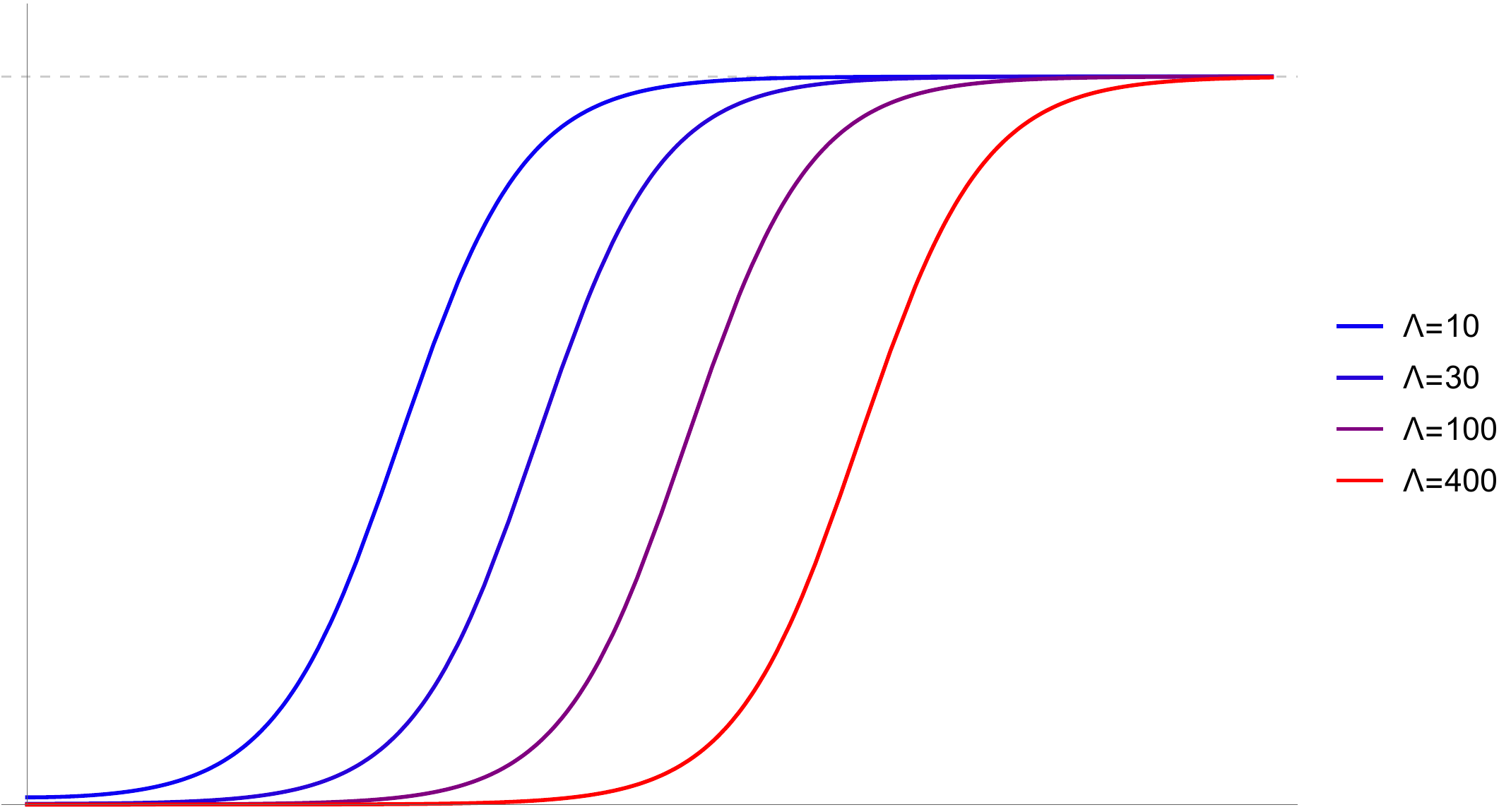}
\put(-2,0){$\f_f$}
\put(10,46){$\f_0$}
\put(0,54){$\f(\bar{\lambda},\chi)$}
\put(86,0){$\chi$}
\end{overpic}
\vskip 1cm

\begin{overpic}
[width=0.8\textwidth]{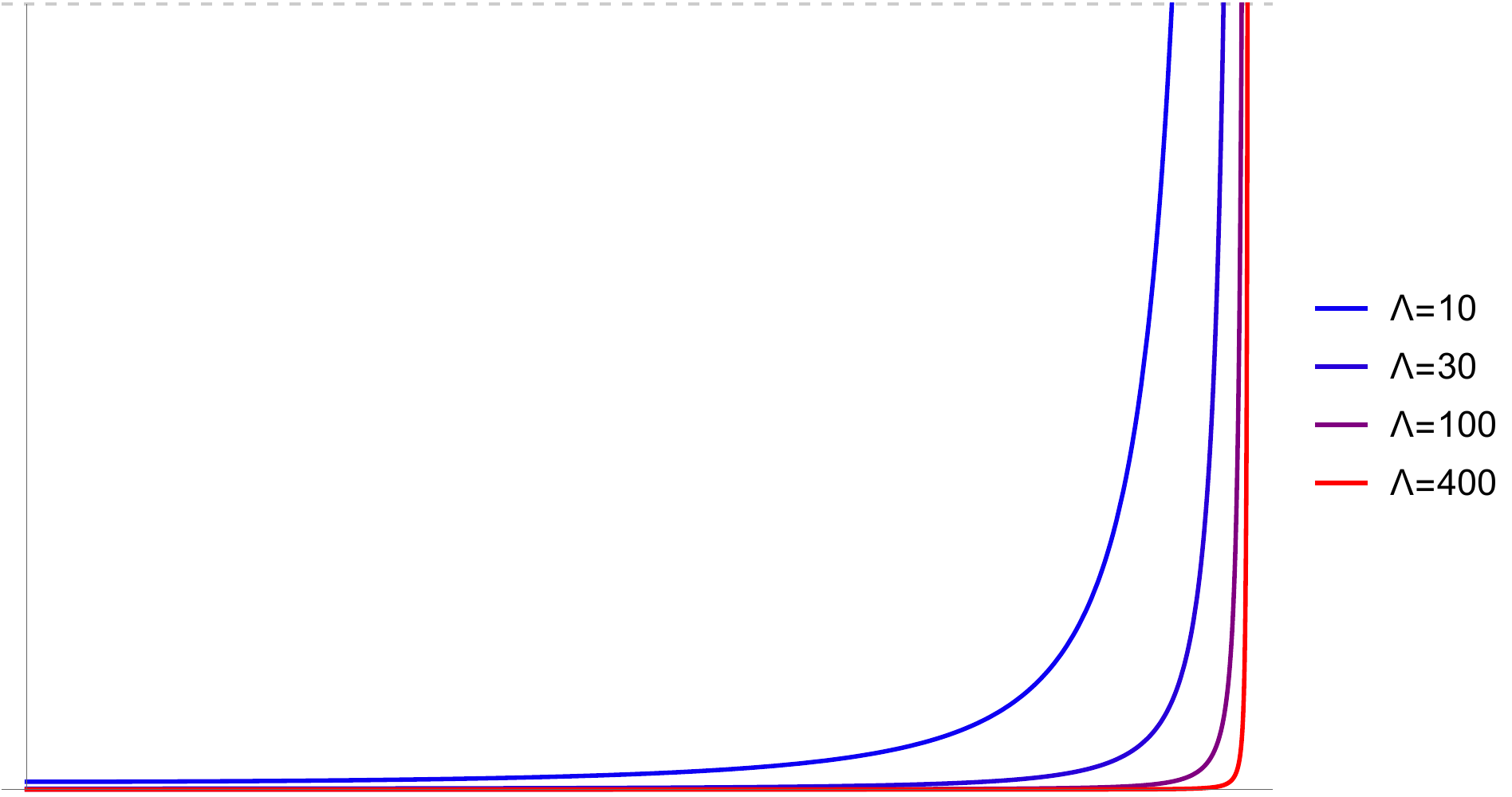}
\put(-2,0){$\f_f$}
\put(10,50){$\f_0$}
\put(0,54){$\f(\bar{\lambda},\psi)$}
\put(86,0){$\psi$}
\end{overpic}
\caption{The different curves illustrate the time-dependent behavior of the
  scalar field $\f$ evaluated on different constant-$\lambda$ cutoff
  hypersurfaces, as a function  of de Sitter time $\chi$ (top
  figure) or the global time $\psi$ (bottom figure). The cutoff in the figure is defined as
  $\ell_f \Lambda\equiv\sinh\bar{\lambda}/\ell_f$. At early times the scalar field
  displays the near-boundary behavior $\f \simeq \f_f +
  \epsilon^{\Delta}(\cosh\chi)^{\Delta}$, where $\epsilon \equiv
\Lambda^{-1}$. }
\label{fig:scalarprofile}
\end{figure}

The situation can be described in an equivalent way, if we trace the
cut-off surface $\lambda=\bar{\lambda}$ in the de Sitter-sliced
geometry, as we show in figures\footnote{These figures are qualitative and not obtained from the full solution since they would require knowing the full coordinate transformation (\ref{cyl-16}). In practice they are obtained by neglecting the backreaction of the scalar field on the metric and using the transformation (\ref{cyl-5}).} \ref{fig:cutoff} and
\ref{fig:scalarprofile}.  As one can see in figure \ref{fig:cutoff}, the cylinder cut-off surface has a trajectory which approaches the center of the bubble. As time moves forwards, on any cut-off surface the value of the scalar field will inevitably depart from the asymptotic value $\f_f$ and will approach the value at the center of the bubble, $\f_0$, as one can  see in figure \ref{fig:scalarprofile}.

In the limit $\Lambda\to \infty$ we can think of the
situation  as if the boundary value of the scalar field  receives an
instantaneous ``kick'' at the global time
$\psi = \pi/2$,  which displaces  the filed  by a finite amount   from its  fixed point value
$\f_f$. This is particularly clear in the bottom figure
\ref{fig:scalarprofile}.
We can think of this as a Dirac $\delta$-like source
concentrated at $\psi=\pi/2$. As such, the cylinder version of the CdL
solution does not describe a solution with the same boundary
conditions obeyed by the global AdS solution: the latter
satisfies source-less AdS boundary conditions on the cylinder for {\em
  all times} $-\infty<\psi < +\infty$.

From the  point of view of the dual field theory on the cylinder, there are two possible ways to
interpret the CdL solution, depending on two possible definitions of
the gravity path integral:
\begin{enumerate}
\item We can define the dual field theory by   imposing on the gravity
  side the usual boundary conditions which accommodate the
global AdS solution, i.e. that the asymptotic boundary is the full
cylinder $R\times S^{d-1}$ with zero source term for the scalar
field. In this case, the CdL solution does not correspond to a state
of the theory, and it does not indicate a pathology of the CFT on the
cylinder.
\item We can instead define the boundary conditions on the gravity
  side  in order to accommodate the CdL geometry as an allowed
  solution. Provided one can do this
consistently\footnote{It is not entirely clear to us how to
  consistently define the boundary condition which allow the CdL
  state in terms of the usual leading/subleading classification of the
  boundary behaviour of the scalar field, because on any cut-off surface the asymptotic
  expansion in terms  breaks down after a finite
  time. There may be a distributional limit for the source term which
  exactly matches the behavior of the CdL solution  as $\Lambda \to
  \infty$, but this is not evident.}, this defines a {\em different} field theory on the
  cylinder, which does not admit the global AdS solution as one its
  states. This theory can only be extended up to a finite global time,
  at which point the time-evolution reaches a singularity as a
  consequence of an infinite instantaneous ``kick'' at $\psi = \pi/2$. This point
  of view is closer to the one taken in
  \cite{Barbon:2010gn}.
\end{enumerate}

Notice that,  although at the initial time (say $\psi=0$, if we
prepare the states using analytic continuation from the Euclidean
solution) the CdL and global AdS states look exactly the same from the
point of view of the boundary sources, this does not
contradict the fact that they may have a different future evolution,
because in AdS the initial condition alone does not uniquely specify
the  state at subsequent times: in order to do that, one has to specify boundary
conditions for all future times. In this language, the two theories
above have different time evolution  because in theory 2 there is an
instantaneous perturbation which occurs at a finite time. Therefore,
the picture above is consistent and describes the evolution of a
stable CFT on the cylinder (case 1) versus an unstable one (case
2). In the second case,  the work of \cite{Barbon:2010gn} constituted
a proposal for identifying this instability as originating from an
unbounded CFT Hamiltonian, and curing it by a stabilizing extra term.   

We have argued that the $O(d+1)$-instanton solution we discussed in this
paper mediates vacuum decay of the field theory on dS$^d$.
This  cannot be mapped  to a solution describing a similar {\em
  spontaneous} vacuum decay process of the same theory on the full  $R\times
S^{d-1}$. One may wonder if such a process can be mediated by
different solutions with $O(d)$-symmetry. If such solutions exist,
they are certainly {\em not} found by looking at holographic RG flows on   $R\times
S^{d-1}$ in which the bulk field evolves only in the radial direction,
of the form:
\be \label{cyl-22}
ds^2 = d\lambda^2 - f^2(\lambda)d\psi^2 + g^2(\lambda) d\Omega_{d-1}^2, \qquad \f =
\f(\lambda).
\ee
Solutions of this form  are expected to exist, which have an IR endpoint
 $\lambda_0$ such that either $f(\lambda_0)=0$, with $g(\lambda_0)$ finite, or
$g(\lambda_0)=0$ with $f(\lambda_0)$  finite. In the first case,  (\ref{cyl-22})
 describes  a black hole.  However, if
we insist that the theory starts in its zero-temperature ground state,
then we must take Euclidean time to be non-compact, and regularity
requires $f(\lambda)$ to be nowhere vanishing. The second case corresponds
to non-trivial RG flows in $R\times S^{d-1}$ in which the $S^{d-1}$
shrinks to zero size at the IR endpoint.\footnote{Similar solutions
  where discussed for general $S^{d_1}\times S^{d_2}$ foliations in
  pure gravity \cite{Aharony:2019vgs}, and recently for $S^2\times S^2$ foliations in
  Einstein-dilaton theory \cite{Kiritsis:2020bds}.} Although these
solutions are expected to exist, they cannot mediate vacuum decay:
since they are time-independent and the time-direction is infinite,
the Euclidean solution cannot have a finite action.

By the above argument,  any finite-action instanton preserving $O(d)$
symmetry must be time-dependent:  it has to reduce to the vacuum for the
Euclidean time $\tau\to -\infty$, for all values of the radial
coordinate. These instantons  must therefore have the more general
form (\ref{cyl-17}). We emphasize however, that they are {\em not} obtained as coordinate transformations of $O(d+1)$-symmetric instantons.

\section*{Acknowledgements}
\addcontentsline{toc}{section}{Acknowledgements}

We would like to thank Jos\'e Barbon, Francesco Bigazzi, Aldo Cotrone,
Roberto Emparan, Daniel Harlow, Javier Mas, Mukund Rangamani, Eliezer
Rabinovici, Nick Tetradis for fruitful discussions. 
We also thank Jos\'e Barbon, Daniel Harlow and Eliezer
Rabinovici for a critical reading of the manuscript. 

This work was supported in part by the European Union via the ERC Advanced Grant SM-GRAV, No 669288. LW also acknowledges support from the European Union via the ERC Starting Grant GEODESI, No 758792. JKG acknowledges the postdoctoral program at ICTS for funding support through the Department of Atomic Energy, Government of India, under project no.~RTI4001 and partial support through the Simons Foundation Targeted Grant to ICTS-TIFR ``Science without Boundaries''.

\newpage
\appendix

\begin{appendix}
\renewcommand{\theequation}{\thesection.\arabic{equation}}
\addcontentsline{toc}{section}{Appendices}
\section*{APPENDIX}

\section{Expansions of $A(u)$ and $\f(u)$ near UV fixed points} \label{App:Pert}
Here we record the asymptotic form for the scale factor $\f(u)$ and $A(u)$ near a UV fixed point that is reached for $u \rightarrow -\infty$. As pointed out in section \ref{sec:CDLasHoloRG}, the near-UV expansions encode the QFT data of the UV fixed point in a precise way. There we focused on the near-UV expansions for the function $W(\f)$ finding two branches of solutions denoted by $(-)$ and $(+)$ respectively. Here, for completeness, we record the corresponding expressions for $\f(u)$ and $A(u)$.

Given the near-UV expressions for $W(\f)$ and $S(\f)$ one can derive the corresponding expressions for $\f(u)$ and $A(u)$ in virtue of \eqref{eq:defWc}. The relevant expansions are given in \cite{Kiritsis:2016kog,Ghosh:2017big} and we refer the reader to these works for details. Without loss of generality we consider a UV fixed point at an extremum of $V$ at $\f=0$ with the potential in the vicinity given by
\begin{align}
&V(\f) = -\frac{d(d-1)}{\ell_f^2} - \frac{\Delta(d-\Delta)}{2 \, \ell_f^2} \f^2 + \mathcal{O}(\f^3) \, \\
\label{eq:Deltaapp} &\textrm{with} \quad \Delta = \frac{d}{2} \left(1 + \sqrt{1 +\frac{4 (d-1)}{d} \frac{V''(0)}{|V(0)|}} \right) \, .
\end{align}
For the $(-)$-branch of solutions one then finds:
\begin{align}
\label{eq:phimsol} \f(u) &= \f_- \ell_f^{(d-\Delta)}e^{(d-\Delta)u / \ell_f} \left[ 1+ \mathcal{O} \left(R^{(\zeta)} \ell_f^2 e^{2u/\ell} \right) + \ldots \right] \\
\nonumber & \hphantom{=} \ + \frac{C d \, |\f_-|^{\Delta / (d-\Delta)}}{(d-\Delta)(2 \Delta -d)} \, \ell_f^{\Delta} e^{\Delta u \ell_f} \left[ 1+ \mathcal{O} \left(R^{(\zeta)} \ell_f^2 e^{2u/\ell_f} \right) + \ldots \right] + \ldots \, , \\
\label{eq:Amsol} A(u) &= -\frac{u}{\ell_f} - \frac{\f_-^2 \, \ell_f^{2 (d-\Delta)}}{8(d-1)} e^{2 (d-\Delta) u / \ell_f}  -\frac{R^{(\zeta)} \ell_f^2}{4d(d-1)} e^{2u/\ell_f} \\
\nonumber & \hphantom{=} \ - \frac{\Delta C |\f_-|^{d/(d-\Delta)} \, \ell_f^d}{d(d-1)(2\Delta -d)}e^{du/\ell_f} +\ldots \, .
\end{align}
where $\f_-$ is an integration constants and the ellipses denote subleading terms for $u \rightarrow -\infty$
On the $(+)$-branch we instead obtain:
\begin{align}
\label{eq:phipsol} \f(u) &= \f_+ \ell_f^{\Delta}e^{\Delta u / \ell_f} \left[ 1+ \mathcal{O} \left(R^{(\zeta)} \ell_f^2 e^{2u/\ell_f} \right) + \ldots \right] + \ldots \, , \\
\label{eq:Apsol} A(u) &= -\frac{u}{\ell_f} - \frac{\f_+^2 \, \ell_f^{2 \Delta}}{8(d-1)} e^{2\Delta u / \ell_f}  -\frac{R^{(\zeta)} \ell_f^2}{4d(d-1)} e^{2u/\ell_f} +\ldots \, ,
\end{align}
where  $\f_+$ is an integration constants.

The QFT data can be read from this as follows. First note the appearance of the boundary curvature $R^{(\zeta)}$ which is the curvature of the manifold on which the QFT is defined. On the $(-)$-branch of solutions we further identify $\f_-$ with the source for the operator $\mathcal{O}$ deforming the UV CFT. The parameter $C$ has the interpretation of the vev of $\mathcal{O}$ in units of the source. In particular:
\begin{align}
\langle \mathcal{O} \rangle_- = (M \ell_f)^{(d-1)} \, \frac{Cd}{d-\Delta} \, |\f_-|^{\Delta / (d-\Delta)} \, ,
\end{align}
where the subscript indicates that this only holds for $(-)$-branch solutions. On the $(+)$-branch the source vanishes and the corresponding flows are purely driven by the vev of $\mathcal{O}$. This is related to the parameter $\f_+$ as
\begin{align}
\langle \mathcal{O} \rangle_+ = (M \ell_f)^{(d-1)} \, (2\Delta-d) \, \f_+ \, .
\end{align}

\noindent \textbf{The instanton action and the cancellation of divergences:} The instanton action as given in \eqref{eq:acdiff} is defined as the difference between the on-shell action for the interpolating (tunnelling) solution $S_{E, \textrm{inter}}$ and the on-shell action evaluated on the false vacuum solution $S_{E, \textrm{false}}$. If the false vacuum at $\f_f$ is an AdS extremum, both individual expressions $S_{E, \textrm{inter}}$ and $S_{E, \textrm{false}}$ are formally divergent, with the divergence arising from the infinite volume of the AdS geometry associated with $\f_f$. For the interpolating solution to describe tunnelling with a non-zero decay rate the divergences have to cancel between $S_{E, \textrm{inter}}$ and $S_{E, \textrm{false}}$ so that the instanton action $B$ gives a finite value. Here we  show that this is the case if the interpolating solution is a $(+)$-branch solution but not if it is a $(-)$-branch solution. To do so, we compute the divergent contributions to $S_{E, \textrm{inter}}$ and $S_{E, \textrm{false}}$ explicitly. In the latter case this can be done as the full expression for $A(u)$ is known, while in the former case the near-boundary expressions recorded above are sufficient to obtain the relevant results.

To be specific, in the following we restrict attention to $D=d+1=3+1$. To regulate the divergences we introduce the quantity
\begin{align}
\label{eq:Lambdadef} \Lambda \equiv \frac{e^{A(u_{\textsc{uv}})}}{\ell_f} \, ,
\end{align}
which diverges as $\Lambda \rightarrow \infty$ for $u_{\textsc{uv}}
\rightarrow -\infty$. This is convenient for making contact with
previous literature on holographic RG flows (see e.g. \cite{F,Ghosh:2020qsx}), where $\Lambda$ has the interpretation as a UV cutoff for the dual field theory.

We begin with $S_{E, \textrm{false}}$. To this end we insert $A(u)$ as given in \eqref{eq:AofuCFTfalse} into \eqref{eq:Son}. Eliminating $u_{\textsc{uv}}$ in favour of $\Lambda$ using \eqref{eq:Lambdadef}, after some algebra one finds:
\begin{align}
\label{eq:SEfalsediv} \frac{S_{E, \textrm{false}}}{(M \ell_f)^2} & = 8 \pi^2 \bigg\{ 1- {\Big( \frac{6 \Lambda^2}{R^{(\zeta)}} +1\Big)}^{3/2} \bigg\} \\
\nonumber & = - 12 \sqrt{6} \pi^2 \bigg\{ \frac{4 \Lambda^3}{(R^{(\zeta)})^{3/2}} + \frac{\Lambda}{(R^{(\zeta)})^{1/2}} - \frac{2}{3 \sqrt{6}} \bigg\} + (\textrm{vanishing for } \Lambda \rightarrow \infty) \, ,
\end{align}
where on the second line we isolated the two manifestly divergent terms for $\Lambda \rightarrow \infty$. We now turn to $S_{E, \textrm{inter}}^{(-)}$ for a solution on the $(-)$-branch, focusing on the divergent terms for $u_{\textsc{uv}} \rightarrow -\infty$. To this end we insert the near-boundary expansion \eqref{eq:Amsol} into \eqref{eq:Son}. Again, eliminating $u_{\textsc{uv}}$ in favour of $\Lambda$ using \eqref{eq:Lambdadef} we find the following:
\begin{align}
\label{eq:SEmdiv} \frac{S_{E, \textrm{inter}}^{(-)}}{(M \ell_f)^2} \underset{\Lambda \rightarrow \infty}{=} & - 12 \sqrt{6} \pi^2 \bigg\{ \hphantom{+} \frac{4 \Lambda^3}{(R^{(\zeta)})^{3/2}} \Big(1 + \mathcal{O} \big( |\f_-|^2 \Lambda^{-2(3-\Delta)} \big) \Big)  \\
\nonumber & \hphantom{- 12 \sqrt{6} \pi^2 \bigg\{} + \frac{\Lambda}{(R^{(\zeta)})^{1/2}} \Big( 1+ \mathcal{O} \big( |\f_-|^2 \Lambda^{-2(3-\Delta)} \big) \Big) \bigg\} \\
\nonumber & + (\textrm{manifestly finite for } \Lambda \rightarrow \infty) \, .
\end{align}
We can repeat the analysis for $S_{E, \textrm{inter}}^{(+)}$, i.e.~the action evaluated on a solution on the $(+)$-branch. Inserting \eqref{eq:Amsol} into \eqref{eq:Son} and re-expressing in terms of $\Lambda$ one now finds:
\begin{align}
\label{eq:SEpdiv} \frac{S_{E, \textrm{inter}}^{(+)}}{(M \ell_f)^2} \underset{\Lambda \rightarrow \infty}{=} & - 12 \sqrt{6} \pi^2 \bigg\{ \hphantom{+} \frac{4 \Lambda^3}{(R^{(\zeta)})^{3/2}} \Big(1 + \mathcal{O} \big( |\f_+|^2 \Lambda^{-2\Delta} \big) \Big)  \\
\nonumber & \hphantom{- 12 \sqrt{6} \pi^2 \bigg\{} + \frac{\Lambda}{(R^{(\zeta)})^{1/2}} \Big( 1+ \mathcal{O} \big( |\f_+|^2 \Lambda^{-2\Delta} \big) \Big) \bigg\} \\
\nonumber & + (\textrm{manifestly finite for } \Lambda \rightarrow \infty) \, .
\end{align}
We are now in a position to check to what extent the divergent terms cancel between $S_{E, \textrm{inter}}^{(\pm)}$ and $S_{E, \textrm{false}}$. We begin with $S_{E, \textrm{inter}}^{(-)}$, i.e.~the action evaluated on a solution on the $(-)$-branch. These solutions only exist if the UV fixed point reached for $u \rightarrow -\infty$ is a maximum of $V$. From \eqref{eq:Deltaapp} this in turn implies that these solutions only exist for $d/2 < \Delta < d$, i.e.~$3/2 < \Delta < 3$ for  $d=3$ as we consider here. It is then easy to see that $S_{E, \textrm{inter}}^{(-)}$ contains divergences which do not appear in $S_{E, \textrm{false}}$ and hence do not cancel between them. In particular, the subleading term on the first line of expression \eqref{eq:SEmdiv} scales as
\begin{align}
\sim (R^{(\zeta)})^{-3/2} \, |\f_-|^2 \, \Lambda^{3 - 2 \Delta} \, ,
\end{align}
which is always divergent for $3/2 < \Delta < 3$. No such divergence exists in $S_{E, \textrm{false}}$ and subtracting $S_{E, \textrm{false}}$ from $S_{E, \textrm{inter}}^{(-)}$ does hence not lead to a finite expression for the instanton action $B$. As a result, $(-)$-branch solutions cannot be understood as describing tunnelling.

In contrast to $(-)$-branch solutions, $(+)$-branch solutions can depart from both maxima and minima of $V$. Therefore, from \eqref{eq:Deltaapp}, it follows that $(+)$-branch solutions can exist for any value $\Delta > d/2$, i.e.~$\Delta > 3/2$ for $d=3$. By inspecting the expressions \eqref{eq:SEpdiv} and \eqref{eq:SEfalsediv} it is then easy to see that $S_{E, \textrm{inter}}^{(+)}$ and $S_{E, \textrm{false}}$ share the same divergent terms. As a result, the instanton action $B$ obtained by subtracting $S_{E, \textrm{false}}$ from $S_{E, \textrm{inter}}^{(+)}$  will be finite, as expected if the $(+)$-branch solution describes tunnelling.

\section{Sufficient condition for tunnelling solutions from AdS minima}
In section \ref{sec:CDLnongeneric} we identified a sufficient condition for a potential to admit $O(D)$-instantons describing tunnelling from a false AdS vacuum. In the mechanical picture introduced there, this condition is that a test particle in the inverted potential $-V$, when released from rest at $\f_t$, will overshoot $\f_f$ when released in that direction. An alternative formulation of that condition is that the potential admits a flat domain-wall solution/ flat-sliced holographic RG flow that has its IR end point at $\f_t$.

Here we record a quantitative version of this condition. This will be phrased in terms of solutions for the function $W(\f)$ introduced in \eqref{eq:defWc}. One observation is that (for $R^{(\zeta)} \geq 0$) the equation of motion \eqref{eq:EOM5} implies that $W(\f) \geq B(\f)$ where $B(\f)$ is defined as $B(\f)=\sqrt{-\tfrac{4(d-1)}{d} V(\f)}$. Furthermore, $W(\f)=B(\f)$ only occurs at critical points of the flow, i.e.~at fixed points or where the flow in $\f$ reverses direction.

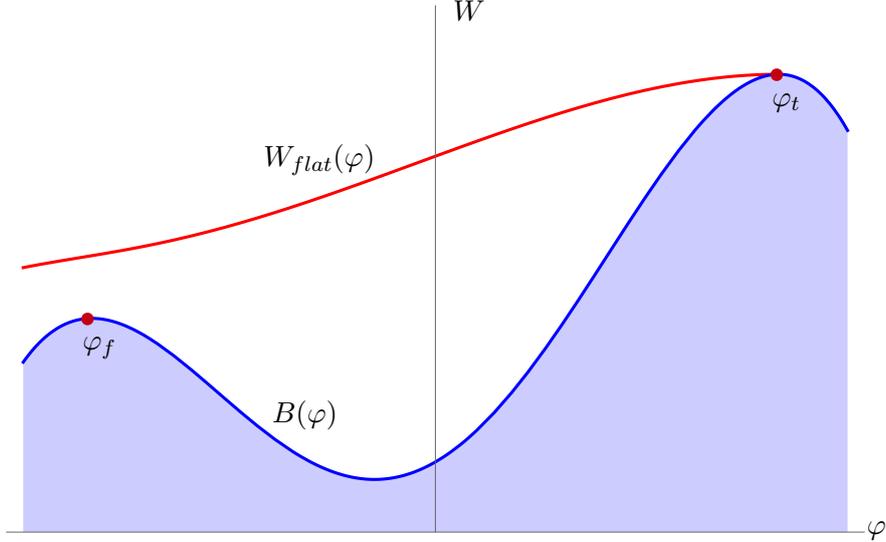
\begin{figure}[t]
\centering
\begin{overpic}
[width=0.75\textwidth]{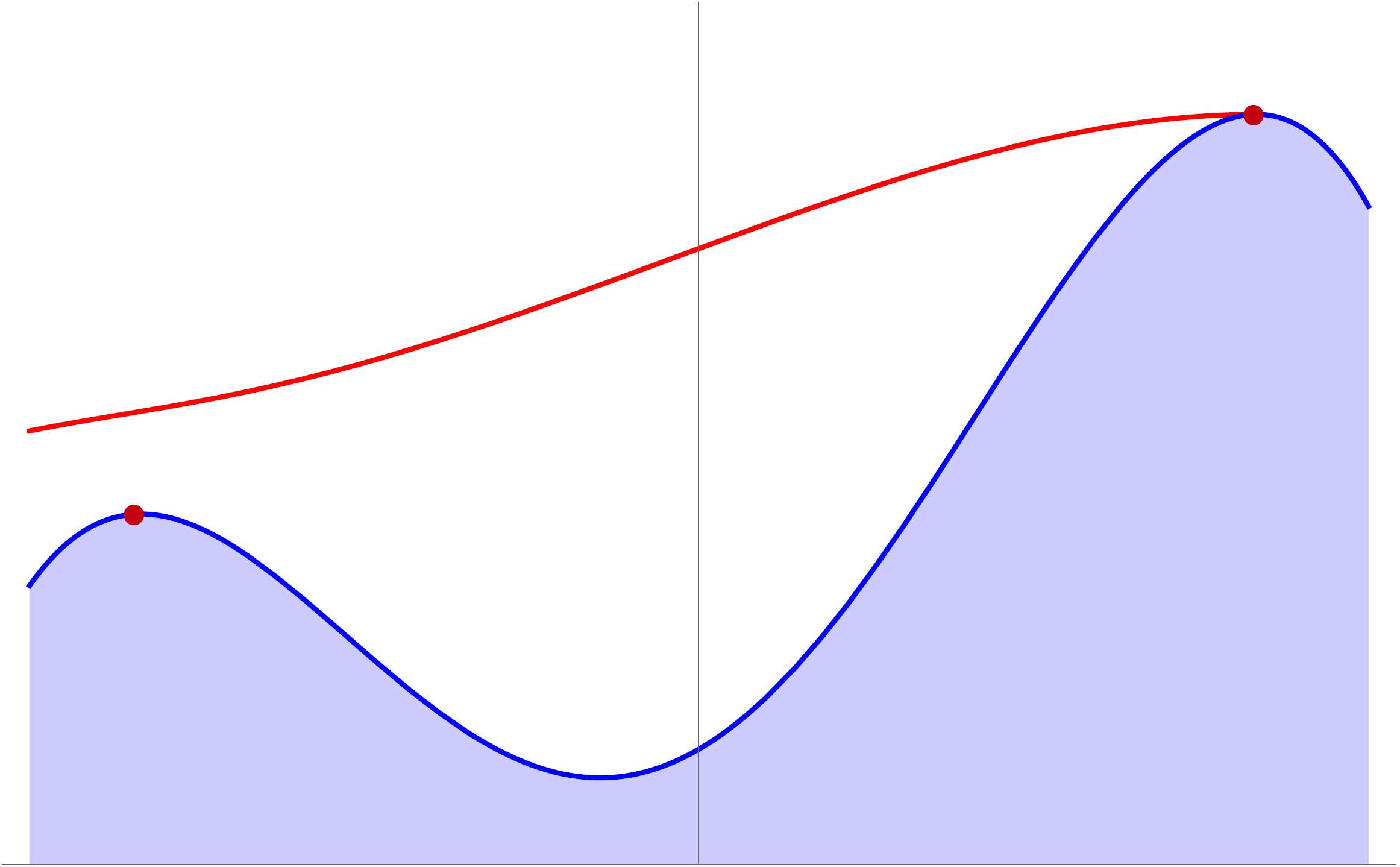}
\put(52,60){$W$}
\put(100,0){$\f$}
\put(9,21.5){$\f_f$}
\put(89,50){$\f_t$}
\put(31,13){$B(\f)$}
\put(30,43){$W_{flat}(\f)$}
\end{overpic}
\caption{If the  $W_{flat}(\f)$ solution from the true vacuum $\f_t$ does not touch  $B(\f)$ at $\f_f$, then there exists a curved $W_+$ solution.  }
\label{Wplusexist}
\end{figure}

Now consider $W_{flat}(\f)$ by which we denote the tentative flat domain-wall solution ending at $\f_t$ whose existence would suffice to prove the existence of CdL instantons in the same potential. In terms of $W$, the statement that the particle overshoots $\f_f$ is equivalent to the statement that $\f_f$ is not a critical point and hence $W_{flat}(\f_f)>B(\f_f)$. This is illustrated in fig.~\ref{Wplusexist}. The conditions for the existence of a CdL instanton are thus
\begin{equation}
W_{flat}(\f_t)= B(\f_t) \, , \quad W_{flat}(\f_f) > B(\f_f) \, .
\end{equation}
In the following, we will exploit that $W_{flat}$ satisfies \eqref{eq:EOM4}--\eqref{eq:EOM6} with $T=0$, which also implies that $S_{flat}=W_{flat}'$. Then we can write:
\begin{align}
& W_{flat}(\f_t)-W_{flat}(\f_f)< B(\f_t)-B(\f_f) \ , \label{con1} \\
& \Rightarrow \int_{\f_f}^{\f_t} d\f W_{flat}'(\f)< 2(d-1)\left( \frac{1}{\ell_t}-\frac{1}{\ell_f} \right) \ , \label{con2}\\
& \Rightarrow \int_{\f_f}^{\f_t} d\f \sqrt{\frac{d}{2(d-1)}W_{flat}^2+2 V}< 2(d-1)\left( \frac{1}{\ell_t}-\frac{1}{\ell_f} \right) \ . \label{con3}
\end{align}
where we have used $B(\f_t)=\frac{2(d-1)}{\ell_t}$, $B(\f_f)=\frac{2(d-1)}{\ell_f}$ and \eqref{eq:EOM5}.

This condition can be used as follows. We can compute $W_{flat}(\f)$ by integrating \eqref{eq:EOM5} starting from $\f_t$, where we implement the boundary condition $W_{flat}(\f_t) =\frac{2(d-1)}{\ell_t}$, all the way to $\f_f$. If inserted into \eqref{con3}, this condition is satisfied, the potential will admit a CdL instanton describing tunnelling from $\f_f$

\section{Flat domain-walls with a thin-wall limit: an alternative approach} \label{app:altthinwall}
In section \ref{sec:flatthinwall}, we constructed analytic flat domain-wall solutions that admitted a thin-wall limit in the sense described in \ref{sec:flatthinwall}. In particular, in the limit of an infinitely thin wall $\dot{\f} \sim \delta$, that is $\dot{\f}$ takes the form of a $\delta$-function. Here, we consider the an alternative case: In particular, here we demand that in the limit of an infinitely thin wall $\ddot{A} \sim \delta$.

The relevant equations of motion are again given in \eqref{c3}--\eqref{c5} with $R^{(\zeta)}=0$ for a flat domain-wall. As in section \ref{sec:flatthinwall} we  seek solutions that:
\begin{itemize}
\item interpolate between two AdS space-times with length scales $\ell_f$ and $\ell_t$, i.e.
\begin{align}
\label{eq:AdotBC} \dot{A}(u \rightarrow -\infty) = -\frac{1}{\ell_f} \, , \qquad \dot{A}(u \rightarrow +\infty) = -\frac{1}{\ell_t} \, ,
\end{align}
with $\ell_t < \ell_f$:
\item interpolate at the same time between the values $\f=\f_f$ and $\f=\f_t$.
\end{itemize}
As before, in the extreme thin-wall-limit we require:
\be
\dot A=\left\{ \begin{array}{lll}
\displaystyle -\frac{1}{\ell_{f}},&\phantom{aa} & u<u_0     ,\\ \\
\displaystyle -\frac{1}{\ell_{t}},&\phantom{aa}& u>u_0    .
\end{array}\right. \, , \quad
\f=\left\{ \begin{array}{lll}
\displaystyle \f_f,&\phantom{aa} & u<u_0     ,\\ \\
\displaystyle \f_t,&\phantom{aa}& u>u_0    .
\end{array}\right.
\label{appc5}
\ee
As stated at the beginning, here we satisfy the above by requiring that in the extreme thin-wall limit we observe $\ddot{A}(u) \sim \delta(u-u_0)$, while in the main text we instead required $\dot{\f}(u) \sim \delta(u-u_0)$.\footnote{Recall that simultaneously requiring $\ddot{A}(u) \sim \delta(u-u_0)$ and $\dot{\f}(u) \sim \delta(u-u_0)$ is not permitted by the equations of motion.}

We therefore propose the following ansatz for $\ddot{A}(u)$:
\begin{align}
\label{c6} \ddot{A}(u) = \frac{a_1}{2 \epsilon L} \frac{1}{\cosh^2 \frac{u-u_0}{\epsilon L}} \, ,
\end{align}
with $a_1$ a parameter to be fixed later by boundary conditions, $L$ is a fiducial length scale and $\epsilon$ another numerical parameter that can be used to realise the thin-wall limit by letting $\epsilon \rightarrow 0^+$. One can confirm that in this limit $\lim_{\epsilon \rightarrow 0+} \ddot{A} = \frac{a_1}{L} \delta(u-u_0)$ as demanded.

Integrating once we obtain:
\begin{align}
\label{c7} \dot{A} &= \frac{a_2}{2 L} + \frac{a_1}{2 L} \tanh \frac{u -u_0}{\epsilon L} \, ,
\end{align}
where $a_2$ is another integration constant. Implementing the boundary conditions \eqref{eq:AdotBC} one then finds:
\begin{align}
\label{c9} a_1 = \frac{L}{\ell_f} - \frac{L}{\ell_t} \, , \qquad  a_2 = -\frac{L}{\ell_f} - \frac{L}{\ell_t} \, .
\end{align}
Integrating $\dot{A}$ once more we obtain
\begin{align}
\label{c10} A(u) = \bar{A} + \frac{a_2}{2} \frac{u}{L} + \frac{a_1 \epsilon}{2}  \ln \Big( \cosh \frac{u -u_0}{\epsilon L} \Big) \, ,
\end{align}
with $\bar{A}$ an integration constant.

We now turn to $\f(u)$. Using the equation of motion \eqref{c3} we can then obtain an expression for $\dot{\f}$, given by
\begin{align}
\label{c11} \dot{\f} = \frac{1}{L} \, \sqrt{-\frac{(d-1)a_1}{\epsilon}} \, \frac{1}{\cosh \frac{u -u_0}{\epsilon L}} \, .
\end{align}
Integrating, this gives
\begin{align}
\label{c12} \f(u) = \f_0 + \sqrt{-(d-1)a_1 \epsilon} \, \arctan \Big( \sinh \frac{u -u_0}{\epsilon L} \Big) \, .
\end{align}

Finally, from \eqref{c4} we obtain an expression for $V$. We can write this as a function of $\f$ by inverting \eqref{c12}. The result is
\begin{align}
\label{c13} V = - \frac{d(d-1) a_2^2}{4 L^2} - \frac{d(d-1) a_1 a_2}{2 L^2} \, \sin \chi - \frac{d(d-1) a_1^2}{4 L^2} \, \sin^2 \chi - \frac{(d-1) a_1}{2 \epsilon L^2} \, \cos^2 \chi \, ,
\end{align}
where we defined
\begin{align}
\label{c14} \chi \equiv \frac{\f - \f_f}{\sqrt{-(d-1) a_1 \epsilon}} - \frac{\pi}{2} \, .
\end{align}
This is a potential with two minima at $\chi = \pm \tfrac{\pi}{2}$ corresponding to $\f=\f_f$ and $\f=\f_t$, respectively, separated by a barrier in-between. In the limit $\epsilon \rightarrow 0$ (with all other parameters fixed) this barrier diverges in height in virtue of the last term in \eqref{c13}. At the same time the two minima at $\f_i$ and $\f_f$ approach one another.


\section{Comparison with thin-wall approximation of Coleman-de Luccia}
\label{app:CDLcompare}
In \cite{CdL} Coleman and de Luccia consider a potential with one true and one false minimum (which for consistency we label again as $\f_f$ and $\f_t$), which can be written as
\begin{align}
\label{eq:CDLPot} U(\f) = U_0(\f) + \epsilon u(\f) \, .
\end{align}
Here $U_0$ is a potential with two \emph{degenerate} minima at $\f_f$ and $\f_t$. The separation in energy into false and true minimum is then achieved by adding $\epsilon u(\f)$ where $\epsilon$ is defined as the potential difference between true and false minimum:
\begin{align}
\epsilon \equiv U(\f_f)-U(\f_t) \, .
\end{align}
According to Coleman and de Luccia the thin-wall approximation applies ``in the limit of a small energy difference between the two vacuums,'' that is as long as $\epsilon$ is small. This statement is problematic, as $\epsilon$ is a dimensionful parameter and can hence only be small compared to another quantity sharing its dimensions. Here we assume that $\epsilon$ is small compared to the barrier separating the minima at $\f_f$ and $\f_t$. To summarise, according to Coleman and de Luccia the thin-wall approximation applies if the potential can be split into a part with degenerate minima and a correction that splits the degeneracy, but is suppressed compared to the degenerate part. The parameter that establishes the hierarchy between the two parts is the ratio between the energy difference of the two minima and the barrier height.

In this work we studied CdL tunnelling in the sextic potential $V$ defined in \eqref{eq:Vnum}. In section \ref{sec:ODthinwall} we then showed that potentials of this type admit thin-walled CdL bubble solutions in the limit $\Delta \rightarrow \infty$, $\Delta \tfrac{\ell_t}{\ell_f} \rightarrow \infty$. In the following, we show that in this thin-wall limit the potential $V$ from \eqref{eq:Vnum} can indeed be brought into the form \eqref{eq:CDLPot}. To split the potential into a part with degenerate minima and a correction, we separate it its symmetric and anti-symmetric parts about $\f=\tfrac{\f_f+\f_t}{2}$, respectively. In particular, we define
\begin{align}
\label{eq:U0def} U_0(\f) &= \frac{1}{2} \bigg(V(\f) + V(\f_f+\f_t-\f) \bigg) + \frac{1}{2} \bigg(V(\f_f) - V(\f_t) \bigg) \, , \\
\label{eq:epsudef} \epsilon u(\f) &= \frac{1}{2} \bigg(V(\f) - V(\f_f+\f_t-\f) \bigg) - \frac{1}{2} \bigg(V(\f_f) - V(\f_t) \bigg) \, .
\end{align}
The constant shifts are included to ensure that $U_0(\f_f)=U_0(\f_t)=V(\f_f)$. Using the explicit expression for $V(\f)$ in \eqref{eq:Vnum} with the help of (\ref{eq:V0num},\ref{eq:vdef}) we can then compute $U_0(\f)$ and $\epsilon u(\f)$ in the limit $\Delta \rightarrow \infty$, $\Delta \tfrac{\ell_t}{\ell_f} \rightarrow \infty$. Using \eqref{eq:Vparameterconstraint} to eliminate $\f_t$ we find, setting again $\f_f=0$ for convenience:
\begin{align}
U_0(\f) = \frac{1}{\ell_f^2} \bigg[ &- d(d-1) + a_2 \, \Delta^2 \Big(1 + \mathcal{O}\big( \tfrac{1}{\Delta} \big) \Big) \f^2 + a_3 \, \Delta^{5/2} \Big(1 + \mathcal{O}\big( \tfrac{1}{\Delta} \big) \Big) \f^3 \\
\nonumber & + a_4 \, \Delta^3 \Big(1 + \mathcal{O}\big( \tfrac{1}{\Delta} \big) \Big) \f^4 + a_5 \, \Delta^{5/2} \Big(1 + \mathcal{O}\big( \tfrac{1}{\Delta} \big) \Big) \f^5 + a_6 \, \Delta^3 \Big(1 + \mathcal{O}\big( \tfrac{1}{\Delta} \big) \Big) \f^6 \bigg] \, , \\
\epsilon u(\f) = \frac{1}{\ell_f^2} \bigg[ & b_2 \, \Delta \Big(1 + \mathcal{O}\big( \tfrac{1}{\Delta} \big) \Big) \f^2 + b_3 \, \Delta^{3/2} \Big(1 + \mathcal{O}\big( \tfrac{1}{\Delta} \big) \Big) \f^3 \bigg] \, ,
\end{align}
where $a_{2,3,4,5,6}$ and $b_{2,3}$ are coefficients depending on $\ell_t/\ell_f$ and $v_0$. The observation now is that the potential $\epsilon u$ is suppressed by one power of $\Delta$ compared to $U_0$. That is, the monomial $\f^2$ in $\epsilon u$ comes with a power of $\Delta$, while in $U_0$ the same monomial is premultiplied by $\Delta^2$. Similarly, the monomial $\f^3$ in $\epsilon u$ is proportional to $\Delta^{3/2}$, while in $U_0$ the cubic term has a prefactor $\Delta^{5/2}$. Recall from section \ref{sec:ODthinwall} that in the thin-wall approximation, the ratio of the energy difference between the minima and the barrier height scales as $\sim 1/ \Delta$, see \eqref{eq:barrierdifferenceratio}. Therefore, $1/\Delta$ behaves exactly as the small parameter in the thin-wall approximation of Coleman and de Luccia. We consequently conclude, that in the thin-wall limit ($\Delta \rightarrow \infty$), the sextic potential $V$ in \eqref{eq:Vnum} can indeed be brought into the schematic form \eqref{eq:CDLPot} proposed by Coleman and de Luccia.

\end{appendix}

\bibliography{ref}
\bibliographystyle{JHEP}

\end{document}